\documentclass[11pt, a4paper]{article}
\usepackage[T1]{fontenc}
\usepackage{a4wide, url}
\usepackage[english]{babel}
\usepackage{graphicx}
\usepackage{array}
\usepackage{natbib}
\usepackage[textfont=sc,labelfont=bf]{caption}
\usepackage{setspace}
\usepackage{multirow}
\usepackage{amsthm}
\usepackage{amsmath}
\usepackage{amsfonts}
\usepackage{amssymb}
\usepackage{bigints}
\usepackage{pgf} 
\usepackage{bm}
\usepackage{paralist}
\usepackage{hyperref}
\usepackage{pdflscape}
\usepackage{xfrac}
\usepackage[linesnumbered,ruled,vlined]{algorithm2e}
\SetKwInput{KwInput}{Input}                
\SetKwInput{KwOutput}{Output}              
\usepackage{enumitem,kantlipsum}
\usepackage{xr}
\usepackage{yfonts}

\usepackage{xcolor}
\hypersetup{
    colorlinks,
    linkcolor={red!50!black},
    citecolor={blue!50!black},
    urlcolor={blue!80!black}
}

\usepackage{comment}
\excludecomment{ext}\excludecomment{comment}

\usepackage[top=1in,bottom=1in,left=.75in,right=.75in]{geometry}

\usepackage{tabu}

\newcolumntype{L}[1]{>{\raggedright\let\newline\\\arraybackslash\hspace{0pt}}m{#1}}
\newcolumntype{C}[1]{>{\centering\let\newline\\\arraybackslash\hspace{0pt}}m{#1}}
\newcolumntype{R}[1]{>{\raggedleft\let\newline\\\arraybackslash\hspace{0pt}}m{#1}}

\DeclareMathAlphabet\mathbfcal{OMS}{cmsy}{b}{n}

\newcommand{\mbf}{\mathbf}
\newcommand{\beq}{\begin{equation}}
\newcommand{\eeq}{\end{equation}}
\newcommand{\bea}{\begin{eqnarray}}
\newcommand{\eea}{\end{eqnarray}}
\newcommand{\ba}{\begin{array}}
\newcommand{\ea}{\end{array}}
\newcommand{\bit}{\begin{itemize}}
\newcommand{\eit}{\end{itemize}}
\newcommand{\ben}{\begin{enumerate}} 
\newcommand{\een}{\end{enumerate}}
\newcommand{\bpm}{\begin{pmatrix}}
\newcommand{\epm}{\end{pmatrix}}
\newcommand{\bbm}{\begin{bmatrix}}
\newcommand{\ebm}{\end{bmatrix}}

\renewcommand{\l}{\left}
\renewcommand{\r}{\right}

\newcommand{\E}[0]{\mathbb{E}}

\newcommand{\nn}{\nonumber}
\newcommand{\wh}{\widehat}
\newcommand{\wt}{\widetilde}

\newtheorem{ass}{Assumption}

\newtheorem{prop}{Proposition}
\newtheorem{rem}{Remark}
\newtheorem{lem}{Lemma}

\graphicspath{{Figures/Submission/}}
\title{\textsc{\large 	
Quasi Maximum Likelihood Estimation of High-Dimensional Factor Models: \\
A Critical Review
}}
\date{ }

\begin{document}
\maketitle

\begin{center}\vspace{-1.5cm}
 Matteo Barigozzi$^\dag$ \\[.1cm]

\small This version: \today
\end{center}

\begin{abstract}

We review Quasi Maximum Likelihood estimation of  factor models for high-dimensional panels of time series.
We consider two cases: (1) estimation when no dynamic model for the factors is specified \citep{baili12,baili16}; (2) estimation based on the Kalman smoother and the Expectation Maximization algorithm thus allowing to model explicitly the factor dynamics \citep{DGRqml,BLqml}. 
Our interest is in approximate factor models, i.e., when we allow for the idiosyncratic components to be mildly cross-sectionally, as well as serially, correlated. Although such setting apparently makes estimation harder, we show, in fact, that factor models do not suffer of the {\it curse of dimensionality} problem, but instead they enjoy a  {\it blessing of dimensionality} property. In particular, given an approximate factor structure, if the cross-sectional dimension of the data, $N$, grows to infinity, we show that: (i) identification of the model  is still possible, (ii) the mis-specification error due to the use of an exact factor model log-likelihood vanishes. Moreover, if we let also the sample size, $T$, grow to infinity, we can also consistently estimate all parameters of the model and make inference. The same is true for estimation of the latent factors which can be carried out by weighted least-squares, linear projection, or Kalman filtering/smoothing. 
We also compare the approaches presented with: Principal Component analysis and the classical, fixed $N$, exact Maximum Likelihood approach. We conclude with a discussion on efficiency of the considered estimators.


\vspace{0.5cm}

\noindent \textit{Keywords:} 
Approximate Dynamic Factor Model; Maximum Likelihood Estimation; Weighted Least Squares; Expectation Maximization Algorithm; Kalman smoother.

\end{abstract}

\renewcommand{\thefootnote}{$\dag$} 
\thispagestyle{empty}

\footnotetext{Department of Economics - Universit\`a di Bologna. Email: matteo.barigozzi@unibo.it \\
Sections \ref{sec:static} and \ref{sec:dynamic} heavily draw on  \citet{BLqml} and  \citet{MBQMLPCA} to which the reader is referred for the original results under a more general setting, and for their proof.\\
\\
I thank Matteo Luciani and Rolf Sundberg for helpful comments.\\
\\
A shorter version of this paper is published as: M. Barigozzi, ``Quasi Maximum Likelihood Estimation of High-Dimensional Factor Models'', in Oxford Encyclopedia of Economics and Finance, Oxford University Press, 2024.
} 

\renewcommand{\thefootnote}{\arabic{footnote}}

\newpage
\tableofcontents

\newpage


\section{Introduction}\label{sec:intro}

Factor analysis is one of the earliest proposed multivariate statistical techniques: it dates back to the studies of \citet{spearman04} in experimental psychology. The idea is to decompose a vector of $N$ observed random variables into two components: (i) a common component driven by $r<N$ latent factors, and (ii) a component which is idiosyncratic, in the sense that it can be due either to measurement errors or to specific features of a single, or a sub-group of variables. As such we can retrospectively consider factor analysis as a pioneering technique in the filed of unsupervised statistical learning.

Over the years, two main strands of applications have emerged: first, in psychometrics in a low-dimensional setting \citep{thomson36,bartlett37,bartlett38,lawley1940,thomson51,joreskog69,lawleymaxwell71,moustaki2000generalized,joreskog2001factor,bartholomew2011latent}, 
second, in econometrics, both in a low-dimensional \citep{SS77,GS81,watsonengle83,stockwatson89,stockwatson91}, and in a high-dimensional setting \citep{chamberlainrothschild83,FHLR00,stockwatson02JASA,bai03}.
Applications in econometrics include:
the analysis of financial markets \citep{connor2006common,ait2017using,kim2019factor,barigozzi2020generalized}, 
the measurement and prediction of macroeconomic aggregates \citep{stockwatson02JBES,de2008forecasting,Nowcasting,OGAP}, 
the study of the dynamic effects of unexpected shocks to the economy \citep{BBE05,FGLR09,fornigambettiJME,BCL,FGLS14,smokinggun,marcellino16,BLL2},
and the analysis of demand systems \citep{stone1945analysis,barigozzi2016identifying}. 
An historical overview is in \citet{BH23}.

The former strand of literature deals with random samples of data, i.e., independent observations for each variable, and estimation is almost exclusively by means of Maximum Likelihood \citep{AR56,AFP87,AA88,tippingbishop99,baili12}. The latter strand of literature deals with time series data, i.e., dependent observations for each variable, and estimation is either by means of Principal Component analysis \citep{FHLR00,stockwatson02JASA,bai03,FLM13} or by means of Quasi Maximum Likelihood \citep{quahsargent93,DGRqml,baili16}. Principal Components and Quasi Maximum Likelihood estimation are the most popular estimation techniques, dating back at least to \citet{hotelling1933analysis} and \citet{lawley1940}, respectively. A third, less popular, approach is the centroid method by \citet{thomson51}.

Our focus is on high-dimensional factor models for time series. Specifically, 
we allow for the dimension of the vector, $N$, to grow to infinity. As a consequence, it is realistic to consider an {\it approximate} factor model where we allow for the idiosyncratic components to be cross-sectionally correlated, rather than an {\it exact} factor model where the idiosyncratic components are not correlated. Furthermore, since we deal with time series, both the factors and the idiosyncratic components are allowed to be autocorrelated, and, if we specify also a model describing the time evolution of the factors, e.g., a VAR, we speak of a {\it dynamic} factor model, while, if we do not, we speak of a {\it static} factor model. The approximate dynamic factor model considered in this paper is a restricted version of the more general model proposed by \citet{scherrer1998structure} and \citet{FHLR00}, which is, in fact,  a representation of an infinite dimensional panel of time series as derived by \citet{fornilippi01} and \citet{hallinlippi13}.

The idea of an approximate factor structure dates back to \citet{chamberlain83} and  \citet{chamberlainrothschild83} who showed that, if we let the number of time series, $N$, to grow, then we are able to identify the common components even when allowing for idiosyncratic correlations. Intuitively, since estimation is based on cross-sectional aggregation, the more series we use, i.e., the more we increase $N$, the more the signal, factor-driven, component emerges from the noisy idiosyncratic component because of an assumed cross-sectional weak Law of Large Numbers. Equivalently, this is seen by noticing that, under a factor model, the gap between the largest $r$ eigenvalues of the covariance matrix of the data and the remaining $N-r$ widens  as $N$ grows. As a consequence of this phenomenon, an approximate factor model does not suffer of the {\it curse of dimensionality} problem, which typically affects high-dimensional parametric models, but  it enjoys instead a  {\it blessing of dimensionality} property. 

The idea of a dynamic factor model goes back to \citet{geweke1977dynamic} and \citet{SS77}, who account for the possibility that the $N$ variables co-move not only contemporaneously but also with some leads or lags, in other words, they can load the factors also with lags. In particular, the setting considered here is the same as the one formulated by \citet{FGLR09}, which consists in writing the model as a linear state-space system, having as latent states the factors. 

In this paper, we review likelihood-based methods for estimating high-dimensional approximate static and dynamic factor models, as considered by \citet{baili16} and \citet{DGRqml}, respectively. On the one hand, \citet{baili16} treat the factors as non-stochastic and follow the classical approach based on first estimating the loadings and the idiosyncratic variances by Quasi Maximum Likelihood and then estimating the factors by Weighted Least Squares. On the other hand, \citet{DGRqml} explicitly model the factors as an autocorrelated stochastic process and propose to use the Expectation Maximization (EM) algorithm together with the Kalman smoother to obtain joint estimates of the loadings and the idiosyncratic variances, which approximate the Quasi Maximum Likelihood estimators, and of the factors. Both approaches are based on the maximization of a mis-specified log-likelihood where we do not take into account any of the idiosyncratic cross- and serial correlations, nevertheless, we show that as $N\to\infty$, the mis-specification error vanishes: yet another instance of the blessing of dimensionality.  A summary of the rates of convergence of the loadings and factors estimators in approximate and exact factor models is in Table \ref{tab:lit}.

\begin{table}[t!]
\caption{Rates of consistency of estimators of factor models}\label{tab:lit}
\centering
\scriptsize
\vskip .2cm

\begin{tabular}{cc|c|c}
\hline
\hline
&\\[-4pt]
&&\multicolumn{2}{c}{Exact static factor model ($\bm\Omega^F=\mbf I_{T}\otimes \bm\Gamma^F$, $\bm\Omega^\xi=\mbf I_{T}\otimes \bm\Sigma^\xi$) }\\[3pt]
\cline{3-4}
&&&\\[-4pt]
&&Loadings ($\bm\lambda_i$) & Factors ($\mbf F_t$)\\
[3pt]
\hline
&&&\\[-4pt]
$T\to\infty$	& $n$ finite	&	$\sqrt T$			& 	-\\
			&			&	\citet[Theorem 2F]{AFP87}, QML	&	\\[3pt]
\hline
&&&\\[-4pt]
$T\to\infty$	& $n\to\infty$	&	$\sqrt T$			& 	$\min(\sqrt n,T)$\\
			&			&	\citet[Theorem 5.2]{baili12}, QML	&	\citet[Theorem 6.1]{baili12}, WLS, LP\\[3pt]
\hline
\hline
&\\[-4pt]
&&\multicolumn{2}{c}{Approximate static factor model ($\bm\Omega^F=\mbf I_{T}\otimes \bm\Gamma^F$, $\bm\Omega^\xi$ unrestricted)}\\[3pt]
\cline{3-4}
&&&\\[-4pt]
&&Loadings ($\bm\lambda_i$) & Factors ($\mbf F_t$)\\
[3pt]
\hline
&&&\\[-4pt]
$T\to\infty$	& $n\to\infty$	&	$\min(n,\sqrt T)$				& 	$\min(\sqrt n,T)$\\
			&			&	\citet[Theorem 2]{bai03}, PC					&	\citet[Theorem 1]{bai03}, PC\\
			&			&	\citet[Theorem 1]{baili16}, QML				&	\citet[Theorem 2]{baili16}, WLS, LP\\[3pt]
\hline
&&&\\[-4pt]
$T$ finite	& $n\to\infty$	&	$-$			& 	$\sqrt n$\\
			&			&		&	\citet[Theorem 1]{zaff19}, PC\\[3pt]
\hline
\hline
&\\[-4pt]
&&\multicolumn{2}{c}{Approximate dynamic factor model ($\bm\Omega^F$, $\bm\Omega^\xi$ unrestricted)}\\[3pt]
\cline{3-4}
&&&\\[-4pt]
&&Loadings ($\bm\lambda_i$) & Factors ($\mbf F_t$)\\
[3pt]
\hline
&&&\\[-4pt]
$T\to\infty$	& $n\to\infty$	&	$\min(n/\sqrt{\log T},\sqrt {T})$		& 	$\min(\sqrt n,T/\sqrt{\log n})$\\
			&			&	\citet[Proposition 2]{BLqml}, EM	&	\citet[Proposition 3]{BLqml}, KS\\
			&			&								&	$\min(\sqrt {n/\log n},T^{1/4})$\\
			&			&								&	\citet[Proposition 1]{DGRqml}, KS\\[3pt]
\hline
\hline
\\[-4pt]
\end{tabular}

\begin{tabular}{p{.95\textwidth}}
QML: Quasi Maximum Likelihood; PC: Principal Components; WLS: Weighted Least Squares; LP: Linear Projection; EM: Expectation Maximization Algorithm; KS: Kalman smoother. Here $\bm\Gamma^F$ and $\bm\Sigma^\xi$  are defined in Section \ref{sec:ass}
and $\bm\Omega^F$ and $\bm\Omega^\xi$ are defined in Section \ref{sec:LLL}. In all cases the factors are estimated using the estimator of the loadings on the same row.
\end{tabular}

\end{table}

\section{The approximate factor model}\label{sec:model}

Let us consider the $T\times N$ double-array $\bm X$ having elements $\{x_{it}\in ,\, i=1,\ldots, N,\, t=1,\ldots,T\}$ of $N$ observed times series over $T$ time periods. We say that $\bm X$ follows an $r$-factor model if
\begin{align}
x_{it}&=\alpha_i+\bm\lambda_i^\prime \mbf F_t + \xi_{it}, \quad i=1,\ldots, N,\quad t=1,\ldots,T. \label{eq:SDFM1R}
\end{align}
where $\bm\lambda_i:=(\lambda_{i1}\cdots\lambda_{ir})^\prime$ and $\mbf F_t:=(F_{1t}\cdots F_{rt})^\prime$ are the $r$-dimensional latent vectors of loadings for series $i$ and factors, respectively. We call $\xi_{it}$ the idiosyncratic component and 
$\chi_{it}:=\bm \lambda_i^\prime \mbf F_t$ the common component.

It is convenient to write model \eqref{eq:SDFM1R}, which is written is scalar notation, also in other  equivalent ways which we list below. 

Let $\mbf x_t:=(x_{1t}\cdots x_{Nt})^\prime$ be the $N$-dimensional vector of observables at time $t$. Then, in vector notation  \eqref{eq:SDFM1R} reads
\[
\mbf x_t =\bm\alpha+ \bm\Lambda\mbf F_t +\bm \xi_t,\quad t=1\ldots, T,
\]
where $\bm\xi_{t}:=(\xi_{1t}\cdots\xi_{Nt})^\prime$  is the corresponding $N$-dimensional vector of idiosyncratic components,  $\bm\Lambda:=(\bm\lambda_1\cdots\bm\lambda_N)^\prime=(\bm\ell_1\cdots \bm\ell_r)$ is the $n\times r$ matrix of factor loadings with rows $\bm\lambda_i^\prime$, $i=1,\ldots, N$ and columns $\bm\ell_j$, $j=1,\ldots ,r$, and $\bm\alpha:=(\alpha_1\cdots \alpha_N)^\prime$. We call $\bm\chi_t:=\bm\Lambda\mbf F_t$ the vector of common components. 

Equivalently, let $\bm x_i:=(x_{i1}\cdots x_{iT})^\prime$ be the $T$-dimensional time series vector of the $i$th observable, in vector notation  \eqref{eq:SDFM1R} also reads
\[
\bm x_i = \alpha_i\bm\iota_T+\bm F\bm\lambda_i + \bm\zeta_i, \quad i=1,\ldots, N,
\]
where $\bm\zeta_i:=(\xi_{i1}\cdots \xi_{iT})^\prime$ is the corresponding $T$-dimensional time series vector of the $i$th idiosyncratic component, $\bm F:=(\mbf F_1\cdots\mbf F_T)^\prime$ is the $T\times r$ matrix of factors, and $\bm\iota_T$ is a $T$-dimensional column vector of ones.

Finally, given that $\bm X=(\mbf x_1\cdots\mbf x_T)^\prime$, in matrix notation \eqref{eq:SDFM1R} reads
\beq
\bm X = \bm\iota_T\bm\alpha^\prime + \bm F\bm\Lambda^\prime +\bm\Xi,\label{eq:SDFM1Rmat}
\eeq
where $\bm \Xi:=(\bm\xi_1\cdots\bm \xi_T)^\prime$ is the $T\times N$ matrix of idiosyncratic components.

\subsection{Structural assumptions}\label{sec:ass}
In this section we provide the minimal set of assumptions required to derive the results given in the rest of the paper. Notice that, although these assumptions do not coincide with those made by \citet{bai03}, \citet{baili12}, and \citet{baili16}, they nest all assumptions made in those papers. More precisely, they nest all necessary assumptions in those papers, since, in fact, some of the assumptions made therein turn out  to be redundant. In the following, we compare our assumptions only with \citet{baili12,baili16}, while for a comparison with \citet{bai03} we refer to \citet{MBPCA}.


We characterize the common component by means of the following assumption.

\begin{ass}[\textsc{common component}]\label{ass:common} $\,$
\begin{compactenum}[(a)]
	\item  $\lim_{N\to\infty}\Vert N^{-1}\bm\Lambda^\prime\bm\Lambda-\bm\Sigma_{\Lambda}\Vert=0$, where $\bm\Sigma_{\Lambda}$ is $r\times r$ positive definite, and, for all $i=1,\ldots, N$ and all $N\in\mathbb N$, $\Vert\bm\lambda_i\Vert\le M_\Lambda$ for some finite positive real $M_\Lambda$ independent of $i$.  
\item For all $t=1,\ldots, T$ and all $T\in\mathbb N$, $\E[\mbf F_{t}]=\mbf 0_r$, $\bm\Gamma^F:=\E[\mbf F_t\mbf F_t^\prime]$ is $r\times r$ positive definite, $\Vert\bm\Gamma^F\Vert\le M_F$ and $\E[\Vert \mbf F_t\Vert^{4}]\le K_F$, for some finite positive reals $M_F$ and $K_F$ independent of $t$.

	\item $\mathrm p\text{-}\!\lim_{T\to\infty}\l\Vert T^{-1}\bm F^\prime\bm F -\bm\Gamma^F\r\Vert=0$.
	

	\item There exists a positive integer $N_0$ such that for all $N\in\mathbb N$ with $N> N_0$, $r$ is a finite positive integer, independent of $N$. 

	\item  For all $i=1,\ldots, N$ and all $N\in\mathbb N$, $\alpha_i\in\mathbb R$.
\end{compactenum}
\end{ass}

%
Parts (a) and (b), are also assumed in \citet[Assumptions A and B]{baili16}, and are similar to what assumed by \citet[Assumptions A and C]{baili12}.
They imply pervasiveness of the factors, i.e., all factors have a non-negligible effect on all observables. Indeed, the loadings matrix has asymptotically maximum column rank $r$ (part (a)),  the factors have a finite full-rank covariance matrix (part (b)), and, because of the upper bound on $\Vert\bm\lambda_i\Vert$ in part (a), for any given $N\in\mathbb N$, 
the contribution of each factor to every series is finite.

Part (c) 
is analogous to what assumed by \citet[Assumption A]{baili12,baili16} in the case of deterministic factors. It is very general, it simply says that the sample covariance matrix of the factors is a consistent estimator of its population counterpart $\bm\Gamma^F$. It is satisfied by linear processes having innovations with finite 4th order moments, or strong mixing processes with finite 4th order moments, or, more in general, by processes having summable 4th order cross-cumulants, which is a necessary and sufficient condition for consistency of the sample covariance matrix (\citealp[pp.209-211]{hannan}). 

Part (d) implies the existence of a finite of factors for all $N,T\in\mathbb N$. In particular, the numbers of common factors, $r$, is identified only for $N\to\infty$. Here $N_0$ is the minimum number of series we need to be able to identify $r$ so that it must be such that $N_0\ge r$. Without loss of generality hereafter when we say ``for all $N\in\mathbb N$'' we always mean that $N>N_0$ so that $r$ can be identified. In practice, we must always work with $N$ such that $r<\min(N,T)$. 

Last, part (e), assumes a $\alpha_i$ to be a fixed idiosyncratic effect. Sometimes \eqref{eq:SDFM1R}  is also  written as \citep[see the remarks to Assumption A]{baili12}
$x_{it}=\alpha_i+\bm\lambda_i^\prime \bar{\mbf F}+\bm\lambda_i^\prime(\mbf F_t-\bar{\mbf F})+\mbf \xi_{it} = \alpha_i^*+\bm\lambda_i^\prime \mbf F_t^*+\xi_{it}$, 
with $\bar {\mbf F}:=T^{-1}\sum_{t=1}^T \mbf F_t$, $\mbf F_t^* := \mbf F_t-\bar{\mbf F}$, and $\alpha_i^*:=\alpha_i+\bm\lambda_i^\prime \bar{\mbf F}$ which is then a random idiosyncratic effect.

Now, because of Assumption \ref{ass:common}(a), we have that 
the eigenvalues of $\bm\Sigma_\Lambda$, denoted as $\mu_j(\bm\Sigma_\Lambda)$, $j=1,\ldots, r$ are such that $m_\Lambda^2\le \mu_j(\bm\Sigma_\Lambda)\le M_\Lambda^2$.
Similarly, because of Assumptions \ref{ass:common}(b) and \ref{ass:common}(c), 
we have that 
 the eigenvalues of $\bm\Gamma^F$, denoted as $\mu_j(\bm\Gamma^F)$, $j=1,\ldots, r$ are such that $m_F\le \mu_j(\bm\Gamma^F)\le M_F$.

Hereafter, denote the covariance matrix of the common component as
$$
\bm\Gamma^\chi:=\E[\bm\chi_t\bm\chi_t^\prime]=\bm\Lambda\bm\Gamma^F \bm\Lambda^\prime,
$$ 
with  $j$th largest eigenvalue $\mu_j^\chi$, $j=1,\ldots, r$. Then, it follows that
(\citealp{MBPCA})
for all $j=1,\ldots,r$, 
\beq\label{eq:lindiv}
\underline C_j:=m_\Lambda^2m_F\le \lim\inf_{N\to\infty} \frac{ \mu_{j}^\chi}N\le\lim\sup_{N\to\infty} \frac{\mu_{j}^\chi}N\le M_\Lambda^2M_F=:\overline{C}_j.
\eeq
This means we consider only strong factors. For weak factors see, e.g., 
\citet{onatski2012asymptotics}, \citet{uematsu2021inference}, \citet{bai2021approximate}, and \citet{freyaldenhoven2022factor}, among many others.

To characterize the idiosyncratic component, we make the following assumptions.

\begin{ass}[\textsc{idiosyncratic component}]\label{ass:idio}
  $\,$
\begin{compactenum}[(a)]
	\item For all $i=1,\ldots, N$, all $t=1,\ldots, T$, and all $N,T\in\mathbb N$, $\E[\xi_{it}]=0$ and $\sigma_i^2:=\E[\xi_{it}^2]\in[ C_\xi,M_\xi]$ for some finite positive reals $C_\xi$ and $M_\xi$ independent of  $i$, $t$, $N$, and $T$.

	\item For all $i,j=1,\ldots, N$, all $t=1,\ldots T$, all $N,T\in\mathbb N$, and all $k\in\mathbb Z$, $\vert \E[\xi_{it}\xi_{j,t-k}]\vert\le \rho^{\vert k\vert} M_{ij}$ for some finite positive reals $\rho$ and $M_{ij}$ independent of $t$  and such that $0\le \rho <1$, $M_{ii}\le M_\xi$, 
	$\sum_{j=1, j\ne i}^N M_{ij}\le M_\xi$, and $\sum_{i=1, i\ne j}^N M_{ij}\le M_\xi$ for some finite positive real $M_{\xi}$ independent of  $i$, $j$, and $N$. 

\item For all $i=1,\ldots, N$, all $t=1,\ldots, T$, and all $N,T\in\mathbb N$, $\E[\vert \xi_{it}\vert^{4}]\le K_\xi$ for some finite positive real $K_\xi$ independent of $i$ and $t$.
\item For all $j=1,\ldots, N$, all $s=1,\ldots, T$, and all $N,T\in\mathbb N$,
\[
\max\l\{\E\l[\l\vert\frac 1{\sqrt{NT}} \sum_{i=1}^N\sum_{t=1}^T\l\{\xi_{it}\xi_{jt}-\E[\xi_{it}\xi_{jt}]\r\} \r\vert^2\r],\E\l[\l\vert\frac 1{\sqrt{NT}}\sum_{i=1}^N\sum_{t=1}^T\l\{\xi_{it}\xi_{is}-\E[\xi_{it}\xi_{is}] \r\}\r\vert^2\r]\r\}\le K_\xi
\]
for some finite positive real $K_\xi$ independent of $j$, $s$, $N$, and $T$.
 \end{compactenum}
 \end{ass}


Part (a), is a minimum requirement for having non-degenerate idiosyncratic components. Although, in principle, the assumption on the lower bound for $\sigma_i^2$  could be removed  \citep[Chatper 3.4]{bartholomew2011latent}, here it is maintained for simplicity, so that the log-likelihood is always well-defined. 
Moreover, from part (a), jointly with Assumption \ref{ass:common}(b)  by which $\E[\chi_{it}]=0$, we see that we are implicitly assuming that the observables are such that $\E[x_{it}]=\alpha_i$, $i=1,\ldots, N$.

Part (b) has a twofold purpose. First, it limits the degree of serial correlation of the idiosyncratic components. Second, it also limits the degree of cross-sectional correlation between idiosyncratic components. In particular, part (b) implies the conditions required by 
\citet[Assumption C3, C4, and E1]{baili16} (see Lemma 1(i)-1(iii) in \citealp{MBPCA}), i.e., 
\begin{align}
&\sup_{N,T\in\mathbb N}\frac 1{NT}\sum_{i,j=1}^N\sum_{t,s=1}^T \vert\E_{}[\xi_{it}\xi_{js}]\vert \le \frac{2M_\xi(1+\rho)}{1-\rho},\nn\\
&\sup_{N,T\in\mathbb N}\max_{t=1,\ldots, T}\frac 1{N}\sum_{i,j=1}^N \vert\E_{}[\xi_{it}\xi_{jt}]\vert \le 2M_{\xi},\label{eq:bai}\\
&\sup_{N,T\in\mathbb N}\max_{i=1,\ldots, N}\frac 1{T}\sum_{t,s=1}^T \vert\E_{}[\xi_{it}\xi_{is}]\vert \le \frac{M_\xi(1+\rho)}{1-\rho},\nn
\end{align}
where $\rho$ and $M_\xi$ are defined in Assumption \ref{ass:idio}(b). Similar results are also in \citet{FLM13}, where sparsity of the idiosyncratic covariance matrix is assumed.

Define the $N\times N$ idiosyncratic covariance matrix and the diagonal matrix of idiosyncratic variances as:
$$
\bm\Gamma^\xi:=\E[\bm\xi_t\bm\xi_t^\prime]\quad \text{and}\quad
\bm\Sigma^\xi:=\text{diag}(\sigma_1^2,\ldots \sigma_N^2),
$$
respectively.
Then, letting $\mu_1^\xi$ be the largest eigenvalue of $\bm\Gamma^\xi$, it follows that (see Lemma 1(v) in \citealp{MBPCA})
\beq\label{eq:evalidio}
 \sup_{N\in\mathbb N}\mu_{1}^\xi  \le M_\xi,
 \eeq
 where $M_\xi$ is defined in Assumption \ref{ass:idio}(b). Condition \eqref{eq:evalidio} was originally assumed by \citet{chamberlainrothschild83} to control the amount of idiosyncratic correlation that we can allow for. Moreover, by setting $k=0$ in Assumption \ref{ass:idio}(b), it follows also that for all $i=1,\ldots, N$ and all $N\in\mathbb N$, $\sigma_i^2 \le M_\xi$. Thus, all idiosyncratic components have finite variance.

Parts (c) and (d) require finite 4th order moments as in \citet[Assumption B]{baili12} and finite sums of 4th order cross-moments. Notice that this is weaker than the requirement of finite 8th order cross-moments assumed  by \citet[Assumption C]{baili16}.
Part (d) also implies that the sample covariance between $\{\xi_{it}\}$ and $\{\xi_{jt}\}$ is a consistent estimator of $\E[\xi_{it}\xi_{jt}]$ (\citealp[pp.209-211]{hannan}). 
 
The basic set of assumption is completed by the following requirement.
\begin{ass}[\textsc{independence}]
\label{ass:ind} For all $N,T\in\mathbb N$, the sequences
$\{\xi_{it},\, i=1,\ldots, N,\, t=1,\ldots, T\}$ and $\{F_{jt},\, j=1,\ldots, r,\, t=1,\ldots, T\}$ are mutually independent. 
\end{ass}

This assumption obviously implies that the common components are independent of the idiosyncratic components at all leads and lags and across all units. Notice that \citet{baili12,baili16} treat the factors as constant parameters, therefore, by construction, they have no relation with the idiosyncratic components. Finally, Assumption \ref{ass:ind} is compatible with a structural macroeconomic interpretation of factor models, according to which the factors driving the common component are independent of the idiosyncratic components representing measurement errors or local dynamics. 

We conclude with an assumption of two Central Limit Theorems which are required to allow us to conduct inference on the loadings and the factors. It coincides with the assumption made by \citet[Assumption F]{baili16}.

\begin{ass}[\textsc{Central limit theorems}]
\label{ass:CLT}
$\,$
\begin{compactenum}[(a)]
\item For all $i=1,\ldots,N$ and all $N\in\mathbb N$, as $T\to\infty$,
	\[
	\frac 1{\sqrt T}\sum_{t=1}^T
	 \mbf F_t \xi_{it}
	\to_d
	\mathcal N\l(\mbf 0_r, \lim_{T\to\infty} \frac{\E[\bm F^\prime\bm\zeta_i\bm\zeta_i^\prime\bm F]}{T}
	\r).
	\]
\item For all $t=1,\ldots, T$ and all $T\in\mathbb N$, as $N\to\infty$,
	\[
	\frac 1{\sqrt N}\sum_{i=1}^N
	 \bm\lambda_i  \xi_{it}
	\to_d
	\mathcal N\l(\mbf 0_r, \lim_{N\to\infty} \frac{\E[\bm\Lambda^\prime\bm\xi_t\bm\xi_t^\prime\bm\Lambda]}{N}
	\r).
	\]
\end{compactenum}
\end{ass}


\begin{rem}
\upshape{
Assumption \ref{ass:CLT}(a) can be derived from more primitive assumptions, for example it follows from \citet[Theorem 1.4]{ibra62} under the assumption of strong mixing factors and idiosyncratic components (see also Section 3 in \citealp{MBPCA}).  Assumption \ref{ass:CLT}(b) of course holds if we assumed cross-sectionally independent idiosyncratic components. In general, to derive it from primitive assumptions we should introduce some notion of ordering of the $N$ cross-sectional units, as, e.g., the notion of mixing random fields \citep{bolthausen1982}. Since a notion of ordering might be unnatural in many contexts, we could apply results on exchangeable sequences, which are instead independent of the ordering, and are in turn obtained by virtue of the Hewitt-Savage-de Finetti theorem (\citealp[Theorem 4]{austern2022limit}, and \citealp{bolthausen1984}).
}
\end{rem}

\subsection{Identification conditions}

Let $\mu_i^x$, $i=1,\ldots, N$, be the $i$th largest eigenvalue of $\bm\Gamma^x:=\E[(\mbf x_t-\bm\alpha)(\mbf x_t-\bm\alpha)^\prime]$.
Then, we have the following result.

\begin{lem}\label{lem:egap}
Under Assumptions \ref{ass:common} through \ref{ass:ind}:
\begin{compactenum}[(i)]
\item for all $j=1,\ldots,r$, 
$\underline C_j\le \lim\inf_{N\to\infty} N^{-1}{ \mu_{j}^x}\le\lim\sup_{n\to\infty} N^{-1}{\mu_{j}^x}\le \overline C_j$;
\item $\sup_{N\in\mathbb N}\mu_{r+1}^x \le M_\xi$;
\end{compactenum}
where $\underline C_j$ and $\overline C_j$ are defined in \eqref{eq:lindiv}, and $M_\xi$ is defined in Assumption \ref{ass:idio}(b).
\end{lem}

The proof is omitted since it is a direct consequence of Weyl's inequality, combined with \eqref{eq:lindiv} which follows from Assumption \ref{ass:common}, \eqref{eq:evalidio} which follows from Assumption \ref{ass:idio}, and since $\bm\Gamma^x=\bm\Gamma^\chi+\bm\Gamma^\xi$ which follows from Assumption \ref{ass:ind}.

Lemma \ref{lem:egap} shows that the spectrum of the population covariance matrix of $\{\mbf x_t\}$ has an eigen-gap which widens as $N$ grows. This is  a necessary condition for identification of the model, and, indeed, virtually all existing methods for estimating the number of factors, $r$, are based on such property (\citealp{baing02}, \citealp{ABC10}, \citealp{onatski10}, \citealp{ahnhorenstein13}, \citealp{trapani2018randomized}). Notice, however, that even if an eigen-gap is displayed by an observed data set, this is not sufficient to say that the underlying data generating process is a factor model. 


For QML estimation it is crucial that the parameters of the model and the factors are identified. However, it is well known that in general the factor loadings, as well as the factors, are not identified. Indeed, all the structures equivalent to \eqref{eq:SDFM1R} can be obtained through an $r\times r$ invertible matrix $\mbf R$, which defines the transformed quantities:
$\mbf F_t^o:=\mbf R^{-1}\mbf F_t$,  $\bm{\Lambda}^{o}:=\bm{\Lambda}\mbf R$,
and $\bm\Gamma^{\xi o}:=\bm\Gamma^{\xi}$.
Under such relationships, using only first- and second-moment information, as it is the case with gaussian QML estimation,
 the model specified by $\mbf F_t^o$, $\bm{\Lambda}^{o}$, and $\bm\Gamma^{\xi o}$, is indistinguishable from the one given by $\mbf F_t$, $\bm{\Lambda}$, and $\bm\Gamma^{\xi}$.

To obtain an identified model,  we then need enough a priori structure to preclude any but the trivial transformation $\mbf R=\mbf I_r$. 
This can be achieved by imposing additional $r^2$ identifying constraints.  In particular, we impose the following  standard identifying conditions.

\begin{ass}[\textsc{loadings and factors identification}]\label{ass:ident}
$\,$
\begin{compactenum}[(a)]
\item For all $N\in\mathbb N$, $\frac{\bm\Lambda^\prime\bm\Lambda}{N}$ diagonal with distinct elements.
\item For all $T\in\mathbb N$ $\frac{\bm F^\prime\bm F}{T}=\mbf I_r$.
\item For all $j=1,\ldots, r$, one of the two following conditions hold:
\begin{inparaenum}[(i)]
\item $\lambda_{1j}> 0$;
\item $F_{j1}> 0$. 
\end{inparaenum}
\end{compactenum}
\end{ass}

Under parts (a) and (b) the loadings and the factors are identified up to a sign  \citep[see conditions PC1 in][]{baing13}, hence, they are only locally identified. To achieve global identification, in part (c) we fix the sign indeterminacy. 

Let now $\bm V$ be the $r\times r$ diagonal matrix of  eigenvalues of 
$(\bm\Sigma_\Lambda)^{1/2}\bm\Gamma^F(\bm\Sigma_\Lambda)^{1/2}$ or equivalently of 
$(\bm\Gamma^F)^{1/2}\bm\Sigma_\Lambda(\bm\Gamma^F)^{1/2}$, or also of $\bm\Sigma_\Lambda\bm\Gamma^F$.
Then, because of Assumption \ref{ass:ident}: \begin{inparaenum}[(A)]
\item $\bm\Sigma_\Lambda=\bm V$;
\item $\bm\Gamma^F=\mbf I_r$.
\end{inparaenum}
These conditions are sometimes assumed directly in place of Assumption \ref{ass:ident}, in which case the model is identified only as $N,T\to\infty$.

Moreover, from parts (a) and (b) it follows also that for all $N\in\mathbb N$, the $r\times r$ matrix $N^{-1}\bm\Lambda^\prime\bm\Lambda$ is the $r\times r$ diagonal matrix having as diagonal entries $N^{-1}\mu_1^\chi,\ldots, N^{-1}\mu_r^\chi$, i.e., the eigenvalues of $\bm\Gamma^\chi$ scaled by $N$, which therefore are assumed to be distinct.
Note that this implies that $\overline C_j<\underline C_{j-1}$, $j=2,\ldots, r$, in \eqref{eq:lindiv} and that also the diagonal elements of  $\bm V$ are distinct. Requiring distinct eigenvalues is useful for identifying the eigenvectors of $\bm\Gamma^\chi$. Although, in principle, this requirement is needed only for PC estimation \citep[Assumption G]{bai03}, it is also needed  to ensure consistency of the numerical maximization algorithms presented below and used for QML estimation, which are initialized with the PC estimators.

The identifying conditions in Assumption \ref{ass:ident} are similar to the set of constraints IC3 considered in \citet{baili12,baili16}, where, however, part (a) is replaced by the requiring $N^{-1}{\bm\Lambda^\prime(\bm\Sigma^\xi)^{-1}\bm\Lambda}$ to be diagonal for all $N\in\mathbb N$. 

\begin{rem}
\upshape{Clearly, Assumption \ref{ass:ident} does not provide any economic meaning to the factors; hence, they can be used only for an exploratory factor analysis. For alternative identification conditions, which allow to interpret the factors, hence to conduct confirmatory factor analysis, see, e.g., the classical approach by \citet{joreskog69}, then partially adopted also by \citet{baili12,baili16} and \citet{Li18}.
In this paper we do not deal with confirmatory factor analysis since in most macroeconomic applications the focus is only the estimated common component, which is always identified, so no interpretation of the factors is required.
}
\end{rem}

\begin{rem}
\upshape{Let $\mbf V^\chi$ be the $N\times r$ matrix having as columns the $r$ normalized eigenvectors corresponding to the $r$ eigenvalues of $\bm\Gamma^\chi$ which are collected in the $r\times r$ diagonal matrix $\mbf M^\chi$.
Then, under Assumption \ref{ass:ident} it can be proved that the true loadings are given by $\bm\Lambda=\mbf V^\chi(\mbf M^\chi)^{1/2}$ and, by linear projection of  $\bm C=(\bm \chi_1\cdots\bm\chi_T)^\prime$ onto $\bm\Lambda$, we see that 
the true factors are given by $\bm F=\bm C\mbf V^\chi (\mbf M^\chi)^{-1/2}$  \citep[Proposition 7]{MBPCA}. In principle this means that the true loadings and factors depend on $N$ and $T$, since $\bm \Gamma^\chi$ depends on $N$ and $\bm C$ depends on $N$ and $T$, and, thus, the loadings depend on $N$ and the factors depend both on $N$ and $T$, i.e., $\bm\Lambda\equiv\bm\Lambda_N$ and $\bm F\equiv \bm F_{NT}$. 
 Now, closer inspection shows that the sequence $\{\bm\Lambda_N, N\in\mathbb N\}$ is nested, so the issue regards only the factors and their dependence on $T$ not on $N$, since $\{\bm F_{NT},N,T\in\mathbb N\}=\{\bm F_{T},T\in\mathbb N\}$ is not nested. Mathematically a simple solution would be to assume directly $\bm\Gamma^F=\mbf I_r$.}\end{rem}

\subsection{Factor dynamics}

Under Assumptions \ref{ass:common}(b)-\ref{ass:common}(d) the factors can be seen either as an $rT$-dimensional sequence of constant parameters (in which case in \ref{ass:common}(b) the expectations would be replaced by averages and in \ref{ass:common}(c) we would have an ordinary limit), or as a realization of an $r$-dimensional stochastic process $\{\mbf F_t,\, t\in\mathbb Z\}$. The former case is commonly considered in classical factor analysis where observations are collected from random samples. The latter case is especially relevant  in time series analysis.
Nevertheless, if we believe the factors to be truly autocorrelated, then it would be desirable to explicitly model the dynamic evolution of the factors. This is the approach originally adopted by \citet{SS77} and \citet{quahsargent93} for dynamic factor analysis and then re-introduced by \citet{FGLR09} and \citet{DGRfilter,DGRqml}. 

In order to keep things simple, but without loss of generality, here we assume the VAR(1) specification:
\begin{align}
\mbf F_{t}&= \mbf A\mbf F_{t-1}+\mbf H\mbf u_t.\label{eq:VARsimple2}
\end{align}
We make the following assumption.
\begin{ass}\label{ass:VAR}
$\,$
\begin{compactenum}
\item [(a)] All roots of $\det(\mbf I_r -\mbf Az)= 0$ are $z_j^*\in \mathbb C$, $j=1,\ldots, r$,  such that $M_A\le \vert z_j^*\vert \le C_A$ for some finite positive reals $M_A$ and $C_A$  independent of $j$ and such that $M_A > 1$.
\item [(b)] $\mbf H$ is lower-triangular, $\text{\upshape rk}(\mbf H)=r$, and $\Vert\mbf H\Vert \le M_H$ for some finite positive real $M_H$. 
\item [(c)] For all $t=1,\ldots, T$ and all $T\in\mathbb N$, $\E[\mbf u_t] = \mbf 0_r$ and $\E[\mbf u_t\mbf u_t^\prime]=\mbf I_r$.
\item [(d)] For all $t=1,\ldots, T$, all $T\in\mathbb N$, and all $k\in\mathbb Z$ with $k\ne 0$, $\E[\mbf u_t\mbf u_{t-k}^\prime]=\mbf 0_{r\times r}$.
\item [(e)] For all $j_1,j_2,j_3,j_4=1,\ldots, r$, all $k\in\mathbb Z^+$, 
and all $T\in\mathbb N$
$$
\frac 1T \sum_{t,s=1}^T \l\vert\E[u_{j_1t}u_{j_2t-k} u_{j_3s}u_{j_4s-k}]]\r\vert \le K_u, \quad
\frac 1T \sum_{t,s=1}^T \l\vert\E[u_{j_1t}u_{j_2t-k}]\vert\,\vert \E[u_{j_3s}u_{j_4s-k}]]\r\vert \le K_u
$$
for some finite positive real $K_u$  independent of $j_1,j_2,j_3,j_4$, $k$, and $T$.
\item [(f)] For all $t\le 0$, $\mbf u_t=\mbf 0_r$.
\end{compactenum}
\end{ass}

Assumption \ref{ass:VAR} is standard in VAR literature. Part (a), requires the factors to have a causal autoregressive representation. If we let $\mbf v_t:=\mbf H\mbf u_t$ and $\bm\Gamma^v:=\E[\mbf v_t\mbf v_t^\prime]$, then, under parts (b) and (c), $\mbf H$  is identified as $\mbf H=(\bm\Gamma^v)^{1/2}$. Part (e) assumes finite and summable 4th order cross-moments of the innovations $\{\mbf u_t\}$. In part (f), we fix the initial conditions.

As a consequence, $\{\mbf F_t\}$ admits the Wold representation $\mbf F_t=\sum_{k=0}^{t-1}\mbf C_k\mbf v_{t-k}$, with coefficients $\mbf C_k:=\mbf A^k$ such that $\sum_{k=0}^\infty \Vert \mbf C_k\Vert^2\le M_C$ for some finite positive real $M_C$ and with innovations process $\{\mbf v_t\}$ which is a zero-mean white noise with finite and summable 4th order cross-moments, and with covariance matrix $\bm\Gamma^v$ finite and positive definite. 
In fact, parts (a) through (e) imply Assumption \ref{ass:common}(b) and, jointly with part (f), they also imply Assumption \ref{ass:common}(c), and both \ref{ass:common}(b) and \ref{ass:common}(c) become redundant in this setting.\footnote{Specifically, it follows that $\E[\mbf F_t]=\sum_{k=0}^{t-1} \mbf C_k\E[\mbf v_{t-k}]=\mbf 0_r$, 
$\bm\Gamma^F= \sum_{k=0}^{t-1}\mbf C_k\mbf H\mbf H^\prime \mbf C_k^\prime\le M_C^2 M_H^2$ and it is positive definite since $\mu_r(\bm\Gamma^F)\ge\mu_{r}(\mbf H\mbf H^\prime)>0$, 
$
\E[\l \Vert\mbf F_t \r\Vert^4] 
\le r^{10}M_C^4M_H^4 K_u,
$, and, last, $\E[\vert T^{-1/2}\sum_{t=1}^T F_{it}F_{jt}-\E[F_{it}F_{jt}]\vert^2]\le  r^{10}M_C^4M_H^4 K_u$, which implies Assumption \ref{ass:common}(c) by Chebychev's inequality.
}

\begin{rem}\label{rem:deistler}
\upshape{Under Assumptions  \ref{ass:ident}(a), \ref{ass:VAR}(a), and \ref{ass:VAR}(b),  the linear system described by \eqref{eq:SDFM1R} and \eqref{eq:VARsimple2}  is controllable and observable, as well as stabilizable and detectable.\footnote{The couple $(\mbf A, \mbf H)$ is controllable if and only if $\text{rk}[\mbf H (\mbf A\mbf H) \cdots (\mbf A^{(r-1)}\mbf H)] = r$
and the couple $(\mbf A,\bm\Lambda)$ is observable if and only if $\text{rk}[\bm\Lambda^\prime (\bm\Lambda\mbf A)^\prime \cdots (\bm\Lambda \mbf A^{(r-1)})^\prime] = r$ , moreover, the linear system is stabilizable if its unstable states are controllable and all uncontrollable states are stable, and it is detectable if its unstable states are observable and all unobservable states are stable \citep[Appendix C, pp.341-342]{AM79}.} This implies that the standard mini-phase condition:
\[
\text{rk}\l(
\ba{cc}
\mbf I_r-\mbf Az & \mbf H\\
\bm\Lambda &\mbf 0_{r\times r}
\ea
\r)=2r,
\]
is satisfied for all $z\in\mathbb C$ such that $\vert z\vert\le 1$, and, thus, \eqref{eq:SDFM1R}-\eqref{eq:VARsimple2} is the minimal state space representation of a dynamic approximate factor model, having as McMillan degree the number of factors $r$ \citep[Section II]{andersondeistler08}. This result, combined with Assumption \ref{ass:ident}, makes the linear system fully identified. 
}
\end{rem}

\begin{rem}
\upshape{
The state space formulation \eqref{eq:SDFM1R}-\eqref{eq:VARsimple2} is a very general representation. Indeed, under the additional mild requirement that $\{\bm\chi_{nt}\}$ has a rational spectral density, it follows that $\{\bm\chi_{nt}\}$ has a state space representation which can be constructed by means of the Kalman-Akaike procedure \citep{akaike74}. }
\end{rem}

\begin{rem}
\upshape{The factor model defined in \eqref{eq:SDFM1R} is called a static factor model, in that the factors are loaded only contemporaneously by the data. If we include also a dynamic model for the factors, such as \eqref{eq:VARsimple2}, we could also write \eqref{eq:SDFM1R} as
\beq\label{eq:gdfm}
x_{it} =\alpha_i+ \bm\lambda_i^\prime\sum_{k=0}^\infty \mbf A^k\mbf H\mbf u_{t-k}+\xi_{it}, \quad i=1,\ldots, N,\quad t=1,\ldots,T,
\eeq
where $\{\mbf u_t\}$ is an orthonormal $r$-dimensional white noise process of so-called dynamic factors, in that they are loaded by the data together with their lags. For this reason the model described by \eqref{eq:SDFM1R}-\eqref{eq:VARsimple2} is called a dynamic factor model, and \eqref{eq:gdfm} is a special case of the Generalized Dynamic Factor model introduced by \citet{FHLR00}. The GDFM is actually more general and it is defined as
\beq\label{eq:gdfmUR}
x_{it} =\alpha_i+ \sum_{k=0}^\infty \mbf b_{ik}^{\prime}\mbf e_{t-k}+\xi_{it}, \quad i=1,\ldots, N,\quad t=1,\ldots,T,
\eeq
for some square-summable sequence of $N\times q$ matrix of coefficients $\{\mbf b_{ik}\}$ and an orthonormal $q$-dimensional white noise process $\{\mbf e_t\}$ with $q<r$. As shown in \citet[Section 2.1]{stockwatson16}, the GDFM admits a state-space representation as \eqref{eq:SDFM1R}-\eqref{eq:VARsimple2}  only if in \eqref{eq:gdfmUR}  we allow for just finite number of lags, and we impose specific zero restrictions on $\mbf A$ and $\mbf H$.
}
\end{rem}

%
%


\section{A review of the Principal Component estimators}
In this section we briefly review Principal Components (PC) estimation of the static approximate factor model \eqref{eq:SDFM1R}. This is arguably the most popular estimation approach and it is a reference estimator also for QML estimation. The PC estimators are non-parametric and, in particular, their implementation does not require to specify any second order structure of the idiosyncratic components as long as Assumption \ref{ass:idio}(b) is satisfied. There are a few different, although asymptotically equivalent, definitions of PCs. We refer to \citet{MBPCA} for a discussion of the various approaches.
 
The idea of PCs was introduced by \citet{pearson1901} as a way to find the the $r$ directions of best fit in an $n$-dimensional space. It was then proposed as a technique to estimate factor models by \citet{hotelling1933analysis}, whose definition we follow in this paper. Specifically, the loadings are estimated as eigenvectors of the sample covariance matrix of the observables and the factors are estimated as the orthonormal PCs of $\bm X$ 
(see also \citealp[Chapter 4]{lawleymaxwell71}, \citealp[Chapter 9.3]{mardia1979multivariate}, \citealp[Chapter 7.2]{jolliffe2002principal}). This is also the definition adopted by \citet{FGLR09}. In particular, 
the PC estimator of the loadings is defined as:
\beq\label{eq:LPC}
\wh{\bm\lambda}_i^{\text{\tiny PC}}:= \l(\wh{\mbf M}^x\r)^{1/2} \wh{\mbf v}_{i}^x, \quad i=1,\ldots, N,
\eeq
where $\wh{\mbf v}_{i}^{x\prime}$ is the $r$-dimensional $i$th row of the $N\times r$ matrix $\wh{\mbf V}^x$ having has columns the $r$ normalized eigenvectors of the $N\times N$ sample covariance matrix $T^{-1}(\bm X-\bm\iota_T\bar{\mbf x}^\prime)^\prime(\bm X-\bm\iota_T\bar{\mbf x}^\prime)$ corresponding to the $r$ largest eigenvalues collected in the diagonal matrix $\wh{\mbf M}^x$. Notice that, by construction, the $N\times r$ matrix of PC estimated loadings $\wh{\bm\Lambda}^{\text{\tiny PC}}:=(\wh{\bm\lambda}_1^{\text{\tiny PC}}\cdots \wh{\bm\lambda}_N^{\text{\tiny PC}})^\prime$ is such that
$N^{-1}\wh{\bm\Lambda}^{\text{\tiny PC}\prime}\wh{\bm\Lambda}^{\text{\tiny PC}}=N^{-1}\wh{\mbf M}^x$, which is diagonal. Hence,  $\wh{\bm\Lambda}^{\text{\tiny PC}}$ satisfies Assumption \ref{ass:ident}(a).

Under the identifying Assumptions \ref{ass:ident}, we have that the PC estimator is asymptotically equivalent to the unfeasible Ordinary Least Squares (OLS) estimator we would get if the factors were observed:
\beq\label{eq:LOLS}
{\bm\lambda}_i^{\text{\tiny OLS}}:= \l(\bm F^\prime\bm F\r)^{-1}\bm F^\prime (\bm x_i-\bar x_i\bm\iota_T), \quad i=1,\ldots, N.
\eeq
In particular, it can be shown that (\citealp[Section 9]{MBPCA})
\beq\label{eq:AELPC}
\Vert\wh{\bm\lambda}_i^{\text{\tiny PC}}-{\bm\lambda}_i^{\text{\tiny OLS}}\Vert=O_{\mathrm P}\l(\frac 1{\sqrt{NT}}\r)+O_{\mathrm P}\l(\frac 1N\r),
\eeq
from which we prove the following result.

\begin{prop}\label{prop:LPCA}
Under Assumptions \ref{ass:common} through \ref{ass:ident}, for any $i=1,\ldots, N$, 
as $N,T\to\infty$, if $\sqrt T/N\to 0$,
\[
\sqrt T(\wh{\bm\lambda}_i^{\text{\tiny \upshape PC}}-\bm\lambda_i)\to_d\mathcal N\l(\mbf 0_r,\bm{\mathcal V}_{i}^{\text{\tiny \upshape OLS}}\r),
\]
where $\bm{\mathcal V}_{i}^{\text{\tiny \upshape OLS}}:=(\bm\Gamma^F)^{-1}
\l\{
\lim_{T\to\infty} \frac{ \E[\bm F^\prime \bm\zeta_i\bm\zeta_i^\prime\bm F]}{T}
\r\}
(\bm\Gamma^F)^{-1}=\lim_{T\to\infty} \frac{ \E[\bm F^\prime \E[\bm\zeta_i\bm\zeta_i^\prime]\bm F]}{T}$ (because of Assumption \ref{ass:ident}(b)).
\end{prop}

Since the PC  estimator is non-parametric the asymptotic covariance matrix $\bm{\mathcal V}_{i}^{\text{\tiny \upshape OLS}}$ has a sandwich form reflecting the fact that we do not take into account possible idiosyncratic serial correlations. The final expression of $\bm{\mathcal V}_{i}^{\text{\tiny \upshape OLS}}$ comes from the law of iterated expectations, and since factors and idiosyncratic components are independent by Assumption \ref{ass:ind}.
 
The PC estimator of the factors is then obtained by linear projection of $\bm X$ onto $\wh{\bm\Lambda}^{\text{\tiny PC}}$:
\beq\label{eq:FPC}
\wh{\mbf F}_t^{\text{\tiny PC}} := \l(\wh{\bm\Lambda}^{\text{\tiny PC}\prime}\wh{\bm\Lambda}^{\text{\tiny PC}}\r)^{-1} \wh{\bm\Lambda}^{\text{\tiny PC}\prime} (\mbf x_{t}-\bar{\mbf x})=(\wh{\mbf M}^x)^{-1/2} \wh{\mbf V}^x(\mbf x_t-\bar{\mbf x}), \quad t=1,\ldots, T.
\eeq
It follows that the $T\times r$ matrix of PC estimated factors $\wh{\bm F}^{\text{\tiny PC}}:=(\wh{\mbf F}_1^{\text{\tiny PC}}\cdots \wh{\mbf F}_T^{\text{\tiny PC}})^\prime$ is such that  $T^{-1}\wh{\bm F}^{\text{\tiny PC}\prime}\wh{\bm F}^{\text{\tiny PC}}=\mbf I_r$. Hence $\wh{\bm F}^{\text{\tiny PC}}$ satisfies Assumption \ref{ass:ident}(b).

Under the identifying Assumptions \ref{ass:ident}, we have that the PC estimator is asymptotically equivalent to the unfeasible OLS estimator we would get if the loadings were observed:
\beq\label{eq:FOLS}
{\mbf F}_t^{\text{\tiny OLS}}:= \l(\bm \Lambda^\prime\bm \Lambda\r)^{-1}\bm \Lambda^\prime (\mbf x_t-\bar {\mbf x}_t), \quad t=1,\ldots, T.
\eeq
In particular, it can be shown that (\citealp[Section 9]{MBPCA})
\beq\label{eq:AEFPC}
\Vert\wh{\mbf F}_t^{\text{\tiny PC}}-{\mbf F}_t^{\text{\tiny OLS}}\Vert=O_{\mathrm P}\l(\frac 1T\r)+O_{\mathrm P}\l(\frac 1{\sqrt {NT}}\r), 
\eeq
from which we prove the following result.

\begin{prop}\label{prop:FPCA}
Under Assumptions \ref{ass:common} through \ref{ass:ident}, for any $t=1,\ldots, T$, as $N,T\to\infty$, if $\sqrt N/T\to 0$,
\[
\sqrt N(\wh{\mbf F}_t^{\text{\tiny \upshape PC}}-\mbf F_t)\to_d\mathcal N\l(\mbf 0_r,\bm{\mathcal W}_{t}^{\text{\tiny \upshape OLS}}\r),
\]
where $\bm{\mathcal W}_{t}^{\text{\tiny \upshape OLS}}:=(\bm\Sigma_{\Lambda})^{-1}
\l\{
\lim_{N\to\infty}\frac {\bm\Lambda^\prime  \E[\bm \xi_t\bm \xi_t^\prime]\bm\Lambda }N
\r\}
(\bm\Sigma_{\Lambda})^{-1}$.
\end{prop}

Since the PC  estimator is non-parametric the asymptotic covariance matrix $\bm{\mathcal W}_{t}^{\text{\tiny \upshape OLS}}$ has a sandwich form reflecting the fact that we do not take into account possible idiosyncratic cross-sectional correlations and heteroskedasticity. Notice that the asymptotic covariance could be time dependent if we had serially heteroskedastic idiosyncratic components, since it would then depend on $\bm\Gamma_t^\xi:=\E[\bm \xi_t\bm \xi_t^\prime]$.

\begin{rem}
\upshape{
An equivalent set of PC estimators is adopted  by \citet{bai03}, where, however, first the   factors are estimated as $\sqrt T$ times the eigenvectors of the $T\times T$ sample covariance $N^{-1}(\bm X-\bm\iota_T\bar{\mbf x}^\prime)(\bm X-\bm\iota_T\bar{\mbf x}^\prime)^\prime$, hence they are the orthonormal PCs of $\bm X^\prime$ rather than of $\bm X$. The estimated loadings are then obtained by linear projection of $\bm X^\prime$ onto the estimated factors.  Such estimators also satisfy the identifying Assumptions \ref{ass:ident}(a) and \ref{ass:ident}(b). This approach is considered also in \citet{FLM13} and \citet{bai2020simpler}.

Finally, \citet{stockwatson02JASA} define the PC estimator of the loadings as $\sqrt N$ times the eigenvectors of the $N\times N$ sample covariance matrix. However, such definition does not satisfy Assumption \ref{ass:ident}(a) and \ref{ass:ident}(b), since now the estimated loadings would have all the same scale, while the corresponding estimated factors would have different scales.
}
\end{rem}


\section{The log-likelihood function}\label{sec:LLL}

First, let us introduce the following notation. Define $\bm {\mathcal X}:=\text{vec}(\bm X^\prime)=(\mbf x_{1}^\prime\cdots \mbf x_{T}^\prime)^\prime$ and $\bm{\mathcal Z}:=\text{vec}(\bm \Xi^\prime)=(\bm\xi_{1}^\prime\cdots \bm\xi_{T}^\prime)^\prime$  as 
the $NT$-dimensional vectors of observations and idiosyncratic components, $\bm{\mathcal A}:=\text{vec}(\bm\alpha\bm\iota_T^\prime)$  as the $NT$-dimensional vector of constants, 
$\bm {\mathcal F}:=\text{vec}(\bm F^\prime)=(\mbf F_{1}^\prime\cdots \mbf F_T^\prime)^\prime$ as the $rT$-dimensional vector of factors, and  $\bm {\mathfrak L}:=\mbf I_T\otimes \bm\Lambda$ as the $NT\times rT$ block-diagonal matrix containing the loadings matrix replicated $T$ times.

Then, by taking the vectorized version of the transposed of \eqref{eq:SDFM1Rmat}, we can write model \eqref{eq:SDFM1R} as:\footnote{Recall that $\text{vec}(\bm A\bm B\bm C)= (\bm C^\prime \otimes \bm A) \text{vec}(\bm B)$.} 
\begin{align}\label{eq:DFM_matrix}
\bm{\mathcal X}= \bm{\mathcal A}
+
\bm {\mathfrak L} \bm{\mathcal F} + \bm{\mathcal Z}.
\end{align}
Let $\bm\Omega^x:=\E_{}[(\bm {\mathcal X}- \bm{\mathcal A})(\bm {\mathcal X}- \bm{\mathcal A})^\prime]$ and $\bm\Omega^\xi:=\E_{}[\bm{\mathcal Z}\bm{\mathcal Z}^\prime]$, be the $NT\times NT$ covariance matrices of the $NT$-dimensional vectors of data and idiosyncratic  components, respectively. Let also $\bm\Omega^F:=  \E_{}[\bm {\mathcal F}\bm {\mathcal F}^{\prime}]$ be the $rT\times rT$ covariance matrix of the $rT$-dimensional factor vector. Then, because of Assumption \ref{ass:ind},
\[
\bm\Omega^x = \bm {\mathfrak L}\,\bm\Omega^F\bm {\mathfrak L}^\prime + \bm\Omega^\xi.
\]
The parameters of the model that need to be estimated are then $\bm\alpha$ and
$$
\bm\varphi:=(\mathrm{vec}(\bm\Lambda)^\prime, \mathrm{vech}(\bm\Omega^{\xi})^\prime,\mathrm{vech}(\bm\Omega^{F})^\prime)^\prime.
$$ 

\subsection{Exact log-likelihood}
The gaussian quasi-log-likelihood computed in a generic value of the parameters, denoted as $\underline{\bm{\alpha}}$ and $\underline{\bm{\varphi}}$, is given by
\begin{align}
\ell_*(\bm {\mathcal X};\underline{\bm{\alpha}},\underline{\bm{\varphi}})
&:=-\frac {NT}2\log (2\pi)-\frac 12 \log\det\l(\underline{\bm\Omega}^x\r) -\frac 12 \l[ \l(\bm {\mathcal X}-\underline{\bm {\mathcal A}}\r)^\prime \l(\underline{\bm\Omega}^x\r)^{-1}\l(\bm {\mathcal X}-\underline{\bm {\mathcal A}}\r)\r]\label{eq:LL00}.
\end{align}
Let $\bar{\bm{\mathcal X}}=\text{vec}(\bar {\mbf x}\bm\iota_T^\prime)$ with $\bar {\mbf x}=T^{-1}\sum_{t=1}^T\mbf x_t$. Then, it holds that
\[
 \l(\bm {\mathcal X}-\underline{\bm {\mathcal A}}\r)^\prime \l(\underline{\bm\Omega}^x\r)^{-1}\l(\bm {\mathcal X}-\underline{\bm {\mathcal A}}\r)=
 \l(\bm {\mathcal X}-\bar{\bm{\mathcal X}}\r)^\prime \l(\underline{\bm\Omega}^x\r)^{-1}\l(\bm {\mathcal X}-\bar{\bm{\mathcal X}}\r)+\l(\bar{\bm{\mathcal X}}-\underline{\bm {\mathcal A}}\r)^\prime \l(\underline{\bm\Omega}^x\r)^{-1}\l(\bar{\bm{\mathcal X}}-\underline{\bm {\mathcal A}}\r),
\]
which 
shows that  \eqref{eq:LL00}  is maximized when $\underline{\bm {\mathcal A}}=\bar{\bm{\mathcal X}}$, i.e., the QML estimator of $\bm\alpha$ is $\wh{\bm\alpha}^{\text{\tiny QML}}:=\bar{\mbf x}$.

The concentrated quasi-log-likelihood computed in a generic value of the parameters $\underline{\bm{\varphi}}$ is then given by:
\begin{align}
\ell(\bm {\mathcal X};\underline{\bm{\varphi}})
&:=-\frac {NT}2\log (2\pi)-\frac 12 \log\det\l(\underline{\bm\Omega}^x\r) -\frac 12 \l[ \l(\bm {\mathcal X}-\bar{\bm{\mathcal X}}\r)^\prime \l(\underline{\bm\Omega}^x\r)^{-1}\l(\bm {\mathcal X}-\bar{\bm{\mathcal X}}\r)\r]\label{eq:LL0}\\
&=-\frac {NT}2\log (2\pi)-\frac 12 \log\det\l(\underline{\bm {\mathfrak L}}\,\underline{\bm\Omega}^F\underline{\bm {\mathfrak L}}^\prime + \underline{\bm\Omega}^\xi\r)-\frac 12 \l[ \l(\bm {\mathcal X}-\bar{\bm{\mathcal X}}\r)^\prime \l(\underline{\bm {\mathfrak L}}\,\underline{\bm\Omega}^F\underline{\bm {\mathfrak L}}^\prime + \underline{\bm\Omega}^\xi\r)^{-1}\l(\bm {\mathcal X}-\bar{\bm{\mathcal X}}\r) \r]. \nn
\end{align}

In general, it might be very hard, if not impossible, to find the QML estimator of $\bm\varphi$, i.e., the maximizer of \eqref{eq:LL0}. This is due to the very large number of parameters that need to be estimated. The QML estimator might not even exist unless other restrictions are imposed on the model. The main problem here is that, under our assumptions, we should estimate both ${\bm\Omega}^\xi$ and ${\bm\Omega}^F$, which are in general full block Toeplitz matrices containing autocovariance matrices up to lag $(T-1)$. Moreover, each of the $T$ blocks of ${\bm\Omega}^\xi$ is in general an $N\times N$ full matrix, so, without imposing further restrictions, ${\bm\Omega}^\xi$ alone contains already $NT(NT+1)/2$ parameters to be estimated, which is more than the $NT$ available data points. 

\subsection{Mis-specified log-likelihoods}
Henceforth, we will be working with a mis-specified log-likelihood where we treat the idiosyncratic components as serially and cross-sectionally uncorrelated. Thus, we consider a mis-specified model for the idiosyncratic 2nd order structure described by: ${\bm\Omega}^\xi=\mbf I_T\otimes {\bm\Sigma}^\xi$, which reduces the parameters in $\bm\Omega^\xi$ to be estimated to just the $N$ elements of  the diagonal matrix  $\bm\Sigma^\xi$. The log-likelihood \eqref{eq:LL0} is then replaced by:
\begin{align}
\ell_{0}(\bm {\mathcal X};\underline{\bm{\varphi}})
=&\,-\frac 12 \log\det\l(\l\{\mbf I_T\otimes \underline{\bm \Lambda}\r\}\underline{\bm\Omega}^F\l\{\mbf I_T\otimes \underline{\bm \Lambda}^\prime\r\} + \big\{\mbf I_T\otimes \underline{\bm\Sigma}^\xi\big\} \r)\nn\\
&-\frac 12 \l[ \l(\bm {\mathcal X}-\bar{\bm{\mathcal X}}\r)^\prime \l(
\l\{\mbf I_T\otimes \underline{\bm \Lambda}\r\}
\underline{\bm\Omega}^F
\l\{\mbf I_T\otimes \underline{\bm \Lambda}^\prime\r\} + \big\{\mbf I_T\otimes \underline{\bm\Sigma}^\xi\big\}\r)^{-1}\l(\bm {\mathcal X}-\bar{\bm{\mathcal X}}\r) \r]. \label{eq:LL0acf}
\end{align}
In general, maximization of \eqref{eq:LL0acf} is still unfeasible because it requires estimating $\bm\Omega^F$ which contains other $rT(rT+1)/2$ parameters. In this paper we consider two approaches. First, we consider a further mis-specification by imposing zero autocorrelation in the factors, i.e., by setting ${\bm\Omega}^F=\mbf I_T\otimes{\bm\Gamma}^F=\mbf I_{rT}$, where we used also Assumption \ref{ass:ident}(b). This is the approach reviewed in Section \ref{sec:static}, which defines the QML estimator $\wh{\bm\varphi}^{\text{\tiny QML,S}}$. Second, we consider a parametric linear model describing the factor dynamics, so that $\bm\Omega^F$ depends on the parameters characterizing such model. In particular, in the case of the VAR(1) specification given in \eqref{eq:VARsimple2}, the $(t,s)$-block of $\bm\Omega^F$ is the $r\times r$ lag-$(t-s)$ autocovariance matrix: $\bm\Gamma_{t-s}^F:=\E[\mbf F_t\mbf F_s^\prime]$ such that  $\text{vec}(\bm\Gamma_{t-s}^F)=(\mbf I_r\otimes \mbf A)^{t-s} (\mbf I_{r^2}- \mbf A\otimes\mbf A)^{-1}\text{vec}(\mbf H\mbf H^\prime)$. This reduces the number of parameters in $\bm\Omega^F$ to be estimated  to $r^2+r(r+1)/2$. This is the approach reviewed in Section \ref{sec:dynamic}, which defines the QML estimator $\wh{\bm\varphi}^{\text{\tiny QML,D}}$.  

\begin{rem}
\upshape{
An alternative QML estimation approach, not covered in this paper, consists in replacing the log-likelihood \eqref{eq:LL0acf} by its Whittle frequency domain asymptotic approximation which is then maximized to estimate the parameters (\citealp{SS77,dunsmuir79,GS81}). This approach is appealing since, in principle, it does not require specifying a parametric model for the dynamics of the factors. However,  its implementation requires to first estimate a large dimensional spectral density matrix, which will produce an additional estimation error depending also on the chosen bandwidth \citep[]{zhang2021}.
}
\end{rem}

\begin{rem}
\upshape{So far, we implicitly considered the idiosyncratic components as stationary, and, in particular, as serially homoskedastic. Nevertheless, under Assumption \ref{ass:idio} we could  in principle  allow also for serial heteroskedasticity. In this case, we would have $\bm\Gamma_t^\xi:=\E[\bm\xi_t\bm\xi_t^\prime]$ to be a time-dependent matrix with diagonal elements $\sigma_{it}^2$ collected into the diagonal matrix $\bm\Sigma_t^\xi$. Then, letting $\bm\Sigma^\xi:=T^{-1}\sum_{t=1}^T\bm\Sigma_t^\xi$, with diagonal entries $\sigma_i^2:=T^{-1}\sum_{t=1}^T \sigma_{it}^2$, we could still consider the same log-likelihood \eqref{eq:LL0acf}, but where now we introduced an additional mis-specification in that we treat the idiosyncratic components as if they were serially homskedastic. In this case, the QML estimators of the diagonal entries of $\bm\Sigma^\xi$, i.e., of $\sigma_{i}^2$, $i=1,\ldots, N$, have to viewed as estimators of the idiosyncratic variances averaged over time. This is the approach suggested by \citet{baili16}. Given that in this context we cannot  properly address the presence of serially  heteroskedastic idiosyncratic components, hereafter, we rule out such possibility.
}
\end{rem}

\subsection{Identification of the QML estimator}
Hereafter, we define the parameter space as the set $\mathcal O\in\mathbb R^Q$ where $Q$ is the number of parameters to be estimated. It is intended that any generic parameter vector $\underline{\bm\varphi}\in\mathcal O$ has elements satisfying all the assumptions given in Section \ref{sec:model}. In particular, notice that, under those assumptions, the dimension of $\mathcal O$  grows with $N$.

Usually in QML estimation it is required to first prove the existence of the QML estimator. However, to this end we cannot rely on compactness of $\mathcal O$ since its dimension increases to infinity. Nevertheless, as shown in the next sections, in the present setting the parameter estimates maximizing the log-likelihood have an asymptotic linear representation, so we can easily bound the estimation errors directly by the distance between the estimated and the true parameter vector without the need of using any Taylor expansion for asymptotic analysis. This allows us to bypass the need to work with a compact parameter space. The only condition we need is for each element of the parameter vector $\bm\varphi$ to belong to a bounded set, and this is guaranteed by Assumptions \ref{ass:common}(a), \ref{ass:idio}(a), \ref{ass:idio}(b), \ref{ass:VAR}(a), and \ref{ass:VAR}(b).


Finally, in order to have a well defined log-likelihood in correspondence of the QML estimator of $\bm\varphi$, it is common to make the following assumption.

\begin{ass}\label{ass:QML}
For all $i=1,\ldots, N$ and all $N\in\mathbb N$, $\wh{\sigma}_i^{2\text{\tiny \upshape QML,S}}\!\!\in[C_\xi^\prime, M_\xi^\prime]$ and $\wh{\sigma}_i^{2\text{\tiny \upshape QML,D}}\!\!\in[C_\xi^\prime, M_\xi^\prime]$, for some finite positive reals $C_\xi^\prime$ and $M_\xi^\prime$ independent of $i$. 
\end{ass}

This assumption is identical to the requirement imposed by \citet[Assumption D]{baili16}. In fact, it is a redundant condition which we assume only for the sake of simplicity. Indeed, it can be easily proved that  in the present context Assumption \ref{ass:QML} is always satisfied \citep[Theorem 3.1]{GGM21}. 



\section{QML estimation of an approximate static factor model}\label{sec:static}

In this section we review the case in which the dynamics of the factors is not explicitly modeled, i.e., when we set $\bm\Omega^F=\mbf I_{rT}$. So we consider an approximate static factor model described by the equation:
\[
x_{it}=\alpha_i+\bm\lambda_i^\prime\mbf F_t+\xi_{it}, \quad i=1,\ldots, N, \quad t=1,\ldots, T,
\]
together with Assumptions \ref{ass:common} through \ref{ass:ident}.
Then, either $\{\mbf F_t\}$ is effectively an uncorrelated process, so that it can also be considered as a deterministic sequence, or we are introducing a further mis-specification on top of the mis-specifications related to the idiosyncratic 2nd order structure introduced in the previous section.

For a static factor model the log-likelihood \eqref{eq:LL0acf} simplifies to 
\beq\label{eq:LL0iid}
\ell_{0,\text{\tiny S}}(\bm {\mathcal X};\underline{\bm\varphi})\simeq-\frac T2 \log\det\l(\underline{\bm \Lambda}\,\underline{\bm \Lambda}^\prime + \underline{\bm\Sigma}^\xi\r)-\frac 12\sum_{t=1}^T \l[ (\mbf x_{t}-\bar{\mbf x})^\prime \l(\underline{\bm \Lambda}\,\underline{\bm \Lambda}^\prime + \underline{\bm\Sigma}^\xi\r)^{-1}(\mbf x_{t}-\bar{\mbf x}) \r].
\eeq
Notice that, up to a scaling by $T$, this is also known in the literature as the Wishart log-likelihood for the sample covariance $T^{-1}(\mbf x_{t}-\bar{\mbf x})(\mbf x_{t}-\bar{\mbf x})^\prime$ \citep{AR56}.  The parameters to be estimated are: ${\bm\varphi}=(\mathrm{vec}(\bm\Lambda)^\prime,\sigma^2_1,\ldots,\sigma^2_N )^\prime$, which is a $Q=(r+1)N$-dimensional vector. We shall denote the maximizer of $\ell_{0,\text{\tiny S}}(\bm {\mathcal X};\underline{\bm\varphi})$ as $\wh{\bm\varphi}^{\text{\tiny QML,S}}$.

\subsection{Loadings}
Despite all introduced simplifications, no closed form solution exists for the elements of the vector of QML estimators, but there exist few numerical ways to compute it. Classical iterative approaches, dealing with cross-sections of small dimension $N$, were proposed by, e.g, \citet{joreskog69} and \citet[Chapter 2]{lawleymaxwell71}, and extended to the large $N$ case by  \citet{sundbergfeldmann16}.  Similarly, \citet{ng15} adapted the Newton-Raphson maximization algorithm to the large $N$ case by exploiting the properties of tri-diagonal blocked matrices. However, all these algorithms are heuristic and their relation to proper QML estimation is unclear. In this respect, an EM algorithm, thus in principle accomplishing QML estimation, was introduced by \citet{RT82}, and considered in the large $N$ setting by \citet{baili12,baili16}. 

Under our assumption of an approximate factor model, \citet[Theorem 1]{baili16} prove that 
\beq\label{eq:AE2}
\l\Vert\wh{\bm\lambda}_i^{\text{\tiny \upshape QML,S}}-{\bm\lambda}_i^{\text{\tiny \upshape OLS}}\r\Vert=O_{\mathrm P}\l(\frac 1{N}\r)+O_{\mathrm P}\l(\frac 1{\sqrt{NT}}\r).
\eeq
Although this result is derived under the alternative identifying condition $N^{-1}\bm\Lambda^\prime(\bm\Sigma^\xi)^{-1}\bm\Lambda$ being diagonal (IC3 in the original papers), it is possible to show that it is applicable also under our identifying conditions. In particular, in \citet[Theorem 5.1]{MBQMLPCA} it is shown that, under the assumptions of this paper, the following holds:
\beq\label{eq:AE3}
\l\Vert\wh{\bm\lambda}_i^{\text{\tiny \upshape QML,S}}-\wh{\bm\lambda}_i^{\text{\tiny \upshape PC}}\r\Vert=O_{\mathrm P}\l(\frac 1{N}\r).
\eeq

Therefore, \eqref{eq:AE3}, together with \eqref{eq:AELPC}, implies \eqref{eq:AE2}, and we immediately see that the QML, PC, and the unfeasible OLS estimators of the loadings are asymptotically equivalent, as long as $N\to\infty$. This is formalized in the following result, of which part (a) is an immediate consequence of  \eqref{eq:AE3} and Proposition \ref{prop:LPCA}, and part (b), which holds regardless of the imposed identification conditions, is proved by \citet[Theorem S.2]{baili16}.

\begin{prop}\label{prop:LBAILI16} For any $i=1,\ldots, N$, under Assumptions \ref{ass:common} through \ref{ass:ident}:
\ben
\item [(a)] if $\sqrt T/N\to 0$, as $N,T\to\infty$, 
\[
\sqrt T(\wh{\bm\lambda}_i^{\text{\tiny \upshape QML,S}}-\bm\lambda_i)\to_d\mathcal N\l(\mbf 0_r,\bm{\mathcal V}_{i}^{\text{\tiny \upshape OLS}}\r),
\]
where $\bm{\mathcal V}_{i}^{\text{\tiny \upshape OLS}}:=(\bm\Gamma^F)^{-1}
\l\{
\lim_{T\to\infty} \frac{ \E[\bm F^\prime \bm\zeta_i\bm\zeta_i^\prime\bm F]}{T}
\r\}
(\bm\Gamma^F)^{-1}=\lim_{T\to\infty} \frac{ \E[\bm F^\prime\E[ \bm\zeta_i\bm\zeta_i^\prime]\bm F]}{T}$ (because of Assumption \ref{ass:ident}(b));
\item [(b)] if also Assumption \ref{ass:QML} holds, as $N,T\to\infty$, $\min(\sqrt T,N)\Vert
\wh{\sigma}_i^{2\text{\upshape \tiny QML,S}}
-\sigma_i^2
\Vert=O_{\mathrm P}(1)$.
\een
\end{prop}

There are three mis-specifications that we introduced in the log-likelihood \eqref{eq:LL0iid}. First, we treat the factors as serially uncorrelated. This is, however, an innocuous mis-specification that has no effect on the asymptotic properties of the loadings. Indeed, the OLS estimator, which is asymptotically equivalent to the QML estimator, is not affected by the autocorrelation of the regressors, i.e., of the factors, as long as their covariance matrix is well defined as required by Assumptions  \ref{ass:common}(b) and \ref{ass:common}(c).

Second, the QML and PC estimators of the loadings are not the most efficient estimators, since they both neglect the possible idiosyncratic serial correlations, hence the sandwich form of the asymptotic covariance matrix, which for the QML estimator is due to the use of a mis-specified log-likelihood, while for the PC estimator is due to its non-parametric nature. Clearly, if the idiosyncratic components were truly serially uncorrelated, then, we would have $\E[\bm\zeta_i\bm\zeta_i^\prime]=\sigma^2_i\mbf I_T$, and the asymptotic covariance matrix would reduce to $\sigma_i^2 (\bm\Gamma^F)^{-1}=\sigma_i^2\mbf I_r$ (because of Assumption \ref{ass:ident}(b)), which is the Gauss-Markov lower bound under serial homoskedasticity. This is the case studied by \citet[Theorem 5.2]{baili12}, and we refer to Section \ref{sec:classic} for more details.

Third, as a consequence of the fact that we are estimating an approximate factor model using the mis-specified log-likelihood \eqref{eq:LL0iid} of an exact factor model, we would not get consistent estimators of the loadings if $N$ were fixed. Indeed, only in the limit $N\to\infty$ we can disentangle the common components, i.e., the loadings and the factors, from the idiosyncratic components (see Lemma \ref{lem:egap}). This is reflected in the asymptotic expansion \eqref{eq:AE2}, where it is clear that if $N$ is fixed then the estimation error is non-vanishing. In other words, the mis-specification error, which we introduce by using a mis-specified log-likelihood, vanishes asymptotically only if $N\to\infty$. So, despite being fully parametric,  the QML estimator does not suffer of the curse of dimensionality, but, in fact, it produces consistent estimates only in a high-dimensional setting, i.e., it enjoys a blessing of dimensionality.



\begin{rem}\label{rem:BTintuition}
\upshape{An  intuitive argument, alternative to the proof in \citet{MBQMLPCA}, for the equivalence between QML and OLS is the following. Consider the decomposition of the log-likelihood \eqref{eq:LL0iid}:
\beq
\ell_{0,\text{\tiny S}}(\bm{\mathcal X};\underline{\bm\varphi})=\ell_{0,\text{\tiny S}}(\bm{\mathcal X}|\bm{\mathcal F};\underline{\bm\varphi}) +\ell_{0,\text{\tiny S}}(\bm{\mathcal F};\underline{\bm\varphi})-\ell_{0,\text{\tiny S}}(\bm{\mathcal F}|\bm{\mathcal X};\underline{\bm\varphi}).\nn
\eeq 
Then, since under our assumptions we obviously have $\Vert\bm{\mathcal F}\Vert=O_{\mathrm P}(\sqrt T)$, $\Vert\bm\Lambda\Vert=O(\sqrt N)$,  and $\Vert\bm{\mathcal X}\Vert=O_{\mathrm P}(\sqrt {NT})$, the, intuitively, we also have $\sup_{\underline{\bm\varphi}\in\mathcal O}\vert \ell(\bm{\mathcal F};\underline{\bm\varphi})\vert =O_{\mathrm P}(T)$ and  $\sup_{\underline{\bm\varphi}\in\mathcal O}\vert \ell(\bm{\mathcal F}|\bm{\mathcal X};\underline{\bm\varphi})\vert =O_{\mathrm P}(T)$, while $\sup_{\underline{\bm\varphi}\in\mathcal O}\vert \ell(\bm{\mathcal X};\underline{\bm\varphi})\vert =O_{\mathrm P}(NT)$ and $\sup_{\underline{\bm\varphi}\in\mathcal O}\vert \ell(\bm{\mathcal X}|\bm{\mathcal F};\underline{\bm\varphi})\vert =O_{\mathrm P}(NT)$, so that
\begin{align}
\sup_{\underline{\bm\varphi}\in\mathcal O}\frac 1{NT}\l\vert\ell(\bm{\mathcal X};\underline{\bm\varphi})-\ell(\bm{\mathcal X}|\bm{\mathcal F};\underline{\bm\varphi}) \r\vert&\le\frac 1{NT} \sup_{\underline{\bm\varphi}\in\mathcal O} \l\vert\ell(\bm{\mathcal F};\underline{\bm\varphi}) \r\vert+\frac 1{NT}
\sup_{\underline{\bm\varphi}\in\mathcal O} \l\vert\ell(\bm{\mathcal F}|\bm{\mathcal X};\underline{\bm\varphi}) \r\vert = O_{\mathrm P}\l(\frac 1{N}\r).\nn
\end{align}
This implies that, for any given $i=1,\ldots, N$, as $N\to\infty$, the value of the loadings maximizing $\ell_{0,\text{\tiny S}}(\bm{\mathcal X};\underline{\bm\varphi})$, i.e., $\wh{\bm\lambda}_i^{\text{\tiny QML,S}}$ coincides asymptotically with the value of the loadings maximizing $\ell_{0,\text{\tiny S}}(\bm{\mathcal X}|\bm{\mathcal F};\underline{\bm\varphi})$, which is ${\bm\lambda}_i^{\text{\tiny OLS}}$. This reasoning was also suggested, but not formally proved, by \citet{BT11}, and used by \citet{DGRqml} in their proofs.
}
\end{rem}

\begin{rem}
\upshape{Not only the full log-likelihood $\ell_{0,\text{\tiny S}}(\bm{\mathcal X};\underline{\bm\varphi})$ 
coincides asymptotically with the conditional log-likelihood $\ell_{0,\text{\tiny S}}(\bm{\mathcal X}|\bm{\mathcal F};\underline{\bm\varphi})$, but also their derivatives with respect to the loadings coincide.
Namely, \citet[Theorem 6.1]{MBQMLPCA} proves that:
\[
\frac 1{\sqrt T}\l\Vert \l.\frac{\partial \ell_{0,\text{\tiny \upshape S}}(\bm{\mathcal X};\underline{\bm\varphi})}{\partial \underline{\bm\lambda}_i^\prime}\r\vert_{\underline{\bm\varphi}={\bm\varphi}}
-\l.\frac{\partial \ell_{0,\text{\tiny \upshape S}}(\bm{\mathcal X}|\bm{\mathcal F};\underline{\bm\varphi})}{\partial \underline{\bm\lambda}_i^\prime}\r\vert_{\underline{\bm\varphi}={\bm\varphi}}
\r\Vert=O_{\mathrm P}\l(\max\l(\frac 1{\sqrt N},\frac {\sqrt T}{N}\r)\r);
\]
\[
\frac 1T\l\Vert \l.\frac{\partial^2 \ell_{0,\text{\tiny \upshape S}}(\bm{\mathcal X};\underline{\bm\varphi})}{\partial \underline{\bm\lambda}_i^\prime\partial \underline{\bm\lambda}_i}\r\vert_{\underline{\bm\varphi}={\bm\varphi}}
-\l.\frac{\partial^2 \ell_{0,\text{\tiny \upshape S}}(\bm{\mathcal X}|\bm{\mathcal F};\underline{\bm\varphi})}{\partial \underline{\bm\lambda}_i^\prime\partial \underline{\bm\lambda}_i}\r\vert_{\underline{\bm\varphi}={\bm\varphi}}
\r\Vert=O_{\mathrm P}\l(\max\l(\frac 1N,\frac 1{\sqrt {NT}}\r)\r).
\]
By denoting the Fisher information and the population Hessian matrices for $\bm\lambda_i$ derived from the full log-likleihood as: 
\begin{align}
\bm{\mathcal I}_i(\bm{\mathcal X};\bm\varphi)&=\lim_{T\to\infty} \E\l[\l(\frac 1{\sqrt T}
\l.\frac{\partial \ell_{0,\text{\tiny S}}(\bm{\mathcal X};\underline{\bm\varphi})}{\partial \underline{\bm\lambda}_i^\prime}\r\vert_{\underline{\bm\varphi}={{\bm\varphi}}}
\r)
\l(\frac 1{\sqrt T}\l.\frac{\partial \ell_{0,\text{\tiny S}}(\bm{\mathcal X};\underline{\bm\varphi})}{\partial \underline{\bm\lambda}_i}\r\vert_{\underline{\bm\varphi}={{\bm\varphi}}}
\r)\,
\r], \nn\\
\bm{\mathcal H}_i(\bm{\mathcal X};\bm\varphi)&=\lim_{T\to\infty} \E\l[
\frac 1T  \l.\frac{\partial^2 \ell_{\text{\tiny E}}(\bm{\mathcal X};\underline{\bm\varphi})}{\partial \underline{\bm\lambda}_i^\prime\partial \underline{\bm\lambda}_i}\r\vert_{\underline{\bm\varphi}={{\bm\varphi}}}
\r],\nn
\end{align} 
respectively, we have that, as $N,T\to\infty$:
\[
\bm{\mathcal V}_i^{\text{\tiny OLS}}= \l(\bm{\mathcal H}_i(\bm{\mathcal X};\bm\varphi)\r)^{-1}
\bm{\mathcal I}_i(\bm{\mathcal X};\bm\varphi)
\l(\bm{\mathcal H}_i(\bm{\mathcal X};\bm\varphi)\r)^{-1}+ o_{\mathrm P}(1).
\]
As expected, the asymptotic covariance matrix of the QML estimator of the loadings coincides asymptotically with the classical QML sandwich form which we have for a linear regression with autocorrelated and heteroskedastic residuals \citep{white80,NW87,andrews91}.
}
\end{rem}

\begin{rem}\label{rem:TB}
\upshape{It is possible to consider QML estimation based on an even simpler and further mis-specified log-likelihood, where we also avoid to model idiosyncratic cross-sectional heteroskedasticites. In this case, known as the spherical case, the log-likelihood \eqref{eq:LL0iid} is simplified by imposing $\bm\Sigma^\xi=\sigma^2\mbf I_{N}$ for some $\sigma^2>0$. This is the approach proposed by \citet{tippingbishop99}. In this case the QML estimator of the loadings has a closed form solution given by
\beq\label{eq:LML}
\wh{\bm\lambda}_i^{\text{\tiny QML,S}_0} := \l(\wh{\mbf M}^x-\wh{\sigma}^{2\text{\tiny QML,S}_0}\mbf I_r\r)^{1/2} \wh{\mbf v}_{i}^x, \quad i=1,\ldots, N,
\eeq
with $\wh{\sigma}^{2\text{\tiny QML,S}_0} :=(N-r)^{-1}\sum_{j=r+1}^N \wh \mu_j^x$, which can be considered as an estimator of $N^{-1}\sum_{i=1}^N \sigma_i^2$. By comparing \eqref{eq:LML} with the PC estimator defined in \eqref{eq:LPC}, we notice a similarity between the PC estimator and the QML estimator. This is a well known fact, but no formal proof exists, essentially because, in general, nothing can be said about the asymptotic behavior of the non-leading eigenvalues $\wh \mu_j^x$, $j=r+1,\ldots, N$ (see, e.g., \citealp[Lemma 4]{trapani2018randomized}). However, since this is a special case of the QML estimation considered above, from \eqref{eq:AE3} it follows also that (see \citealp[Remark 5.1]{MBQMLPCA}): 
\begin{align}\label{eq:AETB}
\l\Vert\wh{\bm\lambda}_i^{\text{\tiny \upshape QML,S}_0}-\wh{\bm\lambda}_i^{\text{\tiny \upshape PC}}\r\Vert=O_{\mathrm P}\l(\frac 1{N}\r).
\end{align}
So, once again, by virtue of \eqref{eq:AELPC} and \eqref{eq:AETB}, the QML estimator coincides asymptotically with the PC and the unfeasible OLS estimators, and so it still satisfies Proposition \ref{prop:LBAILI16}(a) with asymptotic covariance matrix given by $\sigma^2 (\bm\Gamma^F)^{-1}=\sigma^2\mbf I_r$ (because of Assumption \ref{ass:ident}(b)).
}
 \end{rem}

\begin{rem}
\upshape{ 
 The EM algorithm by \citet{RT82} is the method used by \citet{baili12,baili16} to compute the QML estimator of the loadings maximizing the log-likelihood \eqref{eq:LL0iid}. However, no investigation about its convergence to a (global) maximum of the log-likelihood is reported. Nor any indication is given on how to initialize such algorithm. 
 In fact, to the best of our knowledge no formal proof of convergence of such algorithm to the QML estimator exists. Therefore, in this section we followed the literature by studying directly the properties of the QML estimator, thus implicitly taking  for granted the convergence of any numerical procedure to such estimator.
 }
 \end{rem}

\subsection{Factors}\label{sec:FGLS}

In this section we consider estimation of the factors when their dynamics is not specified. 
If the factors are treated as as a sequence of constant parameters, then it is possible to rewrite the mis-specified log-likelihood \eqref{eq:LL0iid}
as (see, e.g., \citealp{AR56} and \citealp{baili12}):
\begin{align}
\ell_{0,\text{\tiny S}} (\bm {\mathcal X};\underline{\bm{\mathcal F}},\underline{\bm\varphi}) 
&\simeq-\frac T2 \log\det(\underline{\bm\Sigma}^\xi) -\frac 12\sum_{t=1}^T \l(\mbf x_{t}-\bar{\mbf x}-\underline{\bm\Lambda}\,\underline{\mbf F}_t\r)^\prime (\underline{\bm\Sigma}^\xi)^{-1}\l(\mbf x_{t}-\bar{\mbf x}-\underline{\bm\Lambda}\,\underline{\mbf F}_t\r).\label{eq:LL0_bis}
\end{align}
It is immediate to see that for known parameters $\bm\varphi$, the estimator of the factors maximizing \eqref{eq:LL0_bis}, is the unfeasible Weighted Least Squares (WLS):
\beq\label{eq:Fgls_comp} 
{\mbf F}_t^{\text{\tiny WLS}} := \l(\bm\Lambda^\prime (\bm\Sigma^{\xi})^{-1}\bm\Lambda\r)^{-1} \bm\Lambda^\prime (\bm\Sigma^{\xi})^{-1}(\mbf x_{t}-\bar {\mbf x}), \quad t=1,\ldots, T.
\eeq
Once we have also the QML estimator of the parameters $\wh{\bm\varphi}^{\text{\tiny QML,S}}$, maximizing the log-likelihood \eqref{eq:LL0iid}, we can compute \eqref{eq:Fgls_comp}, and obtain the feasible WLS:
\beq
\wh{\mbf F}_t^{\text{\tiny WLS}} := \l(\wh{\bm\Lambda}^{\text{\tiny{QML}}\,\prime} (\wh{\bm\Sigma}^{\xi\text{\tiny{QML}}})^{-1}\wh{\bm\Lambda}^{\text{\tiny{QML}}}\r)^{-1} \wh{\bm\Lambda}^{\text{\tiny{QML}}\,\prime} (\wh{\bm\Sigma}^{\xi\text{\tiny{QML}}})^{-1}(\mbf x_{t}-\bar {\mbf x}), \quad t=1,\ldots, T,\label{eq:GLShat}
\eeq
which is nothing else but the classical ``least-squares'' estimator of the factors originally proposed by \citet{bartlett37} (see also \citealp{thomson36}, \citealp{bartlett38}, and \citealp[Chapter 7.2]{lawleymaxwell71}). 
In the large $N$ case, \citet[Theorem 6.1]{baili12} and \citet[Theorem 2]{baili16} prove that:
\beq\label{eq:AEF1}
\l\Vert\wh{\mbf F}_t^{\text{\tiny WLS}}-{\mbf F}_t^{\text{\tiny WLS}}\r\Vert=O_{\mathrm P}\l(\frac 1T\r)+O_{\mathrm P}\l(\frac 1{\sqrt {NT}}\r).
\eeq
The following result follows.
 
\begin{prop}\label{prop:FBAILI16}
For any $t=1,\ldots,T$, under Assumptions \ref{ass:common} through \ref{ass:QML}
if $\sqrt N/T\to 0$, as $N,T\to\infty$,
\[
\sqrt N(\wh{\mbf F}_t^{\text{\tiny \upshape WLS}}-\mbf F_t)\to_d\mathcal N\l(\mbf 0_r,\bm{\mathcal W}_{t}^{\text{\tiny \upshape WLS}}\r),
\]
where $\bm{\mathcal W}_{t}^{\text{\tiny \upshape WLS}}:=(\bm\Sigma_{\Lambda\xi\Lambda})^{-1}
\l\{
\lim_{N\to\infty}\frac {\bm\Lambda^\prime (\bm\Sigma^\xi)^{-1} \E[\bm \xi_t\bm \xi_t^\prime](\bm\Sigma^\xi)^{-1}\bm\Lambda }N
\r\}
(\bm\Sigma_{\Lambda\xi\Lambda})^{-1}$, with $\bm\Sigma_{\Lambda\xi\Lambda}:=\lim_{N\to\infty}\frac{\bm\Lambda^\prime(\bm\Sigma^\xi)^{-1}\bm\Lambda}{N}$.
\end{prop}

If instead we treat the factors as random variables, but we do not model their dynamics, then their optimal (in mean-squared sense) linear estimator is the linear projection of the true factors onto the observed data. Given that we are considering a mis-specified exact static factor model, such linear projection can be approximated by:
 \beq\label{eq:Freg_comp}
{\mbf F}_{t}^{\text{\tiny LP}} := \l(\bm\Lambda^\prime (\bm\Sigma^{\xi})^{-1}\bm\Lambda+\mbf I_r\r)^{-1} \bm\Lambda^\prime (\bm\Sigma^{\xi})^{-1}(\mbf x_{t}-\bar{\mbf x}),\quad t=1,\ldots, T.
\eeq
Once we have also the QML estimator of the parameters $\wh{\bm\varphi}^{\text{\tiny QML,S}}$ maximizing the log-likelihood \eqref{eq:LL0iid} of a static factor model, we obtain the feasible Linear Projection (LP) estimator
\beq\label{eq:Freg_comp_feas}
\wh{\mbf F}_{t}^{\text{\tiny LP}} := \l(\wh{\bm\Lambda}^{\text{\tiny QML,S}\prime} (\wh{\bm\Sigma}^{\xi\text{\tiny QML,S}})^{-1}\wh{\bm\Lambda}^{\text{\tiny QML,S}}+\mbf I_r\r)^{-1} \wh{\bm\Lambda}^{\text{\tiny QML,S}\prime} (\wh{\bm\Sigma}^{\xi\text{\tiny QML,S}})^{-1}(\mbf x_{t}-\bar{\mbf x}),\quad t=1,\ldots, T,
\eeq
which is nothing else but the classical 
``regression'' estimator of the factors proposed by \citet{thomson51} (see also \citealp{scott66}, and \citealp[Chapter 7.4]{lawleymaxwell71}). Clearly, the unfeasible LP estimator \eqref{eq:Freg_comp} has always a smaller MSE than the unfeasible WLS estimator \eqref{eq:Fgls_comp}. Nevertheless, since because of Proposition \ref{prop:LBAILI16}, and Assumptions \ref{ass:common}(a) and \ref{ass:idio}(a), $\Vert\wh{\bm\Lambda}^{\text{\tiny QML,S}}\Vert = O_{\mathrm P}(\sqrt N)$ and $\Vert (\wh{\bm\Sigma}^{\xi\text{\tiny QML,S}})^{-1}\Vert = O_{\mathrm P}(1)$, then, the feasible WLS and LP estimators coincide asymptotically \citep[see also][Proposition 6.1]{baili12}:
\beq\label{eq:AEBLF}
\l\Vert\wh{\mbf F}_{t}^{\text{\tiny WLS}}- \wh{\mbf F}_{t}^{\text{\tiny LP}}\r\Vert =O_{\mathrm P}\l(\frac 1N\r).
\eeq
The following result follows.
 
\begin{prop}\label{prop:FBAILI16bis}
For any $t=1,\ldots,T$, under Assumptions \ref{ass:common} through \ref{ass:QML}
if $\sqrt N/T\to 0$, as $N,T\to\infty$,
\[
\sqrt N(\wh{\mbf F}_t^{\text{\tiny \upshape LP}}-\mbf F_t)\to_d\mathcal N\l(\mbf 0_r,\bm{\mathcal W}_{t}^{\text{\tiny \upshape WLS}}\r),
\]
where $\bm{\mathcal W}_{t}^{\text{\tiny \upshape WLS}}:=(\bm\Sigma_{\Lambda\xi\Lambda})^{-1}
\l\{
\lim_{N\to\infty}\frac {\bm\Lambda^\prime (\bm\Sigma^\xi)^{-1} \E[\bm \xi_t\bm \xi_t^\prime](\bm\Sigma^\xi)^{-1}\bm\Lambda }N
\r\}
(\bm\Sigma_{\Lambda\xi\Lambda})^{-1}$, with $\bm\Sigma_{\Lambda\xi\Lambda}:=\lim_{N\to\infty}\frac{\bm\Lambda^\prime(\bm\Sigma^\xi)^{-1}\bm\Lambda}{N}$.
\end{prop}

The asymptotic covariance matrix of the estimated factors is the same for both the WLS and LP estimator. It has a sandwich form reflecting the mis-specification of the cross-sectional correlation structure of the idiosyncratic components. It differs from the asymptotic covariance of the PC estimator given in Proposition \ref{prop:FPCA}, which is an OLS-type estimator. Notice that, in general, under our assumptions of an approximate factor model, we cannot say if the WLS/LP estimators are more or less efficient than the PC estimator, i.e., the matrix $\bm{\mathcal W}_{t}^{\text{\tiny \upshape OLS}}-\bm{\mathcal W}_{t}^{\text{\tiny \upshape WLS}}$ is neither positive nor negative definite. The only thing we can say is that neither the WLS/LP nor the PC estimator are the most efficient, the efficiency loss of the former being due to the use of a mis-specified log-likelihood, and the efficiency loss of the latter being due to its non-parametric nature. 

Clearly, if the true model were an exact factor model then we would have $\E[\bm\xi_t\bm\xi_t^\prime]=\bm\Sigma^\xi$, and $\bm{\mathcal W}_{t}^{\text{\tiny \upshape WLS}}$ would reduce to $(\bm\Sigma_{\Lambda\xi\Lambda})^{-1}$, which is the Gauss-Markov lower bound under cross-sectional heteroskedasticity. In this case the WLS/LP estimators are always more efficient than the PC estimator, since the latter, being non-parametric does not even account explicitly for idiosyncratic cross-sectional heteroskedasticity.

It is clear from Propositions \ref{prop:FBAILI16} and \ref{prop:FBAILI16bis}, that in the fixed $N$ case no consistency can be proved for the estimated factors. Indeed, for fixed $N$ there is no hope of estimating the whole $rT$-dimensional vector of factors $\bm F$ consistently, simply due to a lack of degrees of freedom. Moreover, for any given $t=1,\ldots,T$, the WLS/LP estimators of the factors in \eqref{eq:Fgls_comp} are the result of the aggregation of data along the cross-sectional dimension, i.e., they are the result of a cross-sectional regression. Hence, for consistency we must require $N\to\infty$.\footnote{Notice that although the PC estimator of the factors proposed by \citet{bai03} is just an eigenvector so no cross-sectional aggregation is required, we still require $N\to\infty$ to prove its consistency. Indeed, in this case we need to compute eigenvectors of a $T\times T$ covariance matrix, which are consistently estimating the true eigenvectors only if $N\to\infty$.} This is another manifestation of the blessing of dimensionality. 

In fact, from \eqref{eq:AEF1} it is clear that we must have also $T\to\infty$ in order to consistently estimate the factors. This is because the feasible WLS/LP estimators in \eqref{eq:GLShat} and \eqref{eq:Freg_comp_feas} require the use of the QML estimator of the loadings and the idiosyncratic variances, which are consistent if and only if $T\to\infty$ (see Proposition \ref{prop:LBAILI16}).


\begin{rem}
\upshape{Two other estimators of the factors are worth recalling. First, if we compute the WLS estimator in \eqref{eq:Fgls_comp} using the PC estimators of the loadings in \eqref{eq:LPC} and of the idiosyncratic variances\footnote{The PC estimator of the idiosyncratic variances is $T^{-1}\sum_{t=1}^T(x_{it}-\wh{\bm\lambda}_i^{\text{\tiny PC}\prime}\wh{\mbf F}_t^{\text{\tiny PC}})^2$.}, we get the Generalized PC estimator of the factors studied by \citet{BT11} and \citet{choi12}, which, by virtue of \eqref{eq:AE3}, is asymptotically equivalent to the feasible WLS estimator in \eqref{eq:GLShat}.
Second, if we further simplify the log-likelihood \eqref{eq:LL0_bis} by treating the idiosyncratic components as cross-sectionally homoskedastic (see Remark \ref{rem:TB}), and we use the QML estimator of the loadings in \eqref{eq:LML} to compute the WLS estimator in \eqref{eq:Fgls_comp}, we obtain a feasible OLS estimator of the factors which, by \eqref{eq:AETB}, is asymptotically equivalent to the PC estimator defined in \eqref{eq:FPC}.
}
\end{rem}

\section{QML estimation of an approximate dynamic factor model}\label{sec:dynamic}

In this section we review the case in which the dynamics of the factors is explicitly modeled, so we consider an approximate dynamic factor model described by the linear system:
\begin{align}
x_{it}&=\alpha_i+\bm\lambda_i^\prime\mbf F_t+\xi_{it}, \quad i=1,\ldots, N, \quad t=1,\ldots, T,\label{eq:SDFM1Rbis}\\
\mbf F_t&=\mbf A\mbf F_{t-1}+\mbf H\mbf u_t,\label{eq:VARsimplebis}
\end{align}
together with Assumptions \ref{ass:common} through \ref{ass:VAR}. In this case, $\{\mbf F_t\}$ is an autocorrelated stochastic process and $\bm\Omega^F$ is a function of $\mbf A$ and $\mbf H$.

For a dynamic factor model the log-likelihood \eqref{eq:LL0acf} can be decomposed as: 
\beq\label{eq:LL0dyn}
\ell_{0,\text{\tiny D}}(\bm {\mathcal X};\underline{\bm\varphi})=\ell_{0,\text{\tiny D}}(\bm {\mathcal X}|\bm{\mathcal F};\underline{\bm\varphi})+\ell_{0,\text{\tiny D}}(\bm {\mathcal F};\underline{\bm\varphi})-\ell_{0,\text{\tiny D}}(\bm {\mathcal F}|\bm{\mathcal X};\underline{\bm\varphi}),
\eeq
where (recall that $\mbf F_0=\mbf 0_r$ because of Assumption \ref{ass:VAR}(g))
\begin{align}
&\ell_{0,\text{\tiny D}}(\bm {\mathcal X}|\bm{\mathcal F};\underline{\bm\varphi})\simeq -\frac T2\log\det(\underline{\bm\Sigma}^\xi)-\frac 12\sum_{t=1}^T
(\mbf x_t-\bar{\mbf x}-\underline{\bm\Lambda}\mbf F_t)^\prime (\underline{\bm\Sigma}^\xi)^{-1}(\mbf x_t-\bar{\mbf x}-\underline{\bm\Lambda}\mbf F_t),\nn\\
&\ell_{0,\text{\tiny D}}(\bm{\mathcal F};\underline{\bm\varphi})\simeq-\frac T2\log\det(\underline{\mbf H}\, \underline{\mbf H}^\prime)-\frac 12\sum_{t=1}^T 
(\mbf F_t-\underline{\mbf A}\mbf F_{t-1})^\prime
(\underline{\mbf H}\, \underline{\mbf H}^\prime)^{-1}
(\mbf F_t-\underline{\mbf A}\mbf F_{t-1}),\label{eq:LL0dyn3}\\
&\ell_{0,\text{\tiny D}}(\bm {\mathcal F}|\bm{\mathcal X};\underline{\bm\varphi})\simeq -\frac 12\sum_{t=1}^T \log\det\l(\mathbb C\text{ov}_{\underline{\bm\varphi}}(\mbf F_t|\bm{\mathcal X})\r)-\frac 12\sum_{t=1}^T
(\mbf F_t-\E_{\underline{\bm\varphi}}[\mbf F_t|\bm{\mathcal X}])^\prime
\l(\mathbb C\text{ov}_{\underline{\bm\varphi}}(\mbf F_t|\bm{\mathcal X})\r)^{\!-1}\!\!
(\mbf F_t-\E_{\underline{\bm\varphi}}[\mbf F_t|\bm{\mathcal X}]),\nn
\end{align}
with $\mathbb C\text{ov}_{\underline{\bm\varphi}}(\mbf F_t|\bm{\mathcal X})=\E_{\underline{\bm\varphi}}[
(\mbf F_t-\E_{\underline{\bm\varphi}}[\mbf F_t|\bm{\mathcal X}])(\mbf F_t-\E_{\underline{\bm\varphi}}[\mbf F_t|\bm{\mathcal X}])^\prime|\bm{\mathcal X}]$. The parameters to be estimated are 
$\bm{\varphi}=(\bm\phi^\prime,\bm\theta^\prime)^\prime$ with $\bm\phi=(\mathrm{vec}(\bm\Lambda)^\prime, \sigma_1^2,\ldots,\sigma_N^2)^\prime$ and $\bm\theta=(\mathrm{vec}(\mbf A)^\prime, \mathrm{vec}(\mbf H)^\prime)^\prime$, so ${\bm\varphi}$ is a $Q=(r+1)N+2r^2$-dimensional vector. We shall denote the maximizer of $\ell_{0,\text{\tiny D}}(\bm {\mathcal X};\underline{\bm\varphi})$ as $\wh{\bm\varphi}^{\text{\tiny QML,D}}$.

We study QML estimation of the approximate dynamic factor model under the following simplifying assumption.

\begin{ass}\label{ass:tails}
$\,$
\begin{compactenum}
\item [(a)] For all $N,T\in\mathbb N$, $(\bm\xi_{1}^\prime\cdots \bm\xi_{T}^\prime)^\prime\sim \mathcal N(\mbf 0_{NT},\bm\Omega^\xi)$.
\item [(b)] For all $T\in\mathbb N$
$(\mbf u_1^\prime\cdots \mbf u_T^\prime)^\prime\sim \mathcal N(\mbf 0_{rT},\mbf I_{rT})$.
\end{compactenum}
\end{ass}

Gaussianity is a standard assumption in this setting, see, e.g., \citet{shumwaystoffer82}, \citet{watsonengle83}, and \citet{DGRqml}. It strengthens the moment conditions made in Assumptions \ref{ass:idio} and \ref{ass:VAR}. The non-Gaussian case is discussed in \citet{quahsargent93} and \citet{BLqml}.
In particular, from this assumption it follows that, for all $N,T\in\mathbb N$ (see Appendix \ref{app:XFZF} for a proof):
\beq\label{eq:gaussXF}
\l(\ba{c}
\bm{\mathcal X}\\
\bm{\mathcal F}
\ea\r)\sim \mathcal N\l(\l(\ba{c}
\bm{\mathcal A}\\
\mbf 0_{rT}
\ea\r),\l(\ba{cc}
 \bm {\mathfrak L}\,\bm\Omega^F\bm {\mathfrak L}^\prime + \bm\Omega^\xi&  \bm {\mathfrak L}\,\bm\Omega^F\\
\bm\Omega^F  \bm {\mathfrak L}^\prime& \bm\Omega^F
\ea\r)\r).
\eeq


%

\subsection{Loadings}\label{sec:EMload}
The classical approach to estimate a dynamic factor model by QML is based on the maximization of the prediction error log-likelihood which depends on the linear prediction of the factors, usually computed via the Kalman filter, which is defined in the next section. This approach is equivalent to maximizing the log-likelihood \eqref{eq:LL0dyn} (\citealp[see, e.g.,][Chapter 4]{hannan2012statistical}). Although there is no analytic solution for the QML estimator of the parameters obtained in this way, numerical
maximization is standard practice at least for low dimensional exact dynamic factor models, see, e.g., \citet{stockwatson89,stockwatson91} and \citet[Chapter 3.4]{harvey90}. In high-dimensions \citet{JK15} propose to maximize the prediction error log-likelihood subject to a  preliminary step in which the data are projected onto a lower dimensional space. It is not clear, however, what are the effects of this first step on the asymptotic properties of the final estimators.

In high-dimensions maximization of the log-likelihood \eqref{eq:LL0dyn} is usually achieved by means of EM algorithm, which is an iterative procedure proposed by \citet{DLR77} as a general way to achieve QML estimation when dealing with incomplete information (see also \citealp{sundberg74,sundberg76}, and \citealp[Chapter 8]{sundberg2019statistical}, for EM and incomplete data
in exponential families). The use of the EM algorithm for estimating small dimensional state space models dates back to 
\citet{shumwaystoffer82}, 
\citet{watsonengle83}, 
 \citet{harveypeters90}, and \citet{GZ96}, while its use to estimate a high-dimensional approximate dynamic factor model was 
first suggested by \citet{quahsargent93}.
The asymptotic properties of the estimated factors were firstly derived by \citet{DGRqml}, and then completed by \citet{BLqml} who derived also the asymptotic properties of the estimated loadings. 

In the present context, the EM algorithm has two main features: (i) it gives closed form solutions for the estimated parameters, and (ii) it runs in few iterations. In a nutshell, it works as follows. Consider a given iteration $k\ge 0$ of the EM algorithm and let us assume to have an estimate of the parameters $\wh{\bm\varphi}^{(k)}$. Then, for each $k$ repeat two steps. 
\ben
[wide, labelwidth=!, labelindent=0pt]
\item [\underline{E-step}] Compute the expected full-information log-likelihood with respect to the conditional distribution of the factors given the data and computed using the estimated parameters $\wh{\bm\varphi}^{(k)}$:
$$
\mathcal Q(\underline{\bm\varphi},\wh{\bm\varphi}^{(k)}):=\E_{\wh{\bm\varphi}^{(k)}}[\ell_{0,\text{\tiny D}}(\bm  {\mathcal X},\bm  {\mathcal F};\underline{\bm\varphi} )|\bm {\mathcal X}];
$$
%

\item [\underline{M-step}] Compute a new estimate of the parameters maximizing the expected full-information log-likelihood:
\[
\wh{\bm\varphi}^{(k+1)}=\arg\max_{\underline{\bm\varphi}} \mathcal Q(\underline{\bm\varphi},\wh{\bm\varphi}^{(k)}).
\]
\een

The algorithm is intialized and terminated as follows.
\ben
[wide, labelwidth=!, labelindent=0pt]
\item [\underline{Initialization}] The loadings 
are inizialized with their PC estimator
defined in \eqref{eq:LPC}, then, given also the PC estimates of the factors and of the idiosyncratic components, we estimate: 
(i) the VAR parameters, and (ii) the idiosyncratic variances.

\item [\underline{Stopping}] The iterations are stopped at iteration $k^*$ such that the log-likelihoods $\ell_{0,\text{\tiny D}}(\bm{\mathcal X};\wh{\bm{\varphi}}^{(k^*+1)})$ and $\ell_{0,\text{\tiny D}}(\bm{\mathcal X};\wh{\bm{\varphi}}^{(k^*)})$ do not differ more than a pre-specified threshold. We define the EM estimator of the parameters as $\wh{\bm{\varphi}}^{\text{\tiny EM}}:=\wh{\bm{\varphi}}^{(k^*+1)}$. 
\een

The rationale for the EM algorithm is the following. By taking the conditional expectations of the log-likelihood \eqref{eq:LL0dyn} and computed using $\wh{\bm\varphi}^{(k)}$, we get:
\begin{align}
\ell_{0,\text{\tiny D}}(\bm {\mathcal X};\underline{\bm\varphi})&=
\E_{\wh{\bm\varphi}^{(k)}}[\ell_{0,\text{\tiny D}}(\bm  {\mathcal X},\bm  {\mathcal F};\underline{\bm\varphi} )|\bm {\mathcal X}]
-\E_{\wh{\bm\varphi}^{(k)}}[\ell_{0,\text{\tiny D}}(\bm {\mathcal F}|\bm {\mathcal X};\underline{\bm\varphi})|\bm {\mathcal X}] \nn\\
&=:\mathcal Q(\underline{\bm\varphi},\wh{\bm\varphi}^{(k)})-\mathcal H(\underline{\bm\varphi},\wh{\bm\varphi}^{(k)}), \; \text{say}.\label{eq:LLbayes_exp}
\end{align}
The QML estimator of $\bm{\varphi}$ is then a maximum of the right hand side of \eqref{eq:LLbayes_exp}. In fact, since, by definition of Kullback-Leibler divergence, $\mathcal H(\underline{\bm\varphi},\wh{\bm\varphi}^{(k)})\le \mathcal H(\wh{\bm\varphi}^{(k)},\wh{\bm\varphi}^{(k)})$ for any pair $(\underline{\bm\varphi},\wh{\bm\varphi}^{(k)})$ and any  $k\ge0$ (see \citealp[Lemma 1]{DLR77}, and \citealp[Theorem 3]{wu83}),  it is enough to maximize $\mathcal Q(\underline{\bm\varphi},\wh{\bm\varphi}^{(k)})$. It follows that the EM algorithm defines a continuous path in the parameter space from the starting point to the stopping point along which the log-likelihood monotonically increases without leaping over valleys.

In general, the implementation of the EM steps might be non-trivial. However, things simplify considerably in the present setup.
First, the full-information expected log-likelihood $\mathcal Q(\underline{\bm\varphi},\wh{\bm\varphi}^{(k)})$ can be further decomposed as:
\begin{align}\label{eq:LLbayes_exp_bis}
\mathcal Q(\underline{\bm\varphi},\wh{\bm\varphi}^{(k)})
=&\,\E_{\wh{\bm\varphi}^{(k)}}[\ell_{0,\text{\tiny D}}(\bm {\mathcal X}|\bm {\mathcal F};\underline{\bm\varphi})|\bm {\mathcal X}]+\E_{\wh{\bm\varphi}^{(k)}}[\ell_{0,\text{\tiny D}}(\bm {\mathcal F};\underline{\bm\varphi})|\bm {\mathcal X}].
\end{align}
Then,  it is clear from \eqref{eq:LL0dyn3} that, in order to compute the expected full information log-likelihood  in the E-step, we just need to compute the conditional moments of the factors: $\E_{\wh{\bm\varphi}^{(k)}}[\mbf F_t| \bm{\mathcal X}]$, $\E_{\wh{\bm\varphi}^{(k)}}[\mbf F_t\mbf F_t^\prime| \bm{\mathcal X}]$, and $\E_{\wh{\bm\varphi}^{(k)}}[\mbf F_t\mbf F_{t-1}^\prime| \bm{\mathcal X}]$. And, in turn, these moments can be obtained by means of the Kalman smoother implemented using the estimated parameters $\wh{\bm\varphi}^{k}$, as detailed in the next section.

The M-step has then a closed form solution. Indeed, from \eqref{eq:LL0dyn3} we immediately see that the final EM estimator of the parameters, $\bm\phi$, in the measurement equation \eqref{eq:SDFM1Rbis} is:
\begin{align}
\wh{\bm\lambda}_i^{\text{\tiny EM}}&=\l(\sum_{t=1}^T \E_{\wh{\bm\varphi}^{(k^*)}}[\mbf F_t\mbf F_t^\prime| \bm{\mathcal X}]\r)^{-1}\l(\sum_{t=1}^T \E_{\wh{\bm\varphi}^{(k^*)}}[\mbf F_t| \bm{\mathcal X}]\,(x_{it}-\bar x_i)\r), \quad i=1,\ldots, N,\label{eq:LEM}\\
\wh{\sigma}_i^{2\,\text{\tiny EM}}&=\frac 1T \sum_{t=1}^T
\l\{(x_{it}-\bar x_i)^2+
\wh{\bm\lambda}_i^{\text{\tiny EM}\prime}
\l(
\E_{\wh{\bm\varphi}^{(k^*)}}[\mbf F_t\mbf F_t^\prime| \bm{\mathcal X}]
\r)
\wh{\bm\lambda}_i^{\text{\tiny EM}}\r.\nn\\
&\qquad \qquad \l.-2(x_{it}-\bar x_i)\wh{\bm\lambda}_i^{\text{\tiny EM}\prime}\E_{\wh{\bm\varphi}^{(k^*)}}[\mbf F_t| \bm{\mathcal X}]
\r\}, \qquad\qquad\qquad\;\;\quad\quad i=1,\ldots, N,\label{eq:SEM}
\end{align}
such that $\wh{\bm\Lambda}^{\text{\tiny EM}}:=(\wh{\bm\lambda}_1^{\text{\tiny EM}}\cdots \wh{\bm\lambda}_N^{\text{\tiny EM}})^\prime$ is the estimated matrix of all loadings and $\wh{\bm\Sigma}^{\xi\text{\tiny EM}}:=\text{diag}(\wh{\sigma}_1^{2\,\text{\tiny EM}},\ldots, \wh{\sigma}_N^{2\,\text{\tiny EM}})$ is the estimated diagonal matrix of idiosyncratic variances.

Notice that, since to compute the estimators in \eqref{eq:LEM} and \eqref{eq:SEM} we use the mis-specified expected conditional log-likelihood \eqref{eq:LLbayes_exp_bis} of an exact factor model, we reduce a potentially hard maximizations problem of estimating a high-dimensional parameter vector $\bm\phi$  
into the maximizations of $N$ expected log-likelihoods, each depending on a finite $r+1$-dimensional parameter vector $(\text{vec}(\bm\lambda_i)^\prime,\sigma_i^2)^\prime$. This can be seen as an extreme form of regularization of the idiosyncratic covariance matrix which makes estimation of the large dimensional parameter vector $\bm\phi$ straightforward.

Closed form expressions are immediately derived also for the final estimators of the parameters, $\bm\theta$, in the state equation \eqref{eq:VARsimplebis}: 
\begin{align}
\wh{\mbf A}^{\text{\tiny EM}}&=\l(\sum_{t=1}^T \E_{\wh{\bm\varphi}^{(k^*)}}[\mbf F_{t}\mbf F_{t-1}^\prime| \bm{\mathcal X}]\r)\l(\sum_{t=2}^T 
\E_{\wh{\bm\varphi}^{(k^*)}}[\mbf F_{t-1}\mbf F_{t-1}^\prime| \bm{\mathcal X}]
\r)^{-1},\label{eq:AEM}\\
\wh{\mbf H}^{\text{\tiny EM}}&=\l[\frac 1T\sum_{t=1}^T\l\{
\E_{\wh{\bm\varphi}^{(k^*)}}[\mbf F_t\mbf F_t^\prime| \bm{\mathcal X}]
-\l(
\E_{\wh{\bm\varphi}^{(k^*)}}[\mbf F_{t-1}\mbf F_t^\prime| \bm{\mathcal X}]
\r)
\wh{\mbf A}^{\text{\tiny EM}}
\r\}\r]^{1/2}.\label{eq:HEM}
\end{align}

For a log-likelihood belonging to the exponential family, as the Gaussian one, it is possible to prove that the EM algorithm produces a sequence of estimators $\{\wh{\bm\lambda}_i^{(k)}\}$ which converges to a local maximum of the log-likelihood \eqref{eq:LL0dyn} as $k\to\infty$, say $\wh{\bm\lambda}_i^{(\infty)}$ \citep[Theorem 6]{wu83}. 


It is then common practice to consider the output of the EM algorithm $\wh{\bm\lambda}_i^{\text{\tiny EM}}$ as equivalent to the QML estimator $\wh{\bm\lambda}_i^{\text{\tiny QML,D}}$. This, however, is not always true, because there are two additional sources of errors: (i) the local maximum might not be the global maximum, and (ii) the EM algorithm runs for a finite number of iterations $k^*$.

The two aforementioned sources of error are addressed by \citet[Proposition 2]{BLqml} who prove that $\wh{\bm\lambda}_i^{\text{\tiny EM}}$ is asymptotically equivalent to the QML estimator $\wh{\bm{\lambda}}_i^{\text{\tiny QML,D}}$ maximizing the full log-likelihood \eqref{eq:LL0dyn}, which, in turn,  is asymptotically equivalent to the QML estimator $\wh{\bm{\lambda}}_i^{\text{\tiny QML,S}}$ obtained by maximizing the log-likelihood of a static factor model \eqref{eq:LL0iid}. Namely,
\beq\label{eq:AEBL1}
\l\Vert \wh{\bm\lambda}_i^{(\infty)} - \wh{\bm\lambda}_i^{\text{\tiny QML,D}}\r\Vert = O_{\mathrm P}\l(\frac 1N\r)+O_{\mathrm P}\l(\frac 1T\r)+O_{\mathrm P}\l(\frac 1{\sqrt{NT}}\r),
\eeq
and (up to logarithmic terms)
\begin{align}
&\l\Vert\wh{\bm\lambda}_i^{\text{\tiny EM}} -\wh{\bm\lambda}_i^{(\infty)}\r\Vert = O_{\mathrm P}\l(\frac 1N\r)+O_{\mathrm P}\l(\frac 1T\r)+O_{\mathrm P}\l(\frac 1{\sqrt{NT}}\r),\label{eq:AEBL2}\\
&\l\Vert \wh{\bm\lambda}_i^{\text{\tiny QML,D}}- \wh{\bm\lambda}_i^{\text{\tiny QML,S}}\r\Vert=O_{\mathrm P}\l(\frac 1N\r). \label{eq:AEBL3}
\end{align}
In particular, \eqref{eq:AEBL1} is a consequence of the fact that we initialize the EM algorithm with the PC estimator which is fully identified under our Assumption \ref{ass:ident}(b), so the sequence $\{\wh{\bm\lambda}_i^{(k)}\}$ actually converges to the global identified maximum $\wh{\bm\lambda}_i^{\text{\tiny QML,D}}$ (see also \citealp[Section 4]{ruud91},  for a similar result in the case of one-to-one mapping from the factors to the data, corresponding to the case of no idiosyncratic component). 
Moreover, \eqref{eq:AEBL2} follows from the fact that the numerical error is proportional to the information loss between using the expected joint log-likelihood 
in \eqref{eq:LLbayes_exp_bis} instead of the
full log-likelihood \eqref{eq:LL0dyn}, i.e., it depends on the fraction of missing information due to the fact that the factors are not observed (\citealp{sundberg76}, \citealp{MR94}, and \citealp[Chapter 3.9]{MLT07}). Last, \eqref{eq:AEBL3} follows from the fact that the exact factor model log-likelihood \eqref{eq:LL0iid} and the approximate factor model log-likelihood \eqref{eq:LL0dyn} are both asymptotically equivalent to the log-likelihood of $\bm{\mathcal X}$ conditional on the factors, which is the same in both cases, i.e., $\ell_{0,\text{\tiny S}}(\bm {\mathcal X}|\bm{\mathcal F};\underline{\bm\varphi})=\ell_{0,\text{\tiny D}}(\bm {\mathcal X}|\bm{\mathcal F};\underline{\bm\varphi})$, so asymptotically their maxima should coincide and they both coincide with the unfeasible OLS estimator (see also Remark \ref{rem:BTintuition}).

From \eqref{eq:AEBL1}, \eqref{eq:AEBL2}, and \eqref{eq:AEBL3}, jointly with Proposition \ref{prop:LBAILI16}, it follows that.

\begin{prop}\label{prop:BLload} For any $i=1,\ldots, N$, under Assumptions \ref{ass:common} through \ref{ass:tails}: 
\ben
\item [(a)] if $\sqrt {T\log T}/N\to 0$, as $N,T\to\infty$, 
\[
\sqrt T(\wh{\bm\lambda}_i^{\text{\tiny \upshape EM}}-\bm\lambda_i)\to_d\mathcal N\l(\mbf 0_r,\bm{\mathcal V}_{i}^{\text{\tiny \upshape OLS}}\r),
\]
where $\bm{\mathcal V}_{i}^{\text{\tiny \upshape OLS}}=(\bm\Gamma^F)^{-1}
\l\{
\lim_{T\to\infty} \frac{ \E[\bm F^\prime \bm\zeta_i\bm\zeta_i^\prime\bm F]}{T}
\r\}
(\bm\Gamma^F)^{-1}=\lim_{T\to\infty} \frac{ \E[\bm F^\prime \bm\zeta_i\bm\zeta_i^\prime\bm F]}{T}$ (because of Assumption \ref{ass:ident}(b));

\item [(b)] as $N,T\to\infty$,
$\min(\sqrt {T/\log N},\sqrt N)\l\Vert
\wh{\sigma}_i^{2\text{\tiny EM}}
-\sigma_i^2
\r\Vert=O_{\mathrm P}(1)$. 
\een
\end{prop}

Up to logarithmic terms, the EM estimator of the loadings has the same asymptotic properties of the QML estimator of a static factor model. Therefore, the same comments made after Proposition \ref{prop:LBAILI16} about efficiency apply also in this case. Moreover, the EM estimator of the loadings has also the same asymptotic properties of the PC estimator. This is not surprising since the EM algorithm is initialized with the PC estimator, and, therefore, $\wh{\bm\lambda}_i^{\text{\tiny EM}}$ can be seen as a one-step estimator, which, by construction, must also be consistent and as efficient as the initial estimator $\wh{\bm\lambda}_i^{\text{\tiny PC}}$ (see, e.g., \citealp[Section 5.7]{VDV2000}, and \citealp[Theorem 4.3]{LC06}).


\subsection{Factors}\label{sec:KFKS}

In a dynamic factor model, the factors are explicitly treated as autocorrelated random variables, thus, their optimal linear predictor is the linear projection of the true factors onto the all available data which is collected into the $NT$-dimensional vector $\bm {\mathcal X}$. For generic values of the parameters and any given $t=1,\ldots,T$, such prediction is given by the linear projection: $\text{Proj}_{\underline{\bm\varphi}}\l({\mbf F}_t|\bm{\mathcal X}\r) =: \bm{\mathcal P}_{\underline{\bm\varphi},t}^\prime\l(\bm{\mathcal X}-\underline{\bm{\mathcal A}}\r)$, where $\bm{\mathcal P}_{\underline{\bm\varphi},t}$ is an $NT\times r$ matrix, which depends on $\underline{\bm\varphi}$, and such that it minimizes the mean-squared error (MSE):
\beq\label{eq:mse}
\bm{\mathcal P}_{\underline{\bm\varphi},t}=\arg\!\!\!\!\!\!\!\!\min_{\underline{\bm{\mathcal P}}_{\,t}\in\mathbb R^{r\times {NT}}} \E_{\underline{\bm\varphi}} \l[\l(\mbf F_t-\underline{\bm{\mathcal P}}_{\,t}^\prime\l(\bm{\mathcal X}-\underline{\bm{\mathcal A}}\r)\r)
\l(\mbf F_t-\underline{\bm{\mathcal P}}_{\,t}^\prime\l(\bm{\mathcal X}-\underline{\bm{\mathcal A}}\r)\r)^\prime \r],\quad t=1,\ldots, T.
\eeq
%
By solving \eqref{eq:mse} for all $t=1,\ldots, T$ and for known parameters $\bm\varphi$, we obtain:
\begin{align}
 \text{Proj}\l(\bm{\mathcal F}|\bm{\mathcal X}\r)&=\bm\Omega^F {\bm {\mathfrak L}}^\prime \l({\bm {\mathfrak L}}\,{\bm\Omega}^{F} {\bm {\mathfrak L}}^\prime+\bm\Omega^\xi\r)^{-1}\l(\bm{\mathcal X}-{\bm{\mathcal A}}\r).\label{eq:brembate}
\end{align}
Now, under Assumption \ref{ass:tails}, it is clear that $\text{Proj}_{\underline{\bm\varphi}}\l(\bm{\mathcal F}|\bm{\mathcal X}\r)=\E_{\underline{\bm\varphi}} [\bm{\mathcal F}|\bm{\mathcal X}]$, so \eqref{eq:brembate} coincides with the optimal predictor of the factors. In this case,  \eqref{eq:brembate} can also be derived by noticing that given the joint distribution of $\bm{\mathcal F}$ and $\bm{\mathcal X}$ in \eqref{eq:gaussXF}, we have (the conditional covariance is the Schur complement of the unconditional covariance):
\begin{align}
\bm{\mathcal F}|\bm{\mathcal X}\sim\mathcal N\l(\bm\Omega^F\bm{\mathfrak L}^\prime\l({\bm {\mathfrak L}}\,{\bm\Omega}^{F} {\bm {\mathfrak L}}^\prime+\bm\Omega^\xi\r)^{-1}(\bm{\mathcal X}-\bm{\mathcal A}),\, \bm\Omega^F-\bm\Omega^F\bm{\mathfrak L}^\prime\l({\bm {\mathfrak L}}\,{\bm\Omega}^{F} {\bm {\mathfrak L}}^\prime+\bm\Omega^\xi\r)^{-1}\bm{\mathfrak L}\bm\Omega^F\r).\label{eq:brembote}
\end{align}

In practice, we know that we can always estimate $\bm{\mathcal A}$ with its QML estimator $\bar{\bm{\mathcal X}}$ and, as in the previous sections, we treat the idiosyncratic components as if they were 
%
serially and cross-sectionally uncorrelated. In this case, from \eqref{eq:brembate}, we get the following estimator of the factors:
\begin{align}
\mbf F_{t}^{\text{\tiny WK}}
&:=\l(\bm\iota_t^\prime\otimes \mbf I_r\r)
\l(
\mbf I_T\otimes\l(
{\bm {\Lambda}}^\prime
( \bm\Sigma^\xi)^{-1}
{\bm {\Lambda}} \r)
+ ({\bm\Omega}^{F})^{-1}\r)^{-1} 
\l(\mbf I_T\otimes {\bm {\Lambda}}^\prime (\bm\Sigma^\xi)^{-1}\r)
\l(\bm{\mathcal X}-\bar{\bm{\mathcal X}}\r)\label{eq:Fsmooth_comp}\\
&=\l(\bm\iota_t^\prime\otimes \mbf I_r\r)\l(
\mbf I_T\otimes\l(
{\bm {\Lambda}}^\prime
( \bm\Sigma^\xi)^{-1}
{\bm {\Lambda}} \r)
+ ({\bm\Omega}^{F})^{-1}\r)^{-1} 
\l(\mbf I_T\otimes {\bm {\Lambda}}^\prime (\bm\Sigma^\xi)^{-1}\r)
\l(\ba{c}
(\mbf x_{1}-\bar{\mbf x})^\prime\\
\vdots\\
(\mbf x_{T}-\bar{\mbf x})^\prime
\ea
\r),\quad t=1,\ldots, T.\nn
\end{align}
This is nothing else but the unfeasible estimator obtained by taking the inverse Fourier transform of the smoother originally proposed by \citet{wiener49} and \citet{kolm41}. 
Now since $\{\mbf F_t\}$ is an autocorrelated process, $\bm\Omega^F$ is not diagonal and at each point in time the estimator of the factors \eqref{eq:Fsmooth_comp} is a weighted average of all $T$ present, past, and future values of all $N$ time series. Hence, differently from the PC, WLS, and LP estimators, this estimator is obtained not only by cross-sectional  aggregation, but also by temporal aggregation of the data, thus effectively taking into account all its 2nd order dependencies. Given that \eqref{eq:Fsmooth_comp} uses all available history of $\{\mbf x_t\}$, it is also called a smoother estimator, rather than a filter which, at a given point in time $t$, would use only past and present information.

Direct implementation of \eqref{eq:Fsmooth_comp} is still unfeasible due to the presence of the $rT\times rT$ full matrix, ${\bm\Omega}^{F}$ which needs to be estimated and inverted.\footnote{Although in this paper we do not consider such possibility, it is also worth noticing that in the case of macroeconomic data we often have factors with a singular spectral density, i.e., of rank $q<r$ \citep[see, e.g., the empirical evidence in][]{dagostinogiannone12}. This implies $\mbox{rk}(\bm\Omega^F)<rT$, so that, in such case, the smoother estimator \eqref{eq:Fsmooth_comp} is not even defined.} 
There are two main solutions to this problem. First, we could ignore the autocorrelation of the factors and impose $\bm\Omega^F=\mbf I_{rT}$, because of Assumption \ref{ass:ident}(b), and then it is easily seen that \eqref{eq:Fsmooth_comp} would become the LP estimator defined in \eqref{eq:Freg_comp_feas}.

Second, in the spirit of this section on estimation of a dynamic factor model, we can use a parametric model for the dynamics of the factors, so that $\bm\Omega^F$ is  parametrized accordingly.
In this case, we can estimate the factors by means of the well-known iterative procedures proposed by \citet{kalman60}, producing the Kalman filter and smoother estimators, together with their conditional second moments. For known parameters $\bm\varphi$, these are nothing else but the linear projections: ${\mbf F}_{t|t}:= \mbox{Proj}_{\bm\varphi}(\mbf F_t|\mbf x_t,\ldots, \mbf x_1)$ and ${\mbf F}_{t|T}:= \mbox{Proj}_{\bm\varphi}(\mbf F_t|\bm{\mathcal X})$, respectively (see also \citealp{rauch63}, and \citealp{dejong89}). Hence, under the chosen parametrization of $\bm\Omega^F$,  the Kalman smoother coincides with the estimator given in \eqref{eq:Fsmooth_comp}.

It is now sensible to compute the Kalman smoother by using the EM estimator of the parameters $\wh{\bm\varphi}^{\text{\tiny EM}}$, which approximate the QML estimators, defined in \eqref{eq:LEM} through \eqref{eq:HEM}. However, those estimators of the parameters depend again on the factors. To break this mutual dependence, we can iterate between the Kalman smoother and the EM algorithm to obtain joint estimates of the parameters and the factors. In particular, at each iteration $k\ge 0$ of the EM algorithm we can implement the Kalman smoother using the parameters $\bm\varphi^{(k)}$ estimated at the previous iteration, thus giving estimated factors, $\mbf F_{t|T}^{(k)}$, their conditional MSE, $\mbf P_{t|T}^{(k)}$, and lag-1 conditional MSE, $\mbf C_{t,t-1|T}^{(k)}$ (see below for the explicit definitions), necessary to compute the expected log-likelihood in the E-step. By means of these moments we can compute new estimates of the parameters $\bm\varphi^{(k+1)}$ in the M-step, as detailed in \eqref{eq:LEM} through \eqref{eq:HEM}.

In practice, since, because of Assumption \ref{ass:tails}, the conditional distribution of the factors given the data is Gaussian, see \eqref{eq:brembote}, then, the conditional mean coincides with the linear projection. Therefore, the sufficient statistics (moments of 1st and 2nd order) needed to compute the final estimators of the parameters given in \eqref{eq:LEM} through \eqref{eq:HEM}, are given by:
\begin{align}
&\E_{\wh{\bm\varphi}^{(k^*)}}[\mbf F_t| \bm{\mathcal X}]= \mbf F_{t|T}^{(k^*)},\label{eq:CONDMEAN}\\
&\E_{\wh{\bm\varphi}^{(k^*)}}[\mbf F_t\mbf F_t^\prime| \bm{\mathcal X}]= \mbf F_{t|T}^{(k^*)}\mbf F_{t|T}^{(k^*)\prime}+\mbf P_{t|T}^{(k^*)}, \quad \E_{\wh{\bm\varphi}^{(k^*)}}[\mbf F_t\mbf F_{t-1}^\prime| \bm{\mathcal X}]= \mbf F_{t|T}^{(k^*)}\mbf F_{t-1|T}^{(k^*)\prime}+\mbf C_{t,t-1|T}^{(k^*)}.\nn
\end{align}
Notice, however, that, since we treat the idiosyncratic components as cross-sectionally uncorrelated, then $\mbf P_{t|T}^{(k^*)}$ is just an approximation of the true MSE of the Kalman smoother. In particular, since the Kalman smoother is unbiased (\citealp[Lemma 4.1]{duncan1972}, and \citealp[Chapter 3.2.3, p.111]{harvey90}), the true MSE coincides with the conditional covariance as given in \eqref{eq:brembote}. Nevertheless, since, given $\wh{\bm\varphi}^*$, both matrices are $O_{\mathrm P}(N^{-1})$ (\citealp[Proposition 1]{BLqml}) this approximation error is negligible.



The final Kalman filter and Kalman smoother estimators are obtained at the end of the EM algorithm, i.e., they are given by $\wh{\mbf F}_t^{\text{\tiny KF}}:= \mbf F_{t|t}^{(k^*+1)}$  and $\wh{\mbf F}_t^{\text{\tiny KS}}:= \mbf F_{t|T}^{(k^*+1)}$. In particular, under the VAR(1) model in \eqref{eq:VARsimplebis}, these are defined by the following iterations:
\begin{align}
\wh{\mbf F}_t^{\text{\tiny KF}}:=&\, 
\wh{\mbf A}^{\text{\tiny EM}} \wh{\mbf F}_{t-1}^{\text{\tiny KF}}\hskip 10cm\;\;  t=t,\ldots,T,\nn\\ 
&+
 \l(\wh{\bm\Lambda}^{\text{\tiny EM}\prime}(\wh{\bm\Sigma}^{\xi\text{\tiny EM}})^{-1}\wh{\bm\Lambda}^{\text{\tiny EM}}+\wh{\mbf P}_{t|t-1}^{-1}\r)^{-1}\wh{\bm\Lambda}^{\text{\tiny EM}\prime}(\wh{\bm\Sigma}^{\xi\text{\tiny EM}})^{-1}\l(\mbf x_t-\bar {\mbf x} -\wh{\bm\Lambda}^{\text{\tiny EM}} 
 \wh{\mbf A}^{\text{\tiny EM}} \wh{\mbf F}_{t-1}^{\text{\tiny KF}}
\r),\label{eq:KFiter}\\
\wh{\mbf F}_t^{\text{\tiny KS}}:=&\,\wh{\mbf F}_t^{\text{\tiny KF}}+\wh{\mbf P}_{t|t}\wh{\mbf A}^{\text{\tiny EM}\prime}\wh{\mbf P}_{t+1|t}^{-1}\l(\wh{\mbf F}_{t+1}^{\text{\tiny KS}}- 
\wh{\mbf A}^{\text{\tiny EM}} \wh{\mbf F}_{t-1}^{\text{\tiny KF}}
\r),\quad\quad\quad\quad\quad\quad\quad\quad\quad\quad\quad\quad t=T,\ldots,1,
\label{eq:KSiter}
\end{align}
where 
$\wh{\mbf P}_{t|t-1} = \wh{\mbf A}^{\text{\tiny EM}}\wh{\mbf P}_{t-1|t-1} \wh{\mbf A}^{\text{\tiny EM}\prime} + \wh{\mbf H}^{\text{\tiny EM}} \wh{\mbf H}^{\text{\tiny EM}\prime}$, 
 with $\wh{\mbf P}_{t|t}$, being the filtered conditional MSEs, which evolves according to the iterations given in Appendix \ref{app:KS}. 
These iterations are initialized by $\wh{\mbf F}_{0}^{\text{\tiny KF}}=\mbf 0_r$, $\wh{\mbf F}_{T+1}^{\text{\tiny KS}}=\wh{\mbf A}^{\text{\tiny EM}} \wh{\mbf F}_{T}^{\text{\tiny KF}}$,   
and $\wh{\mbf P}_{0|0} = c \mbf I_r$, for some finite positive real $c$.


Under the present setting, \citet[Proposition 3]{BLqml} show that the Kalman smoother and the Kalman filter are asymptotically equivalent
 \begin{align}
\l\Vert\wh{\mbf F}_{t}^{\text{\tiny KS}}-\wh{\mbf F}_{t}^{\text{\tiny KF}} \r\Vert = O_{\mathrm P}\l(\frac 1N\r),\label{eq:AEBLKS}
\end{align}
and also (up to logarithmic terms):
 \begin{align}
\l\Vert\wh{\mbf F}_{t}^{\text{\tiny KF}}-\wh{\mbf F}_{t}^{\text{\tiny WLS}} \r\Vert = O_{\mathrm P}\l(\frac 1N\r), \qquad \l\Vert\wh{\mbf F}_{t}^{\text{\tiny KF}}-\wh{\mbf F}_{t}^{\text{\tiny LP}} \r\Vert = O_{\mathrm P}\l(\frac 1N\r),\label{eq:AEBLKS2}
\end{align}
that is, the Kalman filter and the feasible WLS/LP estimators are asymptotically equivalent (see also \citealp[Theorem 3]{baili16} where however it is also required  $T/N^3\to 0$). From \eqref{eq:AEBLKS}, \eqref{eq:AEBLKS2}, and Propositions \ref{prop:FBAILI16} or \ref{prop:FBAILI16bis}, we get the following result.

\begin{prop}\label{prop:BLFac} For any $t=1,\ldots, T$, under Assumptions \ref{ass:common} through \ref{ass:tails}, if $\sqrt {N\log N}/T\to 0$, as $N,T\to\infty$, 
\[
\sqrt N(\wh{\mbf F}_t^{\text{\tiny \upshape KS}}-\mbf F_t)\to_d\mathcal N\l(\mbf 0_r,\bm{\mathcal W}_{t}^{\text{\tiny \upshape WLS}}\r),
\]
where $\bm{\mathcal W}_{t}^{\text{\tiny \upshape WLS}}:=(\bm\Sigma_{\Lambda\xi\Lambda})^{-1}
\l\{
\lim_{N\to\infty}\frac {\bm\Lambda^\prime (\bm\Sigma^\xi)^{-1} \E[\bm \xi_t\bm \xi_t^\prime](\bm\Sigma^\xi)^{-1}\bm\Lambda }N
\r\}
(\bm\Sigma_{\Lambda\xi\Lambda})^{-1}$, with $\bm\Sigma_{\Lambda\xi\Lambda}:=\lim_{N\to\infty}\frac{\bm\Lambda^\prime(\bm\Sigma^\xi)^{-1}\bm\Lambda}{N}$.
\end{prop}

Up to logarithmic terms, the Kalman smoother estimator, computed using the EM estimator of the parameters, has the same asymptotic properties of the feasible WLS/LP estimators. Therefore, the same comments made after Propositions \ref{prop:FBAILI16} and \ref{prop:FBAILI16bis} about efficiency apply also in this case.

%
%

\begin{rem}
\upshape{
The iterations given in \eqref{eq:KFiter}  and \eqref{eq:KSiter} might be computationally challenging because at each point in time they require inverting $\wh{\mbf P}_{t|t-1}$, which has a complex form. To avoid incurring in such numerical problems,  
\citet[Chapter 4.4, pp.70-73]{DK01} provide a definition of the Kalman smoother, equivalent to \eqref{eq:KSiter}, but which does not require the inversions of $\wh{\mbf P}_{t|t-1}$.} 
An alternative numerically efficient approach is represented by the matrix formulation, instead of the recursive formulation, of the Kalman filter and smoother (see, e.g., \citealp{delle2019}).
\end{rem}

\begin{rem}
\upshape{The previous results might lead to think that the Kalman filter and the Kalman smoother are useless in a high-dimensional setting, since they are equivalent to the WLS/LP estimator. But there is at least one good reason for preferring these estimators over the WLS/LP ones. Consider for simplicity the case  of one factor, $r=1$, and known parameters. Then, following \citet[Section 2.3]{PR22}, and denoting $B:=\bm\Lambda^\prime(\bm\Sigma^\xi)^{-1}\bm\Lambda$, we get:
\begin{align}
F_{t}^{\text{\tiny KF}}&=\frac{A}{1+BP}F_{t-1}^{\text{\tiny KF}}+\frac{B P}{1+B P}F_t^{\text{\tiny WLS}}=\frac{A}{1+B P}F_{t-1}^{\text{\tiny KF}}+
\frac{(B+1)P}{1+BP}
F_t^{\text{\tiny LP}},
\label{eq:esther}
\end{align}
where $A$ is the scalar autoregressive coefficient in \eqref{eq:VARsimplebis}, and $P$ is the steady-state solution of the Riccati equation describing the dynamics of the one-step-ahead MSE $P_{t|t-1}$ (see Remark \ref{rem:deistler}).\footnote{The steady-state $P$ is given by (\citealp[Section 1.3]{PR15}): 
$$
P=\l\{
H^2B -1+A^2 
\l[
1+\sqrt{
1+
\frac{4 H^2B }{\l(H^2B-1+A^2\r)^2}}
\r]
\r\}
\frac 1{2B}.
$$
} 
Clearly, $B=O(N)$ because of Assumptions \ref{ass:common}(a) and \ref{ass:idio}(a), so, as predicted from \eqref{eq:AEBLKS2}, the Kalman filter is asymptotically equivalent to the WLS/LP estimators. However, if $\{F_t\}$ is a very persistent process which does not fluctuate much, i.e., $A\lesssim 1$ and $H\gtrsim 0$ so that $P\simeq 0$, the first term on the right-hand-side of \eqref{eq:esther} might be non-negligible. 

Notice that, because of \eqref{eq:AEBLKS} the same argument holds also for the Kalman smoother, which, in fact, has to be preferred over the Kalman filter, since, by being computed upon conditioning on a larger set, has always a smaller variance.



}
\end{rem}
 
\begin{rem}
\upshape{
By means of the Kalman fliter and smoother we could treat any autocorrelated idiosyncratic component as additional latent states, thus introducing a dynamic model for each of them and by adding in the measurement equation \eqref{eq:SDFM1Rbis} an error term with small variance \citep[see, e.g.,][]{banburamodugno14}. This is crucial when the idiosyncratic components are very persistent \citep{OGAP}. In this case,  it can be shown that, in order for the presented asymptotic results to still hold, the number of additional latent states should be limited, and they should share some common driving force.
}
\end{rem}

\begin{rem}
\upshape{The factors can also be estimated by means of the Wiener-Kolmogorov filter, which, as mentioned above, is the frequency domain counterpart of the smoother estimator in \eqref{eq:Fsmooth_comp} (\citealp[Chapter III.7]{hannan}). 
By means of such estimator an EM algorithm can be implemented in the spectral domain by first running the Wiener-Kolmogorov filter in the E-step, and then maximizing the expected Whittle log-likelihood in the M-step \citep{quahsargent93,FGS18}. As for all spectral methods, in order to implement a spectral EM we must first estimate a large dimensional spectral density matrix. 
}
\end{rem}

\begin{rem}
\upshape{A last approach to estimate the factors by accounting of their dynamics is proposed by \citet{baili16}, and it is based on the three following steps. Step 1: compute estimators of $\bm\Lambda$ and $\bm\Sigma^\xi$ via QML, and of the factors via WLS, as described in Section \ref{sec:static}. Step 2: use the estimated factors to compute estimators of $\mbf A$ and $\mbf H$ in \eqref{eq:VARsimplebis}. Step 3: use those estimated parameters to compute the Kalman smoother estimator of the factors. Similar multi-step approaches have been proposed by \citet{DGRfilter}, who replace step 1 with the PC estimators of loadings and factors, and \citet{ng15}, who replace step 1 with an estimator of the loadings obtained by maximizing the log-likelihood \eqref{eq:LL0iid} using the Netwon-Raphson algorithm. However, none of those approaches exploits the mutual feedback from the loadings to the factors and viceversa; hence, at least in finite samples, they might incur in efficiency losses. }
\end{rem}

\section{Classical ML estimation }\label{sec:classic}

In this section, we review classical factor analysis for cross-sectional data, which is characterized by: (i) fixed $N$, (ii) uncorrelated Gaussian, i.e., independent, idiosyncratic components $\bm\Omega^\xi=\mbf I_T\otimes \bm\Sigma^\xi$, and (iii) non-stochastic factors, i.e., $\bm\Omega^F=\mbf I_{T}\otimes \bm\Gamma^F$. This is the setting considered mainly in psychometric applications (see \citealp{lawleymaxwell71} and references therein).

Classical Maximum Likelihood (ML) estimation of the factor model can be carried out in two ways. 
\ben
\item [(A)]
Maximize the, now correctly specified, log-likelihood \eqref{eq:LL0iid} subject to some given identifying constraint, in order to obtain an ML estimator of the loadings and of the idiosyncratic variances using, e.g., the EM algorithm by \citet{RT82}. And, then, use these estimates to compute an estimator of the factors, either by least squares (see \citealp{bartlett37}, and \eqref{eq:GLShat} in Section \ref{sec:FGLS}) or by linear projection (see \citealp{thomson51}, and \eqref{eq:Freg_comp_feas} in Section \ref{sec:FGLS}).  

\item [(B)] Use the formulation of the log-likelihood in \eqref{eq:LL0_bis} and jointly estimate loadings and factors, which, in this case, are always estimated via least squares.
\een

In principle, approach A is considered to be the appropriate one  in the case stochastic factors, while approach B is considered to be the appropriate one in the case non-stochastic factors \citep[p.587]{anderson2003}. However, approach A can be used also with non-stochastic factors.  Indeed, the log-likelihood \eqref{eq:LL0iid} does not depend on the factors, nor on their second moments  if we also assume $\bm\Gamma^F=\mbf I_r$. Hence, approach A is commonly considered the most sensible one.

Another reason to prefer approach A is that the log-likelihood \eqref{eq:LL0_bis}, used in approach B, might diverge to infinity under certain choices of the parameters. To be more precise, this issue is related to just one specific configuration, namely the case in which at least one idiosyncratic component has zero variance (\citealp[Section 7.7]{AR56}, \citealp[p.587]{anderson2003}, \citealp{BT11}). This implies that the ML estimator of the idiosyncratic variances might not be defined. To solve this problem, \citet{lawley_1942} suggested to look for solutions satisfying just the first-order conditions, but, as shown by \citet{solari69}, this does not lead to discovering a maximum, but just a stationary point of the log-likelihood. Notice that, if we instead follow approach A, then the log-likelihood \eqref{eq:LL0iid} can easily accommodate some idiosyncratic components with zero variance, see, e.g., \citet[Chatper 3.4]{bartholomew2011latent}. 

Nevertheless, approach B has the nice feature of allowing us to exploit the mutual dependence of factors and loadings, and, moreover, the log-likelihood \eqref{eq:LL0_bis} has a more tractable form than the log-likelihood \eqref{eq:LL0iid}.

Regardless of the approach considered, when $N$ is fixed ML estimation of factor models suffers of two main problems: (I) inference for the loadings is computationally challenging, and (II) the factors  cannot be estimated consistently and their estimation might even pose an incidental parameter problem for the estimation of the other parameters of the model. We now consider both issues in detail and show how letting $N\to\infty$ helps solving both problems and would allow us to use the estimation approach in B.

Consider first, the estimation of the loadings. Under the assumption of an exact factor model, \citet[Theorems 5.2 and 5.4]{baili12} prove that the ML estimator of the loadings, maximizing \eqref{eq:LL0iid}, is such that:
\beq\label{eq:AE1}
\l\Vert\wh{\bm\lambda}_i^{\text{\tiny \upshape ML}}-{\bm\lambda}_i^{\text{\tiny \upshape OLS}}\r\Vert=O_{\mathrm P}\l(\frac 1{\sqrt{NT}}\r).
\eeq
From this result it is clear that since now the factor model is truly exact, then, the ML estimator of the loadings is $\sqrt T$-consistent, regardless of $N$. However, while if $N\to\infty$, the asymptotic covariance is the one given in Proposition \ref{prop:LBAILI16}, when $N$ is fixed the expression is much more complex since in this case the error coming from \eqref{eq:AE1} is also $O_p(T^{-1/2})$ thus contributes to the asymptotic distribution. Indeed, from the asymptotic expansion \eqref{eq:AE1} it is clear that if $N$ is fixed we still have $\sqrt T$ consistency, but an additional term on top of the OLS estimation error, must be included in the asymptotic distribution. The specific expression of the asymptotic covariance in the fixed $N$ case is given, for example, in \citet[Theorem 12.3]{AR56}, \citet[Theorem 2F]{AFP87}, and \citet[Theorems 1, 2, and 3]{AA88} (see also \citealp[Proposition 10]{MBPCA}). Since for fixed $N$ defining a consistent estimator of such covariance matrix is not easy, inference might become a computationally hard problem. Notice that the same problem exists even if we assumed the simplest possible case of an exact factor model with the unrealistic assumption of cross-sectionally homoskedastic idiosyncratic components, as in  \citet{young40} and \citet{whittle52}. 

%
 
Second, the factors represent an additional $rT$-dimensional vector of parameters that need to be estimated. They can either be estimated after we have an ML estimator of the loadings and the idiosyncratic variances (approach A) or they can be estimated jointly with the other parameters (approach B). Now, when $N$ is fixed none of the two approaches  allows to retrieve the factors consistently. The impossibility to retrieve the factors consistently when $N$ is fixed is precisely the reason why in classical factor analysis the factors are usually identified only indirectly through their associated loadings which, as we have seen above  are $\sqrt T$-consistent. This is the spirit of confirmatory analysis \citep{joreskog69}, an example of which is in \citet[Chapter 9.6]{mardia1979multivariate}, where the factors are identified through a VARIMAX orthogonal rotation of the loadings, providing rotated loadings with a few large entries and as many near-zero entries as possible. Another example is in \citet{terada2014strong}, where the loadings are identified by the reduced $K$-means clustering method.

Moreover, besides the aforementioned problem of zero idiosyncratic variances, approach B, although appealing, poses also an incidental parameter problem. Indeed, we need to simultaneously estimate $rT+(r+1)N$ parameters using $NT$ observations, and we might easily run out of degrees-of-freedom. This is especially true if $N$ is small and $T$ is large, while the larger $N$ gets, the less binding the issue is (see Remark \ref{rem:neyman} for details). 

Now, if we let $N\to\infty$, then we can prove that under approach A the factors are consistently estimated (see Propositions \ref{prop:FBAILI16} and \ref{prop:FBAILI16bis}). 
Furthermore, as shown in Section \ref{rem:neyman} below, when $N\to\infty$ also the incidental parameter problem vanishes asymptotically and we could use the estimation approach B by just iterating between the two sets of estimates. Say we start at iteration $k=0$ with an estimate of the loadings and the idiosyncratic variances, e.g., given by the PC estimator. Then, at any iteration $k\ge 0$ we would have the feasible WLS and OLS estimators:
\begin{align}
\wh{\mbf F}_t^{(k)}&=(\wh{\bm\Lambda}^{(k)\prime}(\wh{\bm\Sigma}^{\xi(k)})^{-1}\wh{\bm\Lambda}^{(k)})^{-1}\wh{\bm\Lambda}^{(k)\prime}(\wh{\bm\Sigma}^{\xi(k)})^{-1} (\mbf x_t-\bar{\mbf x}),\quad t=1,\ldots, T,\\
\wh{\bm \lambda}_i^{(k+1)}&= (\wh{\bm F}^{(k)\prime}\wh{\bm F}^{(k)})^{-1}\wh{\bm F}^{(k)\prime} (\bm x_i-\bar x_i\bm\iota_T), \qquad\qquad\qquad\qquad i=1,\ldots, N.
\end{align}
Clearly since we initialize the algorithm with a consistent estimator of the parameters, then at each iteration $k\ge 0$ both estimators can be shown to be consistent too as $N,T\to\infty$ (see also \citealp{pz23}), but no proof of the convergence of $\wh{\bm \lambda}_i^{(k+1)}$ to the ML estimator is available so far. Approach B would then become feasible and, possibly, more preferable than approach A. Indeed, approach B would allow us to explicitly take into account the mutual dependence of factors and loadings and we would also have closed form expressions for all estimators.


%
%
%
%


\subsection{The incidental parameter problem}\label{rem:neyman}
In this section we formally show that if $N$ is fixed, then the score of the log-likelihood \eqref{eq:LL0_bis} with respect to the loadings does not satisfy the orthogonality condition by \citet{neyman79}. Hence, any estimator the loadings using an estimator of the factors, is going to be biased.
More precisely, for any specific value of the factors, $\wt{\mbf F}_t$, of the loadings, $\wt{\bm\Lambda}$, and of the idiosyncratic variances, $\wt{\bm\Sigma}^\xi$, consider the score  with respect to the loadings $\underline{\bm\lambda}_i$, $i=1,\ldots, N$,
\begin{align}
\bm s_{it}(\mbf x_t;\wt{\mbf  F}_t,\wt{\bm\Lambda},\wt{\bm\Sigma}^\xi)&:=\l.\frac{\partial}{\partial\underline{\bm\lambda}_i^\prime}
\l\{
-\frac 12 \log\det(\underline{\bm\Sigma}^\xi) -\frac 12\l(\mbf x_{t}-\bar{\mbf x}-\underline{\bm\Lambda}\,\underline{\mbf F}_t\r)^\prime (\underline{\bm\Sigma}^\xi)^{-1}\l(\mbf x_{t}-\bar{\mbf x}-\underline{\bm\Lambda}\,\underline{\mbf F}_t\r)
\r\}
\r|_{\substack{\underline{\mbf F}_t=\wt{\mbf F}_t\\ \underline{\bm\Lambda}=\wt{\bm\Lambda},\, \underline{\bm\Sigma}^\xi=\wt{\bm\Sigma}^\xi}}\nn\\
&= \frac{\wt{\mbf F}_t(x_{it}-\wt{\mbf F}_t^\prime \wt{\bm\lambda}_i)}{\wt{\sigma}_i^2}.\nn
\end{align}
Clearly, in the true values of factors and parameters the first-order conditions are satisfied: 
\beq\label{eq:score}
\E[\bm s_{it}(\mbf x_t;{\mbf  F}_t,{\bm\Lambda},{\bm\Sigma}^\xi)]=\frac{\E[\mbf F_t\xi_{it}]}{\sigma_i^2}=\frac{\mbf F_t\E[\xi_{it}]}{\sigma_i^2}=\mbf 0_r,
\eeq
since we treat the factors as constant parameters (but this would hold even for stochastic factors uncorrelated with the idiosyncratic components).

We now ask ourselves what happens if we replace the true factors with an estimator, and let us choose the best possible estimator, which is the unfeasible WLS in \eqref{eq:Fgls_comp}. 
To this end, consider the curve $\bm\gamma:[0,1]\to \mathbb R^r$ connecting $\mbf F_t$ and ${\mbf F}_t^{\text{\tiny WLS}}$, i.e, such that $\bm\gamma(\eta)= {\mbf  F}_t+\eta({\mbf F}_t^{\text{\tiny WLS}}-\mbf F_t)$ for $\eta\in[0,1]$, and consider the $r$-dimensional function $\mbf D_{\eta}: \mathbb R^r\to\mathbb R^r$ describing the variation of the score along the curve $\bm\gamma$, so that for any $\eta\in[0,1]$ we have the $r\times 1$ differential vector
\[
\mbf D_{\eta}\l({\mbf F}_t^{\text{\tiny WLS}}-\mbf F_t\r):=\partial_\eta \l\{
\E \l[ \bm s_i\l(\mbf x_t;{\mbf  F}_t+\eta({\mbf F}_t^{\text{\tiny WLS}}-\mbf F_t),{\bm\Lambda},{\bm\Sigma}^\xi\r) \r]
\r\}.
\]
In order, to have unbiased estimators of the loadings the first-order conditions should be locally insensitive to the value of the factors, which here play the role of nuisance parameters.
A necessary condition for this to happen is that the Gateaux derivative along $\bm\gamma$ should be such that (see also \citealp[Definition 2.1]{chernozhukov18}):
\begin{align}
\mbf D_{\eta}\l.[{\mbf F}_t^{\text{\tiny WLS}}-\mbf F_t]\r |_{\eta=0}&
& = \lim_{\eta\to 0} \frac{\E \l[ \bm s_i\l(\mbf x_t;{\mbf  F}_t+\eta({\mbf F}_t^{\text{\tiny WLS}}-\mbf F_t),{\bm\Lambda},{\bm\Sigma}^\xi\r) \r]
-
\E \l[ \bm s_i\l(\mbf x_t;{\mbf  F}_t,{\bm\Lambda},{\bm\Sigma}^\xi\r) \r]
}{\eta}=\mbf 0_r,\label{eq:neyman}
\end{align}
If condition \eqref{eq:neyman} is satisfied, then we could estimate the loadings by plugging into the log-likelihood \eqref{eq:LL0_bis} noisy estimates of the factors, and then maximize the concentrated log-likelihood $\ell(\bm {\mathcal X};{\bm{\mathcal F}}^{\text{\tiny WLS}},\underline{\bm\varphi})$, without strongly violating the moment conditions \eqref{eq:score}. However, it is easy to see that for fixed $N$ condition \eqref{eq:neyman} is not satisfied. Indeed, by using \eqref{eq:score} and the expression of the WLS estimator \eqref{eq:Fgls_comp} into \eqref{eq:neyman}, we get:
\begin{align}
\mbf D_{\eta}\l.[{\mbf F}_t^{\text{\tiny WLS}}-\mbf F_t]\r |_{\eta=0}&=\lim_{\eta\to 0}\frac {\E\l[\mbf F_t\xi_{it}-\eta \mbf F_t({\mbf F}_t^{\text{\tiny WLS}}-\mbf F_t)^\prime\bm\lambda_i+\eta({\mbf F}_t^{\text{\tiny WLS}}-\mbf F_t)\xi_{it}-\eta^2({\mbf F}_t^{\text{\tiny WLS}}-\mbf F_t)({\mbf F}_t^{\text{\tiny WLS}}-\mbf F_t)^\prime\bm\lambda_i \r]}{\eta\sigma_i^2}\nn\\
&=\frac 1{\sigma_i^2}\E\l[ ({\mbf F}_t^{\text{\tiny WLS}}-\mbf F_t)\xi_{it}\r] -\frac 1{\sigma_i^2} \E\l[\mbf F_t({\mbf F}_t^{\text{\tiny WLS}}-\mbf F_t)\bm\lambda_i\r]\nn\\
&=\frac 1{\sigma_i^2} (\bm\Lambda^\prime(\bm\Sigma^\xi)^{-1}\bm\Lambda)^{-1} \bm\Lambda^\prime (\bm\Sigma^\xi)^{-1}\E[\bm\xi_t\xi_{it}]-
\frac 1{\sigma_i^2}\E[\mbf F_t\bm\xi_t] (\bm\Sigma^\xi)^{-1}\bm\Lambda(\bm\Lambda^\prime(\bm\Sigma^\xi)^{-1}\bm\Lambda)^{-1}\nn\\
&= \frac 1{\sigma_i^2} \l(\frac{\bm\Lambda^\prime(\bm\Sigma^\xi)^{-1}\bm\Lambda}N\r)^{-1}\frac 1N \sum_{j=1}^N \frac{\bm\lambda_j \E[\xi_{jt}\xi_{it}]}{\sigma_j^2},\label{eq:gateaux}
\end{align}
which for fixed $N$ is never zero, not even if the model is exact, in which case the summation will just made of one term when $j=i$. So even if we use the ideal unfeasible WLS estimator of the factors, we would get biased estimates for the loadings. However, we clearly see that in the limit $N\to\infty$, the expression in \eqref{eq:gateaux} is $O_{\mathrm P}(N^{-1})$, because of Assumptions \ref{ass:common}(a), \ref{ass:idio}(a), and \ref{ass:idio}(b), and condition \eqref{eq:neyman} is satisfied. So, as expected, the problem of incidental parameters vanishes asymptotically.

\section{On the mis-specification error and the search for efficient estimators}



Consider the two equivalent ways of writing the factor model \eqref{eq:SDFM1R} when stacking all $NT$ observations in one $NT$ dimensional vector, i.e., by vectorzing the matrix representation \eqref{eq:SDFM1Rmat} or its transposed:
\begin{align}
&\bm{\mathfrak X}:=\text{vec}(\bm X) = \text{vec}(\bm\iota_T\bm\alpha^\prime) +\l(\mbf I_N\otimes \bm F\r) \text{vec}(\bm\Lambda^\prime)+\text{vec}(\bm \Xi)=:\bm{\mathfrak A}+ \bm{\mathfrak F}\bm{\mathcal L} + \bm{\mathfrak E},\label{eq:M1}\\
&\bm{\mathcal X}:=\text{vec}(\bm X^\prime) = \text{vec}(\bm\alpha\bm\iota_T^\prime) +\l(\mbf I_T\otimes\bm\Lambda\r) \text{vec}(\bm F^\prime)+\text{vec}(\bm \Xi^\prime)=:\bm{\mathcal A}+ \bm{\mathfrak L}\bm{\mathcal F}+\bm{\mathcal E}.\label{eq:M3}
\end{align}
Let $\bm\Theta^\xi:=\E[\text{vec}(\bm \Xi)\text{vec}(\bm \Xi)^\prime]=\E[\bm{\mathfrak E}\bm{\mathfrak E}^\prime]$ and $\bm\Omega^\xi:=\E[\text{vec}(\bm \Xi^\prime)\text{vec}(\bm \Xi^\prime)^\prime]=\E[\bm{\mathcal E}\bm{\mathcal E}^\prime]$, which are $NT\times NT$ matrices having the same elements just ordered differently.
%

If the factors were observed, the true, i.e., not mis-specified, log-likelihood \eqref{eq:LL0} could be written also as:
\begin{align}
\ell (\bm {\mathfrak X},{\bm{\mathfrak F}};\underline{\bm{\mathcal L}},\text{vech}(\underline{\bm\Theta}^\xi)) &\simeq
-\frac 12\log\det(\underline{\bm\Theta}^{\xi})-\frac 12\l[\l(\bm {\mathfrak X}-\bar{\bm {\mathfrak X}}-\bm{\mathfrak F}\,\underline {\bm {\mathcal L}} \r)^\prime
  (\underline{\bm\Theta}^{\xi})^{-1}
  \l(\bm {\mathfrak X}-\bar{\bm {\mathfrak X}}-\bm{\mathfrak F}\,\underline{\bm {\mathcal L}} \r)
\r].\label{eq:LL0_ter}
\end{align}
Then, for known $\bm\Theta^{\xi}$ the QML estimator of the loadings, maximizing \eqref{eq:LL0_ter}, would be the unfeasible GLS estimator:
\beq
{\bm {\mathcal L}}^{\text{\tiny{GLS}}} =\text{vec}\l({\bm\Lambda}^{\text{\tiny GLS}\prime}\r):= \l\{\bm{\mathfrak F}^\prime({\bm\Theta}^{\xi})^{-1} \bm{\mathfrak F}\r\}^{-1}
\l\{\bm{\mathfrak F}^\prime({\bm\Theta}^{\xi})^{-1}\l(\bm {\mathfrak X}-\bar{\bm {\mathfrak X}}\r) \r\}.\label{eq:LGLS}
\eeq
This would be the QML estimator of the loadings for an approximate factor model with known factors. 

Likewise, 
for known loadings the true, i.e., not mis-specified, log-likelihood \eqref{eq:LL0} could be written also as:
\beq
\ell (\bm {\mathcal X}, \bm {\mathfrak  L};\underline{\bm{\mathcal F}},\text{vech}(\underline{\bm\Omega}^\xi)) \simeq
-\frac 12\log\det(\underline{\bm\Omega}^{\xi})-\frac 12\l[\l(\bm {\mathcal X}-\bar{\bm {\mathcal X}}-{\bm {\mathfrak L}}\,\underline {\bm {\mathcal F}} \r)^\prime
  (\underline{\bm\Omega}^{\xi})^{-1}
  \l(\bm {\mathcal X}-\bar{\bm {\mathcal X}}-{\bm {\mathfrak L}}\,\underline{\bm {\mathcal F}} \r)
\r].\label{eq:LL0_quater}
\eeq
Then, for known $\bm\Omega^{\xi}$ the QML estimator of the factors, maximizing \eqref{eq:LL0_quater}, would be the unfeasible GLS estimator:
\beq
{\bm {\mathcal F}}^{\text{\tiny{GLS}}} =\text{vec}\l({\bm F}^{\text{\tiny GLS}\prime}\r):= \l\{\bm{\mathfrak L}^\prime({\bm\Omega}^{\xi})^{-1} \bm{\mathfrak L}\r\}^{-1}
\l\{\bm{\mathfrak L}^\prime({\bm\Omega}^{\xi})^{-1}\l(\bm {\mathcal X}-\bar{\bm {\mathcal X}}\r) \r\}.\label{eq:FGLS}
\eeq
This would be the QML estimator of the factors for an approximate factor model with known loadings.

Now, even if the factors or the loadings were known, the GLS estimators \eqref{eq:LGLS} and \eqref{eq:FGLS} would still require the unfeasible task to estimate and invert $\bm\Theta^\xi$ or $\bm\Omega^\xi$. As done in Section \ref{sec:static}, we can then consider a mis-specified version of the log-likleihoods \eqref{eq:LL0_ter} or \eqref{eq:LL0_quater}, where we do not account for idiosyncratic correlations and we set $\bm\Theta^\xi=\bm\Sigma^\xi\otimes \mbf I_T$ and $\bm\Omega^\xi=\mbf I_T\otimes\bm\Sigma^\xi$. Then, it is easy to see that the maximizers of such mis-specified log-likelihoods are the unfeasible OLS estimator of the loadings and the unfeasible WLS estimator of the factors. Indeed, in this case \eqref{eq:LGLS} would simplify as follows:
\begin{align}
{\bm {\mathcal L}}^{\text{\tiny{OLS}}}=\text{vec}\l({\bm\Lambda}^{\text{\tiny OLS}\prime}\r) &:= 
\l\{\bm{\mathfrak F}^\prime(\bm\Sigma^\xi\otimes \mbf I_T)^{-1} \bm{\mathfrak F}\r\}^{-1}
\l\{\bm{\mathfrak F}^\prime(\bm\Sigma^\xi\otimes \mbf I_T)^{-1}\l(\bm {\mathfrak X}-\bar{\bm {\mathfrak X}}\r) \r\}\nn\\
&=\l\{({\bm\Sigma}^{\xi})^{-1}\otimes \bm{F}^\prime\bm F\r\}^{-1}
\l\{\l(({\bm\Sigma}^{\xi})^{-1}\otimes \bm{F}^\prime\r)\l(\bm {\mathfrak X}-\bar{\bm {\mathfrak X}}\r) \r\}\nn\\
&=\l\{ {\bm\Sigma}^{\xi}\otimes \l(\bm{F}^\prime\bm F\r)^{-1}\r\}
\l\{\text{vec}\l(\bm F^\prime (\bm {X}-\bar{\bm { X}}) (\bm\Sigma^\xi)^{-1}\r) \r\}\nn\\
&=\text{vec}\l( \l(\bm F^\prime \bm F\r)^{-1} \bm F^\prime (\bm {X}-\bar{\bm { X}})\r),\label{eq:LOLS0}
\end{align}
with components $\bm\lambda_i^{\text{\tiny OLS}}$ given in \eqref{eq:LOLS}.
And, similarly, \eqref{eq:FGLS} would simplify as follows:
\begin{align}
{\bm {\mathcal F}}^{\text{\tiny{WLS}}} =\text{vec}\l({\bm F}^{\text{\tiny WLS}\prime}\r)&:= \l\{\bm{\mathfrak L}^\prime(\mbf I_T\otimes\bm\Sigma^\xi)^{-1} \bm{\mathfrak L}\r\}^{-1}
\l\{\bm{\mathfrak L}^\prime(\mbf I_T\otimes\bm\Sigma^\xi)^{-1}\l(\bm {\mathcal X}-\bar{\bm {\mathcal X}}\r) \r\}\nn\\
&=\l\{\mbf I_T \otimes \bm\Lambda^\prime(\bm\Sigma^\xi)^{-1}\bm\Lambda\r\}^{-1}
\l\{\l(\mbf I_T\otimes  \bm\Lambda^\prime(\bm\Sigma^\xi)^{-1}\r)\l(\bm {\mathcal X}-\bar{\bm {\mathcal X}}\r)  \r\}\nn\\
&=\l\{\mbf I_T\otimes  \l(\bm\Lambda^\prime(\bm\Sigma^\xi)^{-1}\bm\Lambda\r)^{-1}\bm\Lambda^\prime(\bm\Sigma^\xi)^{-1}\r\}\l(\bm {\mathcal X}-\bar{\bm {\mathcal X}}\r)\nn\\
&=\text{vec}\l(\l(\bm\Lambda^\prime(\bm\Sigma^\xi)^{-1}\bm\Lambda\r)^{-1}\bm\Lambda^\prime(\bm\Sigma^\xi)^{-1} \l(\bm { X}-\bar{\bm { X}}\r)^\prime \r),
\label{eq:FWLS}
\end{align}
with components $\mbf F_t^{\text{\tiny WLS}}$ given in \eqref{eq:Fgls_comp}. 

As shown in Section \ref{sec:static}, these unfeasible OLS and WLS estimators are asymptotically equivalent to the QML and feasible WLS estimators studied in Propositions \ref{prop:LBAILI16} and \ref{prop:FBAILI16}, respectively.
So to understand what is the effect of the mis-specifications introduced in the log-likelihood \eqref{eq:LL0iid}, it is enough to study how different the unfeasible OLS and WLS estimators in \eqref{eq:LOLS0} and \eqref{eq:FWLS} are from the most efficient estimators which are the unfeasible GLS estimators in \eqref{eq:LGLS} and \eqref{eq:FGLS}, respectively.

On average the full and the mis-specified idiosyncratic covariances are asymptotically equivalent since, for all $N,T\in\mathbb N$,
\begin{align}
\frac 1{NT}&\l\Vert\bm\Theta^\xi-(\bm\Sigma^\xi\otimes \mbf I_T) \r\Vert_F=\frac 1{NT}\l\Vert\bm\Omega^\xi-(\mbf I_T\otimes \bm\Sigma^\xi) \r\Vert_F\nn\\
&\le \frac 1 {\sqrt{NT}} \max_{i=1,\ldots, N}\max_{t=1,\ldots, T}\sum_{j=1}^N\sum_{s=1}^T  \vert\E_{}[\xi_{it}\xi_{js}]\vert\le \frac 1  {\sqrt{NT}} \max_{i=1,\ldots, N}\max_{t=1,\ldots, T}\sum_{j=1}^N\sum_{s=1}^T  M_{ij} \rho^{\vert t-s\vert}\nn\\
&\le \frac {M_\xi}{(1-\rho)\sqrt{NT}},\label{eq:miserror}
\end{align}
because of Assumption \ref{ass:idio}(b) and since $M_\xi$ and $\rho$ do not depend on $i,j,t,s$.\footnote{Recall the inequalities: $\l\Vert\bm\Theta^\xi\r\Vert_F\le \sqrt{NT}\l\Vert\bm\Theta^\xi\r\Vert\le \sqrt{NT}\l\Vert\bm\Theta^\xi\r\Vert_1$ and  $\l\Vert\bm\Omega^\xi\r\Vert_F\le \sqrt{NT}\l\Vert\bm\Omega^\xi\r\Vert\le \sqrt{NT}\l\Vert\bm\Omega^\xi\r\Vert_1$.}
 Nevertheless, the difference between the two sets of estimators depends on the inverse idiosyncratic covariances, i.e., on
\begin{align}
\l\Vert(\bm\Theta^\xi)^{-1}-(\bm\Sigma^\xi\otimes \mbf I_T)^{-1} \r\Vert &\le \l\Vert(\bm\Theta^\xi)^{-1}\r\Vert\,
\l\Vert\bm\Theta^\xi-(\bm\Sigma^\xi\otimes \mbf I_T) \r\Vert\, 
\l\Vert (\bm\Sigma^\xi\otimes \mbf I_T)^{-1} \r\Vert,\nn\\
\l\Vert(\bm\Omega^\xi)^{-1}-(\mbf I_T\otimes \bm\Sigma^\xi)^{-1} \r\Vert&\le \l\Vert(\bm\Omega^\xi)^{-1}\r\Vert\,
\l\Vert\bm\Omega^\xi-(\mbf I_T\otimes \bm\Sigma^\xi) \r\Vert\, 
\l\Vert (\mbf I_T\otimes \bm\Sigma^\xi)^{-1} \r\Vert,\nn
\end{align}
which, even by assuming that $\bm\Theta^\xi$ and $\bm\Omega^\xi$ are positive definite, are still finite for all $N$ and $T$, but do not vanish. Therefore, as expected, the GLS estimators are in principle better than the unfeasible QML and WLS estimators described in the previous section.


In light of this discussion, it is natural to consider extensions of QML estimation of the static approximate factor model where we model explicitly some of the serial and cross-sectional idiosyncratic correlations. This would allow us to compute new estimators of the loadings and the factors which are closer to the unfeasible GLS estimators and thus are more efficient than the OLS and WLS.

In particular, the second best that we can hope for are the following pseudo-GLS estimators. Consider, first, estimation of the loadings when we do not model either the idiosyncratic cross- or cross-autocorrelations, i.e., when imposing $\E[\bm\zeta_i\bm\zeta_j^\prime]=\mbf 0_{T\times T}$ for $i\ne j$. So we maximize the log-likelihood \eqref{eq:LL0_ter} when $\bm\Theta^\xi$ is block-diagonal with $N$ blocks $\bm\Delta_i^\xi:=\E[\bm\zeta_i\bm\zeta_i^\prime]$, $i=1,\ldots, N$, each being a $T\times T$ matrix, with entries $\E[\xi_{it}\xi_{is}]$, $t,s=1,\ldots, T$, hence, under serial homoskedasticity, $\bm\Delta_i^\xi$ is a Toeplitz matrix. The QML estimator of the loadings is then the pseudo-GLS estimator:
\beq
{\bm\lambda}_i^{\text{\tiny GLS}_0}:= \l(\bm F^\prime(\bm\Delta_i^\xi)^{-1} \bm F\r)^{-1} \bm F^\prime(\bm\Delta_i^\xi)^{-1} (\bm x_i-\bar x_i\bm\iota_T), \qquad i=1,\ldots, N,\label{eq:LQGLScomp}
\eeq
This estimator has an asymptotic covariance matrix which attains the Gauss-Markov lower bound: $\lim_{T\to\infty} T(\E[\bm F^\prime (\bm \Delta_i^\xi)^{-1}\bm F])^{-1}$. 

However, even for known factors, computing \eqref{eq:LQGLScomp} is in general unfeasible, since it requires estimation and inversion of a large matrix $\bm\Delta_i^\xi$. A common approach is to assume a parametric expression for $\bm\Delta_i^\xi$. For example, let $\xi_{it}=\alpha_i\xi_{i,t-1}+\nu_{it}$ with $\vert \alpha_i\vert<1$ and $\nu_{it}\sim (0,\omega_i^2)$ and uncorrelated across $i$ and $t$. Then, both $\bm\Delta_i^\xi$ and its inverse depend only on $\alpha_i$ and $\omega_i^2$.\footnote{
Indeed, we have that the $(t,s)$ entry of $\bm\Delta_i^\xi$ is $\frac{\omega_i^2\alpha_i^{|t-s|}}{1-\alpha_i^2}$ and
\[
\l(\bm\Delta_i^\xi\r)^{-1}\!\!=\frac{\omega_i^2}{(1-\alpha_i^2)^2}
\l(\ba{ccccccc}
1&-\alpha_i&0&\ldots&0&0&0\\
-\alpha_i&1+\alpha_i^2&-\alpha_i&\ldots&0&0&0\\
0&-\alpha_i&1+\alpha_i^2&\ldots&0&0&0\\
\vdots&\vdots&\vdots&\ddots&\vdots&\vdots&\vdots\\
0&0&0&\ldots&-\alpha_i&1+\alpha_i^2&-\alpha_i\\
0&0&0&\ldots&0&-\alpha_i&1\\
\ea
\r).
\]
}
In practice, we could first estimate the loadings by QML, then the factors by WLS, and last the idiosyncratic components, from which, by fitting an AR(1), we could get estimates of $\alpha_i$ and $\omega_i^2$ and thus of $\bm\Delta_i^\xi$ and its inverse. By using all these estimates we can compute a feasible version of the pseudo-GLS estimator of the loadings in \eqref{eq:LQGLScomp}. This is similar to the approach used in PC estimation by \citet{BT11}. and, for known factors, it coincides with the classical feasible GLS estimator of the loadings, which is asymptotically equivalent to their QML estimator when considering the mis-specified log-likelihood \eqref{eq:LL0_bis}, but for the filtered data $(1-\alpha_iL) x_{it}$, $i=1,\ldots, N$, and, thus, with $\bm\Sigma^\xi$ replaced by a diagonal matrix with entries $\omega_i^2$, $i=1,\ldots, N$ \citep{CU49}. 

Similarly, in order to explicitly model the autocorrelation of the idiosyncratic components, the EM algorithm presented in Section \ref{sec:EMload} can also be modified. 
Namely, in the M-step of any iteration $k\ge 0$, we could iterate between estimates of the loadings conditional on $\alpha_i$ and $\omega_i$, and estimates of $\alpha_i$ and $\omega_i$ conditional on the loadings, until convergence, and then move to iteration $k+1$ of the EM algorithm. This is the procedure adopted by \citet{reiswatson10} and it is an instance of an Expectation Conditional Maximization (ECM) algorithm \citep{MR93}. 

Second, consider estimation of the factors when we do not model either idiosyncratic auto- and cross-autocorrelations, i.e., when we impose $\E[\bm\xi_t\bm\xi_s^\prime]=\mbf 0_{N\times N}$ for $t\ne s$. So we maximize the log-likelihood \eqref{eq:LL0_quater} when $\bm\Omega^\xi=\mbf I_T\otimes \bm\Gamma^\xi$, where $\bm\Gamma^\xi:=\E[\bm\xi_t\bm\xi_t^\prime]$ is a $N\times N$ matrix, with entries $\E[\xi_{it}\xi_{jt}]$, $i,j=1,\ldots, N$, hence, under serial homoskedasticity, it is independent of $t$. The QML estimator of the factors is then the pseudo-GLS estimator:
\beq
{\mbf F}_t^{\text{\tiny GLS}_0}:=\l(\bm\Lambda^\prime(\bm\Gamma^\xi)^{-1}\bm\Lambda\r)^{-1}\bm\Lambda^\prime(\bm\Gamma^\xi)^{-1} (\mbf x_t-\bar{\mbf x}),\qquad t=1,\ldots, T.\label{eq:FQGLScomp}
\eeq
This estimator has an asymptotic covariance matrix which attains the Gauss-Markov lower bound: $\lim_{N\to\infty} N(\bm \Lambda^\prime (\bm \Gamma^\xi)^{-1}\bm \Lambda)^{-1}$. 

However, even for known loadings computing \eqref{eq:FQGLScomp} is in general unfeasible, since it requires estimation and inversion of a large matrix $\bm\Gamma^\xi$. To this end, \citet{bailiao16}  propose an EM algorithm to 
maximize the log-likelihood \eqref{eq:LL0iid} when replacing $\bm\Sigma^\xi$ with the full matrix $\bm\Gamma^\xi$ but subject to an $\ell_1$ penalty imposed on its off-diagonal entries. They then use the QML estimators of the loadings and the idiosyncratic covariance obtained in this way to compute a feasible version of the pseudo-GLS estimator of the factors in \eqref{eq:FQGLScomp}. Similar approaches are also in \citet{wang2019penalized}, and \citet{poignard2020statistical}.  Generalizations of the EM algorithm, e.g., by considering penalized M-steps in order to account for cross-sectional idiosyncratic correlations, are in principle possible too. This is the spirit of the approach by \citet{LM02} who assume a sparse VAR model for the idiosyncratic components,  thus accounting also for cross-autocorrelation, and they embed into the EM algorithm a penalized M-step.

Nevertheless, in general, the pseudo-GLS estimators \eqref{eq:LQGLScomp} and \eqref{eq:FQGLScomp} are not necessarily more efficient than the QML and WLS estimators studied in Proposition \ref{prop:LBAILI16} and \ref{prop:FBAILI16}, since, they do not address possible cross-autocorrelations, so they are still maximizers of mis-specified log-likelihoods. We might have gains in efficiency only if the cross-autocorrelations are small enough, i.e., if the $T\times T$ off-diagonal   blocks of $\bm\Theta^\xi$ or the $N\times N$ off-diagonal blocks of $\bm\Omega^\xi$  are sparse enough. Moreover, the feasible pseudo-GLS estimators discussed above my suffer if we do not correctly specify the autoregressive order of the idiosyncratic components or we do not select correctly the degree of penalization when thresholding the idiosyncratic covariance matrix. For these reasons such estimators are hardly considered in empirical econometric applications.

\begin{rem}\label{rem:hetero2}
\upshape{Lets us briefly consider the case in which we had serial idiosyncratic heteroskedasticity. Then, regarding estimation of the loadings, if we do not model any idiosyncratic cross- or cross-autocorrelations, the Gauss-Markov lower bound would be $\lim_{T\to\infty} T(\E[\bm F^\prime (\bm H_i^\xi)^{-1}\bm F])^{-1}$, where $\bm H_i^\xi:=\E[\bm\zeta_i\bm\zeta_i^\prime]$ is a $T\times T$ matrix with entries $\E[\xi_{it}\xi_{is}]$, $t,s=1,\ldots, T$, which now depend on $t$ and $s$, so it is no more a Toeplitz matrix. Regarding the estimated factors, 
if we do not model any idiosyncratic auto- or cross-autocorrelations, the Gauss-Markov lower bound would become $\lim_{N\to\infty} N(\bm \Lambda^\prime (\bm \Gamma_t^\xi)^{-1}\bm \Lambda)^{-1}$, where $\bm\Gamma_t^\xi:=\E[\bm\xi_t\bm\xi_t^\prime]$ is a $N\times N$ matrix with entries $\E[\xi_{it}\xi_{jt}]$, $t,s=1,\ldots, T$, which now depend on $t$.  As a consequence, the pseudo-GLS estimators \eqref{eq:LQGLScomp} and \eqref{eq:FQGLScomp}  become even less efficient since they do not account also for serial heteroskedasticity. 
Notice, however, that estimating and inverting $\bm H_i^\xi$ and $\bm\Gamma_t^\xi$ is a very complex task since both matrices have entries which are time dependent, hence, generalizations of the approaches by \citet{BT11} or \citet{bailiao16}, described above, are not straightforward. 
}
\end{rem}

\section{Simulations}

Throughout, we let $N\in\{20, 50, 100,200\}$, $T=100$, and $r=2$, and, for all $i=1,\ldots, N$ and $t=1,\ldots, T$, we simulate the data according to 
\begin{align}
x_{it}&={\bm\ell}_{i}^\prime {\bm f}_t + \phi_i \xi_{it}, \quad {\bm f}_t ={\bm A}  {\bm f}_{t-1} + \mbf u_t,\quad \xi_{it}=\delta_i\xi_{it-1}+e_{it},\nn
\end{align}
where $\bm\ell_i$ and $\bm f_t$ are $r$-dimensional vectors. Specifically,
\begin{inparaenum}
	\item []  $\bm\ell_i$ has entries ${\ell}_{ij}\stackrel{iid}{\sim}\mathcal{N}(1,1)$, $i=1,\ldots, N$, $j=1,\ldots, r$; 
	\item [] ${\bm A}=0.9 \check{\bm A} \Vert{\bm A}\Vert^{-1}$, where $\check{\bm A} $ has diagonal entries $\check{a}_{jj}\stackrel{iid}{\sim} U[0.5,0.8]$, $j=1,\ldots, r$ and off-diagonal entries $\check{ a}_{jk}\stackrel{iid}{\sim} U[0,0.3]$, $j,k=1,\ldots, r$, $j\ne k$;
	\item [] $\mbf u_{t}\stackrel{iid}{\sim}\mathcal N(\mbf 0_r,\mbf I_r)$;
	\item [] $\bm e_{t}\stackrel{iid}{\sim}\mathcal N(\mbf 0_N,\bm\Gamma^e)$, where $\bm\Gamma^e$ has diagonal entries 
	 $\sigma_{ei}^2\sim U[0.5, 1.5]$, $i=1,\ldots, N$, 
	 and off-diagonal entries $\sigma_{e,ij}=\tau^{\vert i-j\vert }\mathbb I(\vert i-j\vert \le 10)$, $i,j=1,\ldots, N$, $i\ne j$, with $\tau\in\{0,0.5\}$;
	\item  []  $\delta_i\stackrel{iid}{\sim}\mathcal{U}(0,\delta)$, and $\delta\in\{0,0.5\}$;
	\item [] $\phi_i=\sqrt{\theta_i (\sum_{t=1}^T \chi_{it}^2)/(\sum_{t=1}^T \xi_{it}^2)}$,
	and 
	$\theta_i\stackrel{iid}{\sim}\mathcal{U}(0.25,0.5)$.
\end{inparaenum}
The parameters  $\tau$ and $\delta$ control the degrees of
cross-sectional and serial idiosyncratic correlation in the idiosyncratic components. The the noise-to-signal ratio for series $i$ is given by $\theta_i$. 

Finally, in order for the simulated loadings and factors to satisfy Assumptions \ref{ass:ident}(a) and \ref{ass:ident}(b) we proceed as follows. Given the common components is generated as $\chi_{it}={\bm\ell}_{i}^\prime {\bm f}_t$, let $\bm\chi_t=(\chi_{1t}\cdots \chi_{Nt})^\prime$ and compute $\wt{\bm\Gamma}^\chi=\frac 1T\sum_{t=1}^T \bm\chi_t\bm\chi_t^\prime$. Collect its $r$ non-zero eigenvalues into the $r\times r$ diagonal matrix $\wt{\mbf M}^{\chi}$ and the corresponding normalized eigenvectors as the columns of the $N\times r$ matrix $\wt{\mbf V}^{\chi}$, with rows $\wt{\mbf v}_i^{\chi\prime}:=(\wt{v}_{i1}^{\chi}\cdots \wt{v}_{ir}^{\chi})$, $i=1,\ldots,N$. 
The loadings and factors are then simulated as 
\[
\bm \lambda_i:=(\wt{\mbf M}^{\chi})^{1/2}\wt{\bm{\mathcal S}}\wt{\mbf v}_i^{\chi}\; \text{ and }\;  \mbf F_t := (\wt{\mbf M}^{\chi})^{-1/2} \wt{\bm{\mathcal S}}\wt{\mbf V}^{\chi\prime}\bm\chi_t,
\] 
with $\wt{\bm{\mathcal S}}$ being an $r\times r$ diagonal matrix having entries $\pm 1$ according to $\wt{\bm{\mathcal S}}_{jj} :=\mathbb I(\wt{v}^{\chi}_{1j}> 0)-\mathbb I(\wt{ v}^{\chi}_{1j}\le  0)$, $j=1,\ldots, r$, so that Assumption \ref{ass:ident}(c) is also satisfied. It is shown in Appendix \ref{app:LF} that such transformation is equivalent to applying a linear invertible transformation to $\bm \ell_i$ and $\mbf F_t$.

We simulate the model described above $B=500$ times, and at each replication, $b=1,\ldots, B$, we estimate the loadings by means of: (i) unfeasible  OLS using the true factors, (ii) PC as in \citet{MBPCA}, (iii) QML implemented via the EM algorithm in \citet{baili12,baili16}, and (iv) EM algorithm as in \citet{DGRqml} and \citet{BLqml}. Both the QML and EM estimators are obtained by using the PC estimator to initialize the iterations. We estimate the factors in the following ways: (i) unfeasible OLS using the true loadings, (ii) PC using the PC estimator of the loadings as in \citet{MBPCA}, (iii) LP and WLS using the QML estimators of the loadings and  idiosyncratic variances as in \citet{baili12,baili16}, and (iv)  Kalman smoother  using the EM estimators of the loadings and  idiosyncratic variances as in \citet{DGRqml} and \citet{BLqml}.

 In Table \ref{tab:MSEL}  we report the Mean-Squared-Error (MSE) for each column, $j=1,\ldots, r$, of the considered loadings estimators, averaged over the $B$ replications (with standard deviations in parenthesis).


 \begin{table}[h!]
 \caption{MSEs - Loadings}\label{tab:MSEL}
 \centering
\vskip .1cm
\scriptsize
 \begin{tabular}{cccc | cccc | cccc}
 \hline
 \hline
 &&&&&&& \\[-8pt]
 $n$ & $T$ &$\tau$ &$\delta$ &  ${\bm\Lambda}_{\cdot 1}^{\text{\tiny \upshape OLS}}$ & $\wh{\bm\Lambda}_{\cdot 1}^{\text{\tiny \upshape PC}}$ & $\wh{\bm\Lambda}_{\cdot 1}^{\text{\tiny \upshape QML,S}}$&$\wh{\bm\Lambda}_{\cdot 1}^{\text{\tiny \upshape EM}}$ &  ${\bm\Lambda}_{\cdot 2}^{\text{\tiny \upshape OLS}}$ & $\wh{\bm\Lambda}_{\cdot 1}^{\text{\tiny \upshape PC}}$  & $\wh{\bm\Lambda}_{\cdot 2}^{\text{\tiny \upshape QML,S}}$&$\wh{\bm\Lambda}_{\cdot 1}^{\text{\tiny \upshape EM}}$ \\
 \hline
 &&&&&&& \\[-8pt]
 20 & 100 & 0 & 0 &		0.0103	&	0.0123	&	0.0116	&	0.0118	&	0.0099	&	0.0153	&	0.0118	&	0.0131	\\[-3pt]
&&&&\tiny(0.0032)&\tiny(0.0060)&\tiny(0.0053)&\tiny(0.0060)&\tiny(0.0034)&\tiny(0.0069)&\tiny(0.0053)&\tiny(0.0060)\\
 50 & 100 & 0 & 0 &	0.0102	&	0.0109	&	0.0108	&	0.0108	&	0.0102	&	0.0116	&	0.0107	&	0.0112	\\[-3pt]
&&&&\tiny(0.0021)&\tiny(0.0023)&\tiny(0.0023)&\tiny(0.0023)&\tiny(0.0022)&\tiny(0.0025)&\tiny(0.0023)&\tiny(0.0024)\\
 100 & 100 & 0 & 0 &	0.0100	&	0.0103	&	0.0103	&	0.0103	&	0.0100	&	0.0105	&	0.0104	&	0.0104	\\[-3pt]
&&&&\tiny(0.0015)&\tiny(0.0016)&\tiny(0.0015)&\tiny(0.0015)&\tiny(0.0015)&\tiny(0.0016)&\tiny(0.0016)&\tiny(0.0016)\\
 200 & 100 & 0 & 0 &	0.0101	&	0.0102	&	0.0102	&	0.0102	&	0.0101	&	0.0104	&	0.0103	&	0.0103	\\[-3pt]
&&&&\tiny(0.0011)&\tiny(0.0011)&\tiny(0.0011)&\tiny(0.0011)&\tiny(0.0011)&\tiny(0.0011)&\tiny(0.0011)&\tiny(0.0011)\\
 \hline
 &&&&&&& \\[-8pt]
  20 & 100 & 0.5 & 0.5 &			0.0180	&	0.0239	&	0.0230	&	0.0234	&	0.0134	&	0.0270	&	0.0274	&	0.0287	\\	[-3pt]
&&&&\tiny(0.0081)&\tiny(0.0115)&\tiny(0.0110)&\tiny(0.0112)&\tiny(0.0055)&\tiny(0.0188)&\tiny(0.0199)&\tiny(0.0209)\\	
  50 & 100 & 0.5 & 0.5 &		0.0183	&	0.0201	&	0.0199	&	0.0200	&	0.0135	&	0.0166	&	0.0160	&	0.0168	\\	[-3pt]
&&&&\tiny(0.0054)&\tiny(0.0060)&\tiny(0.0060)&\tiny(0.0060)&\tiny(0.0037)&\tiny(0.0050)&\tiny(0.0047)&\tiny(0.0050)\\	
  100 & 100 & 0.5 & 0.5 &		0.0182	&	0.0190	&	0.0189	&	0.0190	&	0.0134	&	0.0146	&	0.0145	&	0.0146	\\	[-3pt]
&&&&\tiny(0.0038)&\tiny(0.0041)&\tiny(0.0041)&\tiny(0.0041)&\tiny(0.0027)&\tiny(0.0031)&\tiny(0.0030)&\tiny(0.0031)\\	
  200 & 100 & 0.5 & 0.5 &		0.0184	&	0.0187	&	0.0187	&	0.0187	&	0.0135	&	0.0141	&	0.0141	&	0.0141	\\	[-3pt]
&&&&\tiny(0.0027)&\tiny(0.0028)&\tiny(0.0028)&\tiny(0.0028)&\tiny(0.0021)&\tiny(0.0022)&\tiny(0.0022)&\tiny(0.0023)\\
\hline
\hline
 \end{tabular}
 \end{table}

 \begin{table}[h!]
 \caption{MSEs - Factors}\label{tab:MSEF}
 \centering
\vskip .1cm
\scriptsize
 \begin{tabular}{cccc | ccccc | ccccc}
 \hline
 \hline
 &&&&&&&& \\[-8pt]
 $n$ & $T$ &$\tau$ &$\delta$ &  ${\bm F}_{\cdot 1}^{\text{\tiny \upshape OLS}}$ & $\wh{\bm F}_{\cdot 1}^{\text{\tiny \upshape PC}}$ & $\wh{\bm F}_{\cdot 1}^{\text{\tiny \upshape WLS}}$&$\wh{\bm F}_{\cdot 1}^{\text{\tiny \upshape LP}}$&$\wh{\bm F}_{\cdot 1}^{\text{\tiny \upshape KS}}$ &  ${\bm F}_{\cdot 2}^{\text{\tiny \upshape OLS}}$ & $\wh{\bm F}_{\cdot 2}^{\text{\tiny \upshape PC}}$ & $\wh{\bm F}_{\cdot 2}^{\text{\tiny \upshape WLS}}$&$\wh{\bm F}_{\cdot 2}^{\text{\tiny \upshape LP}}$&$\wh{\bm F}_{\cdot 2}^{\text{\tiny \upshape KS}}$  \\
 \hline
 &&&&&&&& \\[-8pt]
 20 & 100 & 0 & 0 &	0.0311	&	0.0310	&	0.0301	&	0.0293	&	0.0273	&	0.1576	&	0.1444	&	0.1539	&	0.1329	&	0.1305	\\	[-3pt]
&&&&\tiny(0.0054)&\tiny(0.0069)&\tiny(0.0064)&\tiny(0.0061)&\tiny(0.0062)&\tiny(0.0939)&\tiny(0.0766)&\tiny(0.0950)&\tiny(0.0672)&\tiny(0.0639)\\	
 50 & 100 & 0 & 0 &	0.0121	&	0.0122	&	0.0116	&	0.0115	&	0.0111	&	0.0576	&	0.0560	&	0.0549	&	0.0521	&	0.0518	\\	[-3pt]
&&&&\tiny(0.0019)&\tiny(0.0020)&\tiny(0.0018)&\tiny(0.0018)&\tiny(0.0018)&\tiny(0.0208)&\tiny(0.0200)&\tiny(0.0198)&\tiny(0.0178)&\tiny(0.0176)\\	
 100 & 100 & 0 & 0 &	0.0062	&	0.0062	&	0.0059	&	0.0059	&	0.0057	&	0.0275	&	0.0274	&	0.0260	&	0.0258	&	0.0257	\\	[-3pt]
&&&&\tiny(0.0009)&\tiny(0.0010)&\tiny(0.0009)&\tiny(0.0009)&\tiny(0.0009)&\tiny(0.0086)&\tiny(0.0085)&\tiny(0.0082)&\tiny(0.0080)&\tiny(0.0079)\\	
 200 & 100 & 0 & 0 &	0.0031	&	0.0031	&	0.0029	&	0.0029	&	0.0029	&	0.0135	&	0.0136	&	0.0129	&	0.0128	&	0.0128	\\	[-3pt]
&&&&\tiny(0.0005)&\tiny(0.0005)&\tiny(0.0005)&\tiny(0.0005)&\tiny(0.0005)&\tiny(0.0036)&\tiny(0.0036)&\tiny(0.0034)&\tiny(0.0034)&\tiny(0.0034)\\
 \hline
 &&&&&&&& \\[-8pt]
  20 & 100 & 0.5 & 0.5 &		0.0653	&	0.0638	&	0.0685	&	0.0671	&	0.0632	&	0.1495	&	0.1674	&	0.2227	&	0.1980	&	0.1924	\\	[-3pt]
&&&&\tiny(0.0134)&\tiny(0.0132)&\tiny(0.0134)&\tiny(0.0131)&\tiny(0.0129)&\tiny(0.0826)&\tiny(0.1295)&\tiny(0.1827)&\tiny(0.1509)&\tiny(0.1571)\\	
  50 & 100 & 0.5 & 0.5 &	0.0275	&	0.0270	&	0.0275	&	0.0272	&	0.0264	&	0.0558	&	0.0597	&	0.0652	&	0.0619	&	0.0620	\\	[-3pt]
&&&&\tiny(0.0053)&\tiny(0.0054)&\tiny(0.0052)&\tiny(0.0052)&\tiny(0.0051)&\tiny(0.0229)&\tiny(0.0267)&\tiny(0.0306)&\tiny(0.0279)&\tiny(0.0275)\\	
  100 & 100 & 0.5 & 0.5 &	0.0140	&	0.0139	&	0.0140	&	0.0139	&	0.0137	&	0.0273	&	0.0286	&	0.0290	&	0.0287	&	0.0289	\\	[-3pt]
&&&&\tiny(0.0024)&\tiny(0.0025)&\tiny(0.0025)&\tiny(0.0024)&\tiny(0.0025)&\tiny(0.0091)&\tiny(0.0099)&\tiny(0.0102)&\tiny(0.0100)&\tiny(0.0101)\\	
  200 & 100 & 0.5 & 0.5 &	0.0070	&	0.0070	&	0.0070	&	0.0070	&	0.0069	&	0.0135	&	0.0141	&	0.0139	&	0.0139	&	0.0140	\\	[-3pt]
&&&&\tiny(0.0012)&\tiny(0.0012)&\tiny(0.0013)&\tiny(0.0013)&\tiny(0.0013)&\tiny(0.0040)&\tiny(0.0042)&\tiny(0.0042)&\tiny(0.0041)&\tiny(0.0042)\\
\hline
\hline
 \end{tabular}
 \end{table}

\section{An application on a large euro area dataset}

We analyze a new macroeconomic dataset of the euro area (EA) of $N=116$ quarterly series observed in the period 2000:Q1-2019:Q4. Missing values are imputed using the EM algorithm by \citet{stockwatson02JASA}. This data is available from \citet{BLEA}.

After transforming data to stationarity, the criterion by \citet{baing02} suggests the presence of $r=4$ common factors, explaining on averaged 63\% of total variance in the data. 

\begin{figure}[h!]
\caption{}\label{fig:EA}
\centering
\begin{tabular}{cc}
\scriptsize EM plus Kalman filter&\scriptsize QML plus WLS\\
\includegraphics[width = .35\textwidth]{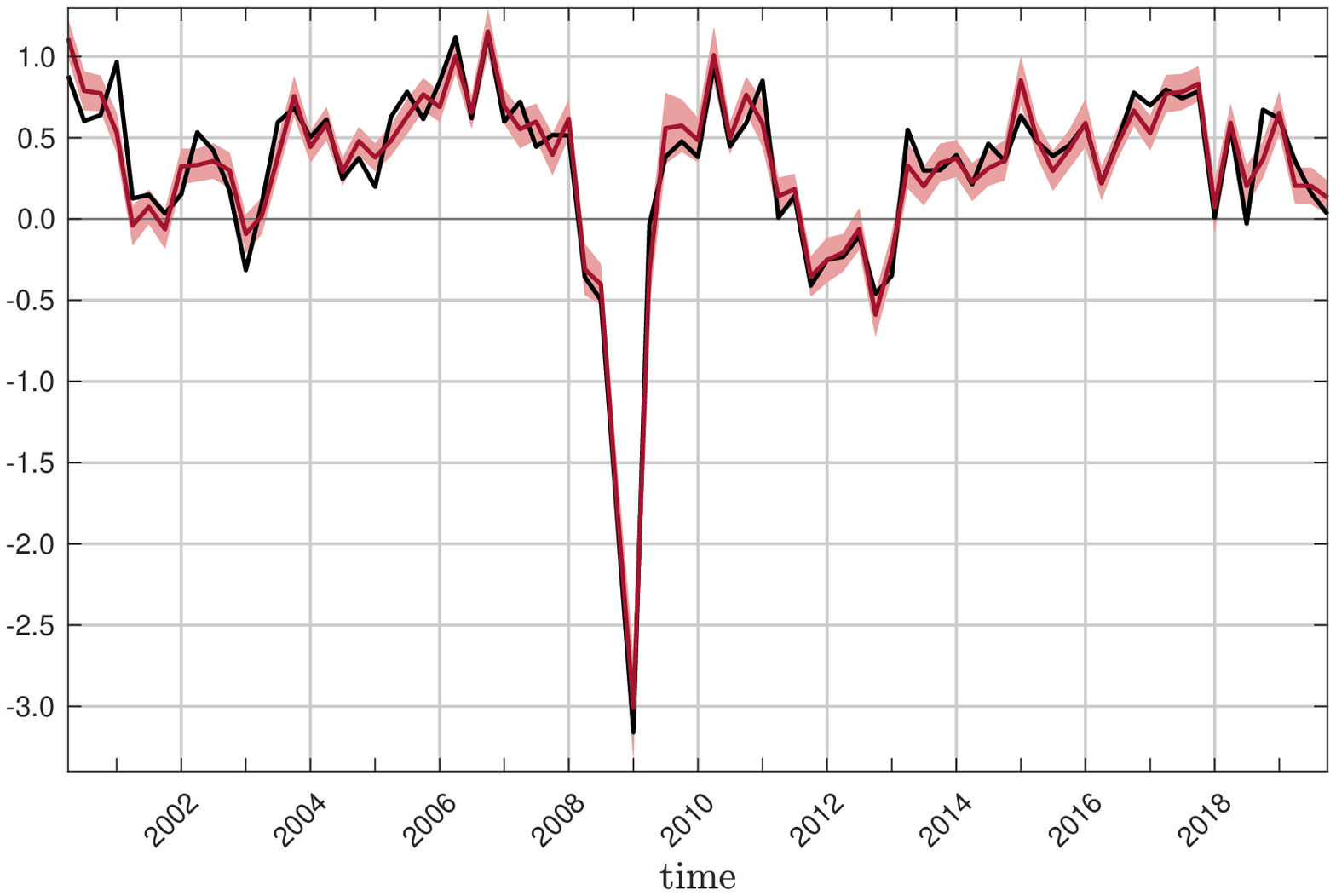}&\includegraphics[width = .35\textwidth]{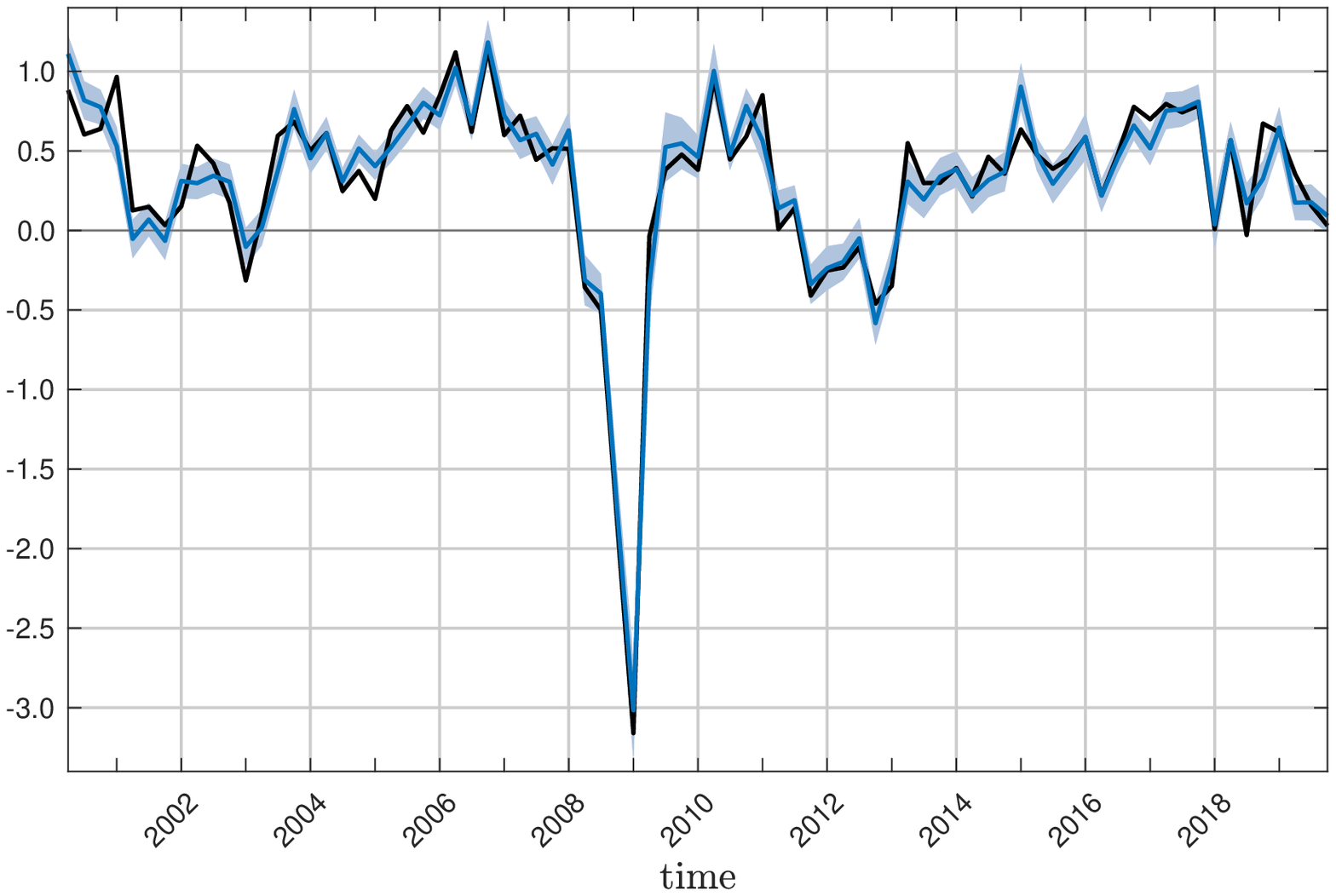}\\[-5pt]
\scriptsize GDP growth rate&\scriptsize GDP growth rate\\
\includegraphics[width = .35\textwidth]{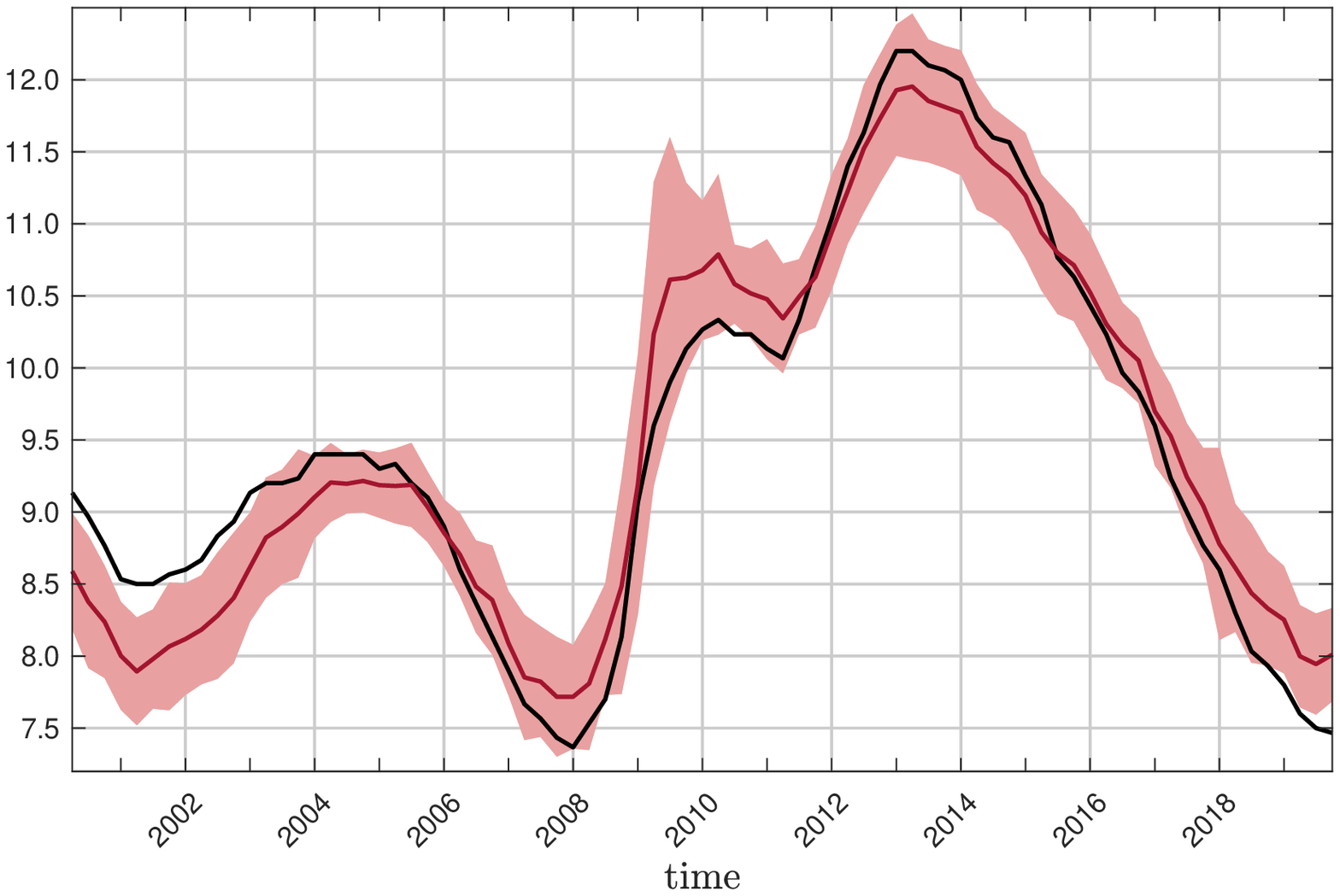}&\includegraphics[width = .35\textwidth]{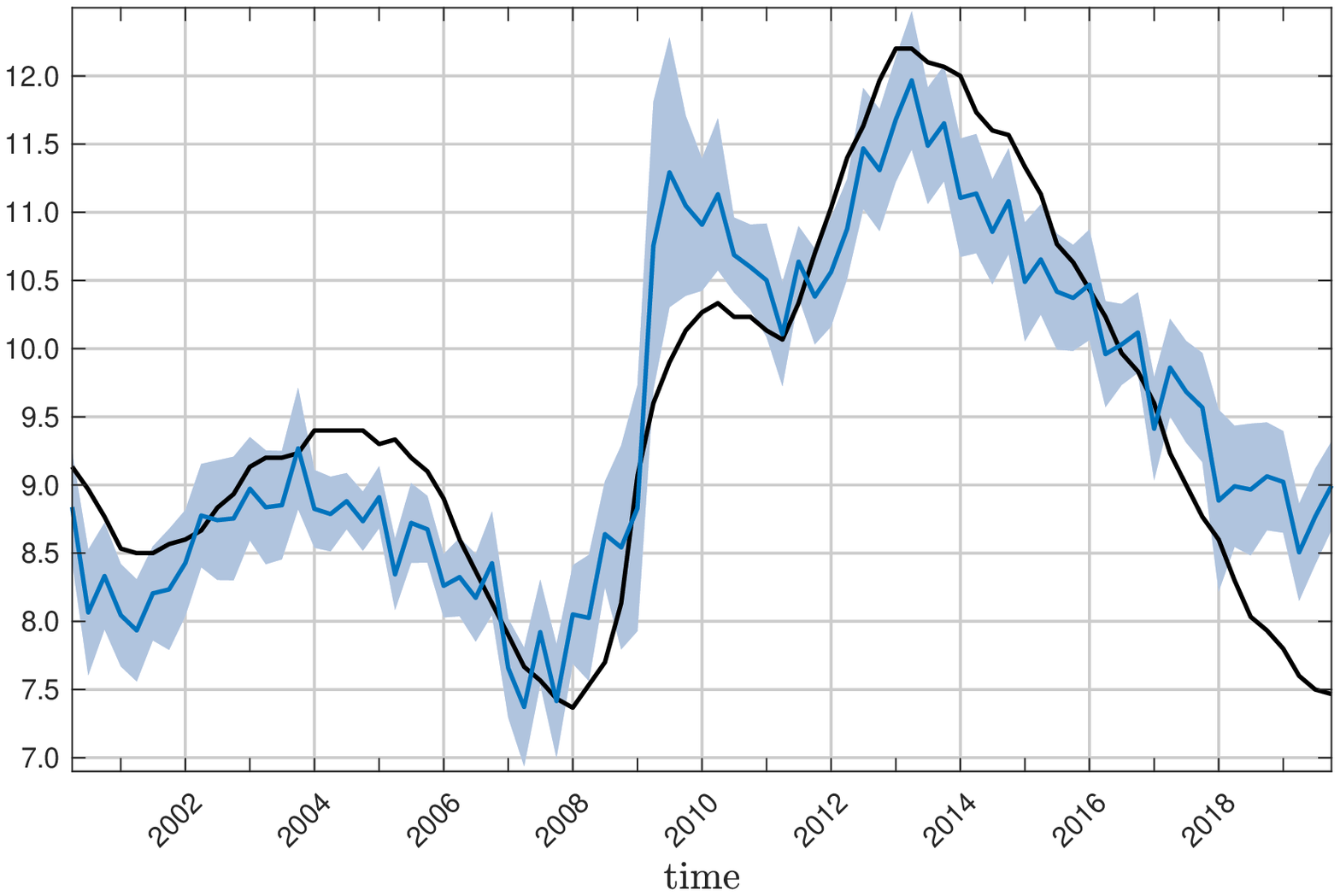}\\[-5pt]
\scriptsize unemployment rate&\scriptsize unemployment rate\\
\includegraphics[width = .35\textwidth]{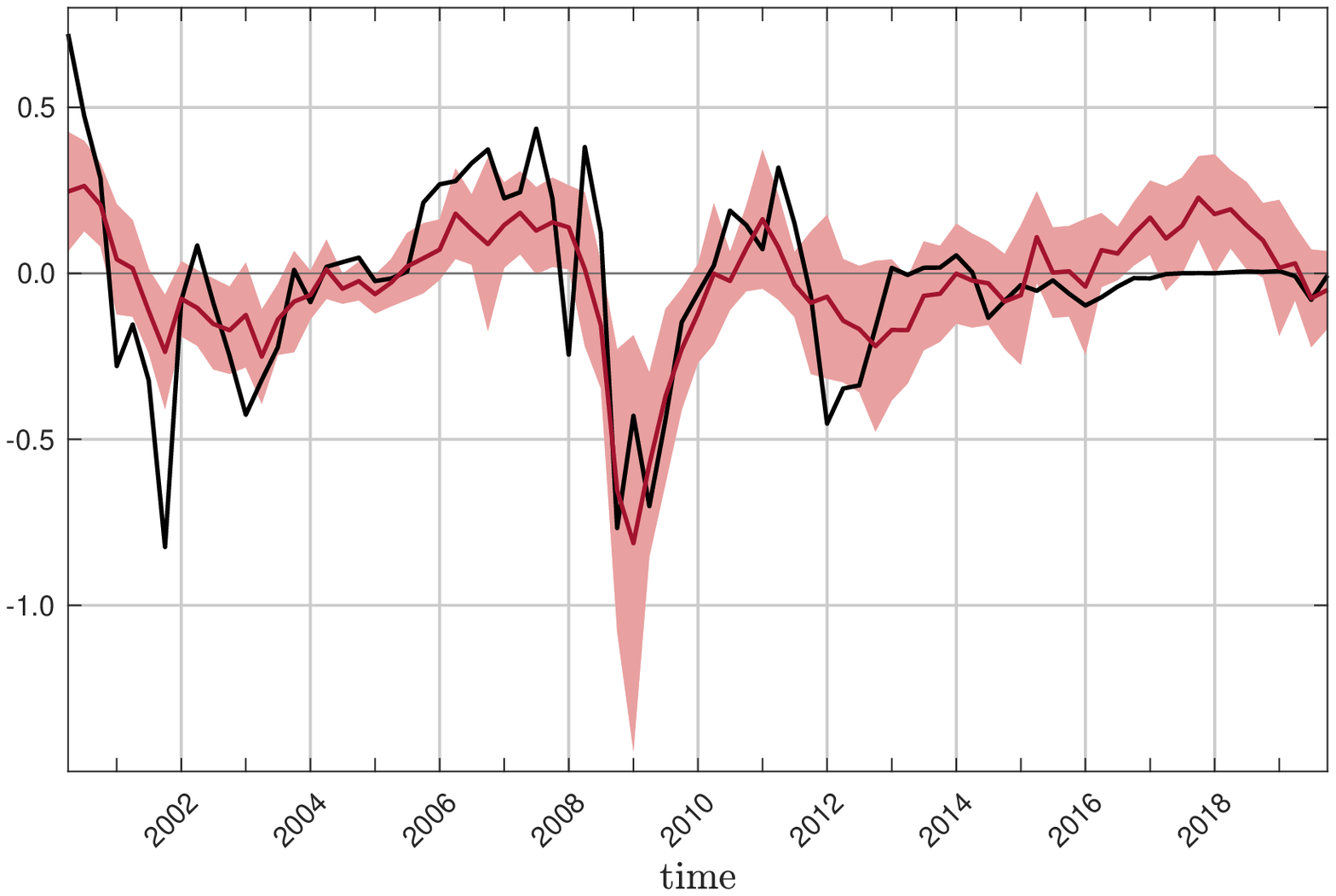}&\includegraphics[width = .35\textwidth]{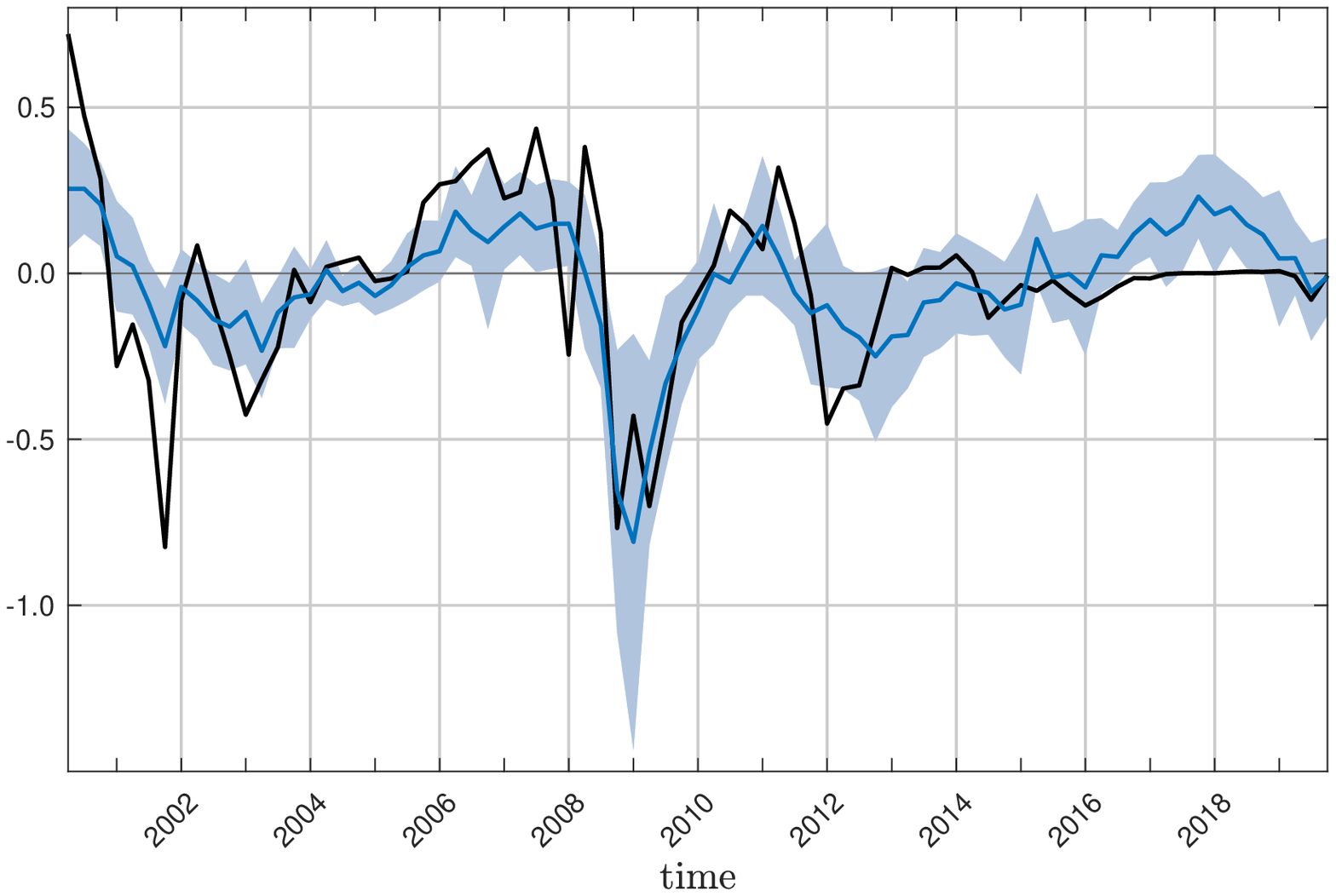}\\[-5pt]
\scriptsize 3-months interest rate&\scriptsize 3-months interest rate\\
\includegraphics[width = .35\textwidth]{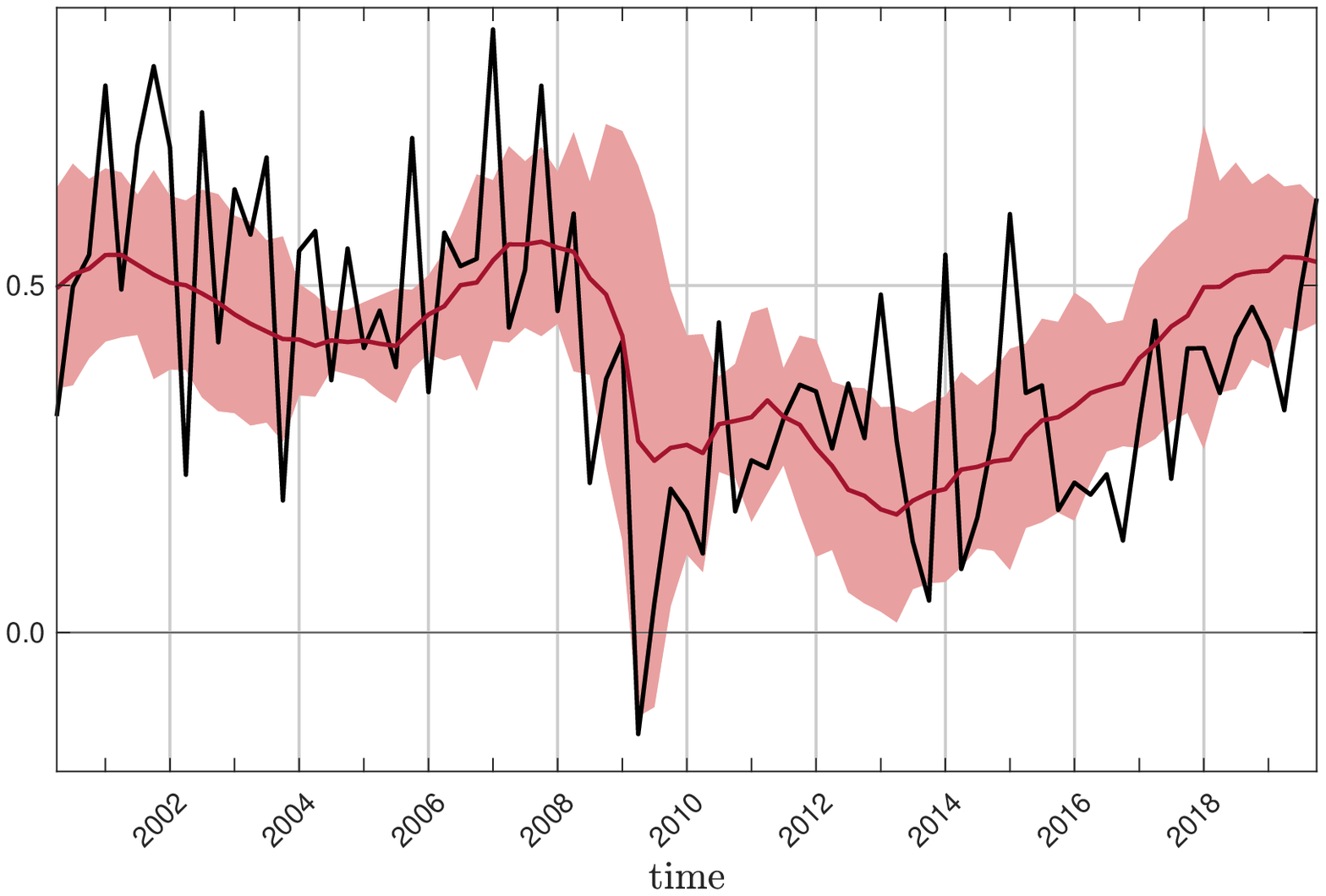}&\includegraphics[width = .35\textwidth]{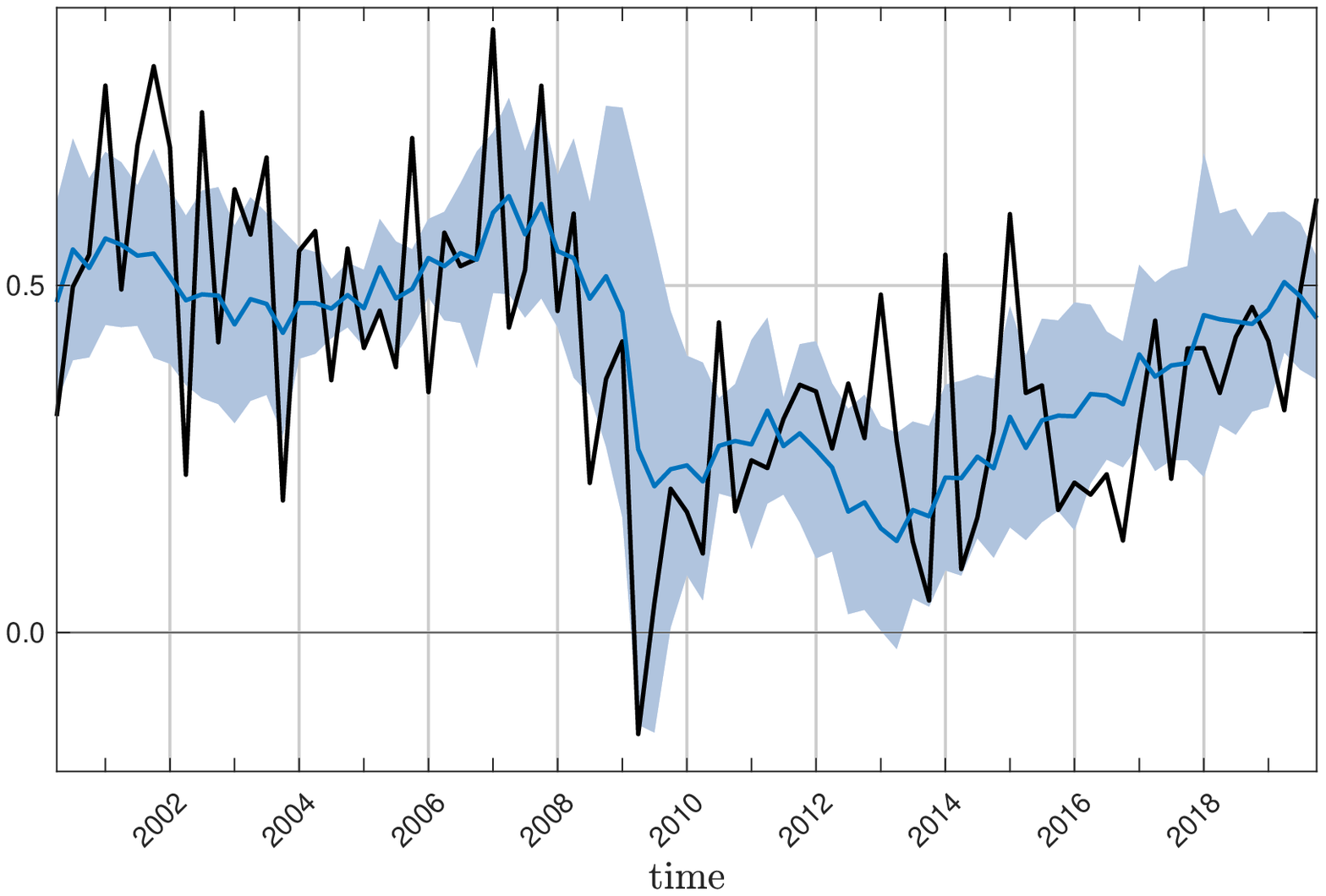}\\[-5pt]
\scriptsize GDP deflator inflation rate&\scriptsize GDP deflator inflation rate\\
\includegraphics[width = .35\textwidth]{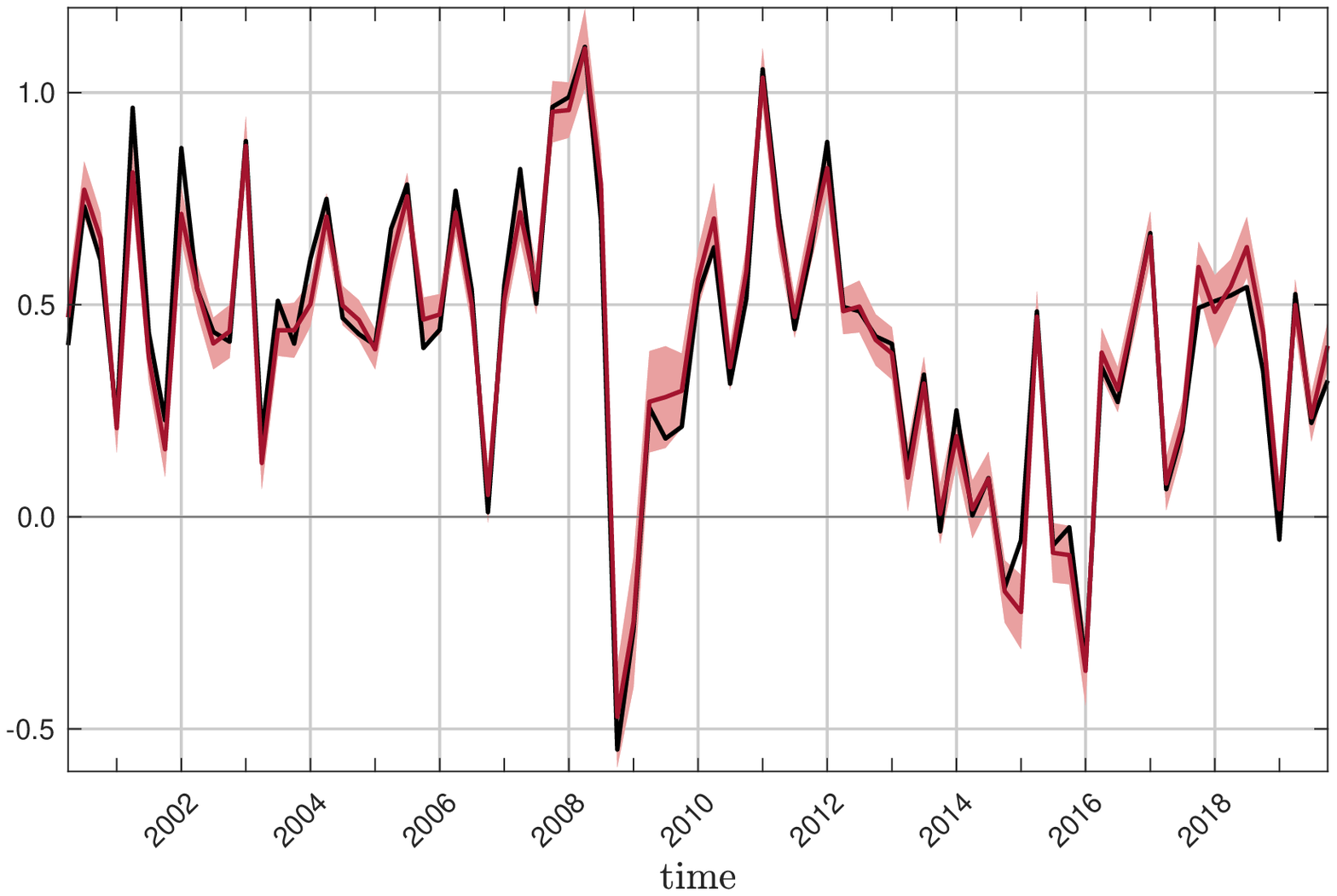}&\includegraphics[width = .35\textwidth]{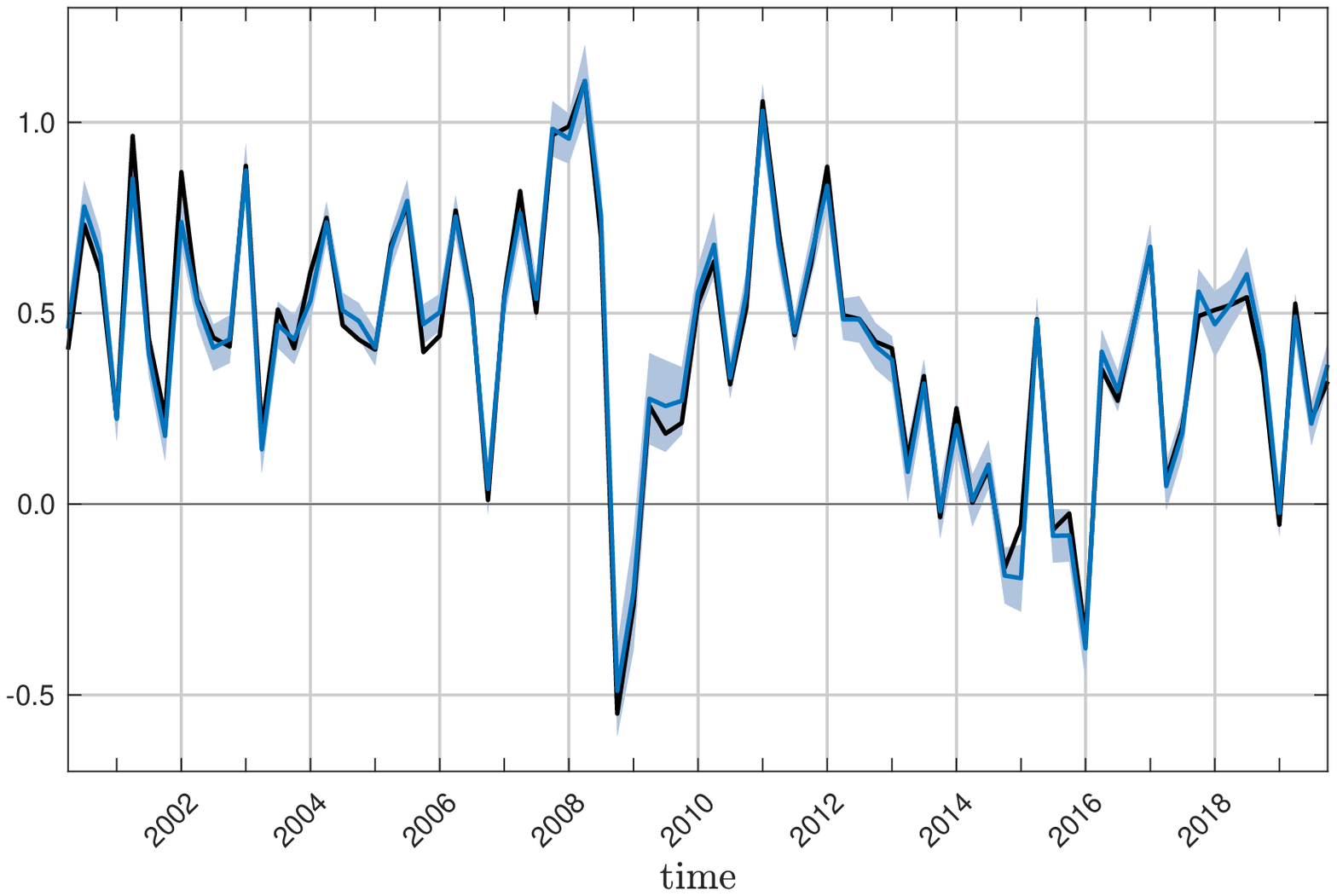}\\[-5pt]
\scriptsize HICP inflation rate&\scriptsize HICP inflation rate\\
\end{tabular}
\end{figure}

Figure \ref{fig:EA}  shows the  common component of: GDP growth rate, unemployment rate, 3-months interest rate, and inflation rates (measured as the growth rate of the GDP deflator and of the Harmonized Index of Consumer Prices). For each variable we consider the estimates obtained by means of PC analysis as in \citet{bai03}, QML plus WLS as in \citet{baili16}, and EM algorithm plus Kalman smoother as in \citet{DGRqml}.
Given Propositions 
\ref{prop:LPCA}, \ref{prop:FPCA},
\ref{prop:LBAILI16}, \ref{prop:FBAILI16},
\ref{prop:BLload}, and \ref{prop:BLFac},  each estimated common components satisfy the following CLTs, as $N,T\to\infty$,
\begin{align}
&\frac 1 {\sqrt{\l(\frac 1T {\mbf F}_t^\prime{\bm{\mathcal V}}_i^{\text{\tiny OLS}} {\mbf F}_t+\frac 1N{\bm\lambda}_i^\prime{\bm{\mathcal W}}_t^{\text{\tiny OLS}}{\bm\lambda}_i\r)}}\l(\wh{\bm\lambda}_i^{\text{\tiny PC}\prime}\wh{\mbf F}_t^{\text{\tiny PC}}-{\bm\lambda}_i^\prime{\mbf F}_t\r)\to_d\mathcal N(0,1), \nn\\
&\frac 1 {\sqrt{\l(\frac 1T {\mbf F}_t^\prime{\bm{\mathcal V}}_i^{\text{\tiny OLS}} {\mbf F}_t+\frac 1N{\bm\lambda}_i^\prime{\bm{\mathcal W}}_t^{\text{\tiny WLS}}{\bm\lambda}_i\r)}}\l(\wh{\bm\lambda}_i^{\text{\tiny QML,S}\prime}\wh{\mbf F}_t^{\text{\tiny WLS}}-{\bm\lambda}_i^\prime{\mbf F}_t\r)\to_d\mathcal N(0,1), \nn\\
&\frac 1 {\sqrt{\l(\frac 1T {\mbf F}_t^\prime{\bm{\mathcal V}}_i^{\text{\tiny OLS}} {\mbf F}_t+\frac 1N{\bm\lambda}_i^\prime{\bm{\mathcal W}}_t^{\text{\tiny WLS}}{\bm\lambda}_i\r)}}\l(\wh{\bm\lambda}_i^{\text{\tiny EM}\prime}\wh{\mbf F}_t^{\text{\tiny KS}}-{\bm\lambda}_i^\prime{\mbf F}_t\r)\to_d\mathcal N(0,1).\nn
\end{align}
The 95\% confidence bands are computed accordingly, and by estimating $\bm{\mathcal V}_i$ using the classical HAC estimator \citep{andrews91} and $\bm{\mathcal W}_t$  using a similar Cross-Sectional-HAC estimator \citep{baing06}.

\section{Concluding remarks}
In this paper, we have shown two main results. First, when estimating an approximate static or dynamic factor model by QML, possibly via the EM algorithm, and using a mis-specified log-likelihood for an exact factor model, the estimated loadings are asymptotically equivalent to the loadings estimated via PC analysis. Why should then we be using the QML estimator rather than the PC estimator of the loadings? The main reason put forward by the literature seems to be the possibility of easily imposing constraints on the loadings, e.g., incorporating prior beliefs.

Second, we have also shown that the factors estimated via WLS or via the Kalman filter/smoother, are all asymptotically equivalent, and these estimators should be always preferred over the PC estimators as they address the heteroskedasticity of the idiosyncratic components so they might even be more efficient (depending on the amount of neglected cross-sectional correlations). Furthermore, in a time series setting the Kalman filter/smoother should be preferred as it allows to easily deal with missing values and given its dynamic aggregation property it can handle very persistent or even non-stationary data without any modification.

Summing up, although the QML plus WLS estimation approach, proposed by \citet{baili16}, is asymptotically equivalent to the EM plus Kalman smoother approach, proposed by \citet{DGRqml}, the latter is likely to produce estimated factors which enjoy better final sample properties and is a more flexible approach for dealing with complex economic datasets. Relevant examples of the use of this estimation technique are:
\begin{inparaenum} 
\item [(i)] counterfactual analysis \citep{GRS06,GLR19};
\item [(ii)] conditional forecasts \citep{banburagiannonelenza15};
\item [(iii)] nowcasting (\citealp{GRS04,Nowcasting,GMR16,BGMR13,modugno2013now});
\item [(iv)] dealing with data irregularly spaced in time (\citealp{marianomurasawa03,JKVW2011,banburamodugno14});
\item [(v)] imposing constraints on the loadings to account for smooth cross-sectional dependence in the case of ordered units (\citealp{koopman13,JKVW2014}) or for a block-specific factor structure (\citealp{CGM16,altavilla2017,DCGF2021,BCGM21});
\item [(vi)] building indicators of the economic activity (\citealp{reiswatson10,OGAP,ng2023constructing});
\item [(vi)] impulse response analysis (\citealp{juvenalpetrella2015,smokinggun});
\item [(vii)] the analysis of stock markets (\citealp{Linton21});
\item [(viii)] firm-level or household-level repeated cross-sections of data (\citealp{barigozzi2023multidimensional}).
\end{inparaenum}

\newpage
\appendix
\small
\section{Kalman filter and Kalman smoother}\label{app:KS}

%
The following iterations are stated assuming that the true value of the parameters $\bm\varphi$ is given and for given initial conditions $\mbf F_{0|0}$ and $\mbf P_{0|0}$. 


The Kalman filter is based on the forward iterations for $t=1,\ldots, T$:
\begin{align}
&\mbf F_{t|t-1} = \mbf A \mbf F_{t-1|t-1},\nn\\
&\mbf P_{t|t-1} = \mbf A\mbf P_{t-1|t-1} \mbf A^\prime + \mbf H\mbf H^\prime,\nn\\
&\mbf F_{t|t} =\mbf F_{t|t-1}+\mbf P_{t|t-1}\bm\Lambda^\prime(\bm\Lambda\mbf P_{t|t-1}\bm\Lambda^\prime+\bm\Sigma^\xi)^{-1}(\mbf x_{t}-\bar{\mbf x}-\bm\Lambda\mbf F_{t|t-1}),\nn\\
&\mbf P_{t|t} =\mbf P_{t|t-1}-\mbf P_{t|t-1}\bm\Lambda^\prime(\bm\Lambda\mbf P_{t|t-1}\bm\Lambda^\prime+\bm\Sigma^\xi)^{-1}\bm\Lambda\mbf P_{t|t-1}.\nn
\end{align}
Alternatively for $t=1,\ldots, T$ (by Woodbury formula)
\begin{align}
&\mbf F_{t|t} =\mbf F_{t|t-1}+(\bm\Lambda^\prime (\bm\Sigma^\xi)^{-1}\bm\Lambda+\mbf P_{t|t-1}^{-1})^{-1}\bm\Lambda^\prime (\bm\Sigma^\xi)^{-1}(\mbf x_{t}-\bar{\mbf x}-\bm\Lambda\mbf F_{t|t-1}),\nn\\
&\mbf P_{t|t} =\mbf P_{t|t-1}-(\bm\Lambda^\prime (\bm\Sigma^\xi)^{-1}\bm\Lambda+\mbf P_{t|t-1}^{-1})^{-1}\bm\Lambda^\prime (\bm\Sigma^\xi)^{-1}\bm\Lambda\mbf P_{t|t-1}.\nn
\end{align}


The Kalman Smoother is then based on the backward iterations for $t=T,\ldots, 1$:
\begin{align}
\mbf F_{t|T} &=\mbf F_{t|t}+\mbf P_{t|t}\mbf A^\prime\mbf P_{t+1|t}^{-1}(\mbf F_{t+1|T}-\mbf F_{t+1|t}),\nn\\
\mbf P_{t|T}&=\mbf P_{t|t} + \mbf P_{t|t} \mbf A^\prime \mbf P_{t+1|t}^{-1}
(\mbf P_{t+1|T}-\mbf P_{t+1|t})\mbf P_{t+1|t}^{-1} \mbf A \mbf P_{t|t}.\nn
\end{align}
Alternatively for $t=T,\ldots, 1$ (see, e.g., \citealp[Chapter 4.4, pp.87-91]{DK01})
\begin{align}
&\mbf F_{t|T}=\mbf F_{t|t-1}+\mbf P_{t|t-1}\mbf r_{t-1},\nn\\
&\mbf r_{t-1}=\bm\Lambda^\prime(\bm\Lambda\mbf P_{t|t-1}\bm\Lambda^\prime+\bm\Sigma^\xi)^{-1}(\mbf x_t-\bar{\mbf x}-\bm\Lambda\mbf F_{t|t-1})+\mbf L^\prime_t\mbf r_t,\nn\\
&\mbf P_{t|T}=\mbf P_{t|t-1}(\mbf I_r-\mbf N_{t-1}\mbf P_{t|t-1}),\nn\\
&\mbf N_{t-1}=\bm\Lambda_n^\prime(\bm\Lambda\mbf P_{t|t-1}\bm\Lambda^\prime+\bm\Sigma^\xi)^{-1}\bm\Lambda+\mbf L_t^\prime\mbf N_t\mbf L_t,\nn\\
&\mbf L_t= \mbf A-\mbf A \mbf P_{t|t-1} \bm\Lambda^\prime (\bm\Lambda\mbf P_{t|t-1}\bm\Lambda^\prime+\bm\Sigma^\xi)^{-1} \bm\Lambda,\nn\\
&\mbf C_{t,t+1|T}=\mbf P_{t|t-1}\mbf L_t^\prime (\mbf I_r-\mbf N_{t}\mbf P_{t+1|t}), \qquad \mbf C_{t,t-1|T}=\mbf C_{t,t+1|T}^\prime,\nn
\end{align}
where $\mbf r_T=\mbf 0_{r}$, $\mbf N_T=\mbf 0_{r}$. 


\newpage
\section{Joint distribution of $\pmb{\mathcal X}$ and $\pmb{\mathcal F}$}\label{app:XFZF}

First, notice that, by Assumptions \ref{ass:VAR}(a) and \ref{ass:VAR}(g), for any $t=1,\ldots, T$, $\mbf F_t=\sum_{k=0}^{t-1} \mbf A^k\mbf H\mbf u_{t-k}$, which, under Assumption \ref{ass:tails}(b), is a linear combination of $t$ jointly Gaussians. Therefore, for all  $T\in\mathbb N$,
\beq\label{eq:GaussF}
\bm{\mathcal F}\sim \mathcal N(\mbf 0_{rT},\bm\Omega^F).
\eeq
Moreover, by Assumption \ref{ass:tails}(a) and \eqref{eq:GaussF}, and since $\bm {\mathcal Z}$ and $\bm{\mathcal F}$ are independent by Assumption \ref{ass:ind}, the joint density of $\bm {\mathcal Z}$ and $\bm{\mathcal F}$ is such that, for all $N,T\in\mathbb N$:
\begin{align}
f&(\bm {\mathcal Z},\bm {\mathcal F})= f(\bm {\mathcal Z})f(\bm {\mathcal F})\nonumber\\
&= \frac{1}{(2\pi)^{NT}\sqrt{\det(\bm\Omega^\xi)}}\exp\l\{-\frac 12\bm{\mathcal Z}^\prime(\bm\Omega^\xi)^{-1}\bm{\mathcal Z}\r\}
\frac{1}{(2\pi)^{rT}\sqrt{\det(\bm\Omega^F)}}\exp\l\{-\frac 12\bm{\mathcal F}^\prime(\bm\Omega^F)^{-1}\bm{\mathcal F}\r\}\label{eq:fZF}\\
&= \frac{1}{(2\pi)^{(N+r)T}\sqrt{\det(\bm\Omega^\xi)\det(\bm\Omega^F)}}\exp\l\{-\frac 12
\l(\ba{c}
\bm{\mathcal Z}\\
\bm{\mathcal F}
\ea\r)^\prime
\l(\ba{cc}
\bm\Omega^\xi&\mbf 0_{NT\times rT}\\
\mbf 0_{rT\times NT}&\bm\Omega^F
\ea
\r)^{-1}
\l(\ba{c}
\bm{\mathcal Z}\\
\bm{\mathcal F}
\ea\r)
\r\}\nn\\
&= \frac{1}{(2\pi)^{(N+r)T}\sqrt{\det(\bm\Omega^\xi)\det(\bm\Omega^F)}}\exp\l\{-\frac 12
\l(\ba{c}
\bm{\mathcal X}-\bm{\mathcal A}-\bm{\mathfrak L}\bm{\mathcal F}\\
\bm{\mathcal F}
\ea\r)^\prime
\l(\ba{cc}
\bm\Omega^\xi&\mbf 0_{NT\times rT}\\
\mbf 0_{rT\times NT}&\bm\Omega^F
\ea
\r)^{-1}
\l(\ba{c}
\bm{\mathcal X}-\bm{\mathcal A}-\bm{\mathfrak L}\bm{\mathcal F}\\
\bm{\mathcal F}
\ea\r)
\r\},\nn
\end{align}
which is equivalent to saying:
\beq\label{eq:gaussZF}
\l(\ba{c}
\bm{\mathcal Z}\\
\bm{\mathcal F}
\ea\r)\sim \mathcal N\l(\l(\ba{c}
\mbf 0_{NT}\\
\mbf 0_{rT}\ea\r),
\l(\ba{cc}
\bm\Omega^\xi& \mbf 0_{NT\times rT}\\
\mbf 0_{rT\times NT} & \bm\Omega^F
\ea\r)\r).
\eeq
Clearly, we have $\E[\bm {\mathcal X}|\bm {\mathcal F}]=\bm{\mathcal A}+ \bm{\mathfrak L}\bm {\mathcal F}+\E[\bm {\mathcal Z}|\bm {\mathcal F}]=\bm{\mathcal A}+ \bm{\mathfrak L}\bm {\mathcal F}+\E[\bm {\mathcal Z}]=\bm{\mathcal A}+ \bm{\mathfrak L}\bm {\mathcal F}$, by Assumption \ref{ass:ind}, and $\mathbb C\text {ov}(\bm {\mathcal X}|\bm {\mathcal F})=\bm\Omega^\xi$. Thus, from the last line of \eqref{eq:fZF} we see that $f(\bm{\mathcal X}|\bm{\mathcal F})=f(\bm{\mathcal Z})$. 
It follows that the joint density of $\bm {\mathcal X}$ and $\bm{\mathcal F}$ is such that:
\beq\label{eq:fXF0}
f(\bm {\mathcal X},\bm {\mathcal F})=f(\bm {\mathcal X}|\bm {\mathcal F})f(\bm {\mathcal F}) = f(\bm {\mathcal Z})f(\bm {\mathcal F}) =f(\bm {\mathcal Z},\bm {\mathcal F}),
\eeq
and the right-hand side is Gaussian because of \eqref{eq:gaussZF}. 
To verify \eqref{eq:fXF0} directly, notice that the distribution given in \eqref{eq:gaussXF} is  is equivalent to \eqref{eq:gaussZF}. Indeed, on the one hand, from the last line of \eqref{eq:fZF} we have:
\begin{align}
f(\bm{\mathcal Z},\bm{\mathcal F})= \frac{1}{(2\pi)^{(N+r)T}\sqrt{\det(\bm\Omega^\xi)\det(\bm\Omega^F)}}\exp\Bigg\{
-\frac 12\bigg[&(\bm{\mathcal X}-\bm{\mathcal A})^\prime(\bm\Omega^\xi)^{-1}(\bm{\mathcal X}-\bm{\mathcal A})-(\bm{\mathcal X}-\bm{\mathcal A})^\prime(\bm\Omega^\xi)^{-1}\bm{\mathfrak L}\bm{\mathcal F}\nonumber\\
& - \bm{\mathcal F}^\prime \bm{\mathfrak L}^\prime(\bm\Omega^\xi)^{-1}(\bm{\mathcal X}-\bm{\mathcal A})+\bm{\mathcal F}^\prime \bm{\mathfrak L}^\prime(\bm\Omega^\xi)^{-1} \bm{\mathfrak L}\bm{\mathcal F} + \bm{\mathcal F}^\prime(\bm\Omega^F)^{-1} \bm{\mathcal F}
\bigg]
\Bigg\},\nn
\end{align}
On the other hand, letting $\bm{\mathfrak P}:=\l((\bm\Omega^F)^{-1}+\bm{\mathfrak L}^\prime(\bm\Omega^\xi)^{-1}\bm{\mathfrak L}\r)$,
 from
\eqref{eq:gaussXF} we have
\begin{align}
f(\bm {\mathcal X},\bm {\mathcal F}) = \frac{1}{(2\pi)^{(N+r)T}\sqrt{\det\l(\ba{cc}
\bm\Omega^x& \bm{\mathfrak L}\bm\Omega^F\\ 
\bm\Omega^F\bm{\mathfrak L}^\prime&\bm\Omega^F
\ea
\r)}}\exp\Bigg\{-\frac 12\;\;& \l(\ba{c}
\bm{\mathcal X}-\bm{\mathcal A}\\
\bm{\mathcal F}
\ea\r)^\prime
\l(\ba{cc}
\bm\Omega^x& \bm{\mathfrak L}\bm\Omega^F\\ 
\bm\Omega^F\bm{\mathfrak L}^\prime&\bm\Omega^F
\ea\r)^{-1}
\l(\ba{c}
\bm{\mathcal X}-\bm{\mathcal A}\\
\bm{\mathcal F}
\ea\r)
\Bigg\}\nn\\
=\frac{1}{(2\pi)^{(N+r)T}\sqrt{\det(\bm\Omega^\xi)\det(\bm\Omega^F)}}
\exp\Bigg\{-\frac 12\bigg[&
(\bm{\mathcal X}-\bm{\mathcal A})^\prime(\bm\Omega^\xi)^{-1}(\bm{\mathcal X}-\bm{\mathcal A})-(\bm{\mathcal X}-\bm{\mathcal A})^\prime(\bm\Omega^\xi)^{-1}\bm{\mathfrak L}\bm{\mathcal F}\nn\\
&-\bm{\mathcal F}^\prime \l(\bm\Omega^F-\bm\Omega^F\bm{\mathfrak L}^\prime (\bm\Omega^x)^{-1}\bm{\mathfrak L}\bm\Omega^F\r)^{-1}
\bm\Omega^F\bm{\mathfrak L}^\prime (\bm\Omega^x)^{-1}(\bm{\mathcal X}-\bm{\mathcal A})\nn\\
&+\bm{\mathcal F}^\prime
\l(\bm\Omega^F-\bm\Omega^F\bm{\mathfrak L}^\prime (\bm\Omega^x)^{-1}\bm{\mathfrak L}\bm\Omega^F \r)^{-1}
\bm{\mathcal F}
\bigg]\Bigg\}\nn\\
=\frac{1}{(2\pi)^{(N+r)T}\sqrt{\det(\bm\Omega^\xi)\det(\bm\Omega^F)}}
\exp\Bigg\{-\frac 12\bigg[&
(\bm{\mathcal X}-\bm{\mathcal A})^\prime(\bm\Omega^\xi)^{-1}(\bm{\mathcal X}-\bm{\mathcal A})-(\bm{\mathcal X}-\bm{\mathcal A})^\prime(\bm\Omega^\xi)^{-1}\bm{\mathfrak L}\bm{\mathcal F}\nn\\
&-\bm{\mathcal F}^\prime\bm{\mathfrak P}\bm{\mathfrak P}^{-1}
\bm{\mathfrak L}^\prime (\bm\Omega^\xi)^{-1}(\bm{\mathcal X}-\bm{\mathcal A})+\bm{\mathcal F}^\prime\bm{\mathfrak P}\bm{\mathcal F}\bigg]\Bigg\}\nn\\
=\frac{1}{(2\pi)^{(N+r)T}\sqrt{\det(\bm\Omega^\xi)\det(\bm\Omega^F)}}
\exp\Bigg\{-\frac 12\bigg[&
(\bm{\mathcal X}-\bm{\mathcal A})^\prime(\bm\Omega^\xi)^{-1}(\bm{\mathcal X}-\bm{\mathcal A})-(\bm{\mathcal X}-\bm{\mathcal A})^\prime(\bm\Omega^\xi)^{-1}\bm{\mathfrak L}\bm{\mathcal F}\nn\\
&-\bm{\mathcal F}^\prime
\bm{\mathfrak L}^\prime (\bm\Omega^\xi)^{-1}(\bm{\mathcal X}-\bm{\mathcal A})+\bm{\mathcal F}^\prime
\bm{\mathfrak L}^\prime(\bm\Omega^\xi)^{-1}\bm{\mathfrak L}
\bm{\mathcal F}+\bm{\mathcal F}^\prime
(\bm\Omega^F)^{-1}\bm{\mathcal F}\bigg]\Bigg\}\nn\\
= f(\bm{\mathcal Z},\bm{\mathcal F}).\qquad\qquad\qquad\qquad\qquad\qquad\qquad\;\;\;
\label{eq:fXF}
\end{align}
To derive \eqref{eq:fXF} we used 
the formula of the determinant of a block matrix:
\[
\det\l(\ba{cc}
\bm\Omega^x& \bm{\mathfrak L}\bm\Omega^F\\ 
\bm\Omega^F\bm{\mathfrak L}^\prime&\bm\Omega^F
\ea
\r)=\det\l(\bm\Omega^x-\bm{\mathfrak L}\bm\Omega^F(\bm\Omega^F)^{-1}\bm\Omega^F\bm{\mathfrak L}^\prime\r) \det(\bm\Omega^F)=
\det(\bm\Omega^\xi) \det(\bm\Omega^F);
\]
the formula for inversion of a block-matrix:
\begin{align}
&\l(\ba{cc}
\bm\Omega^x& \bm{\mathfrak L}\bm\Omega^F\\ 
\bm\Omega^F\bm{\mathfrak L}^\prime&\bm\Omega^F
\ea
\r)^{-1}\nn\\
&=\l(\ba{cc}
\l(\bm\Omega^x-\bm{\mathfrak L}\bm\Omega^F(\bm\Omega^F)^{-1}\bm\Omega^F\bm{\mathfrak L}^\prime\r)^{-1}&\mbf 0_{NT\times rT}\\
\mbf 0_{rT\times NT} &\l(\bm\Omega^F-\bm\Omega^F\bm{\mathfrak L}^\prime(\bm\Omega^x)^{-1}\bm{\mathfrak L}\bm\Omega^F\r)^{-1}
\ea
\r)
\l(\ba{cc}
\mbf I_{NT}&-\bm{\mathfrak L}\bm\Omega^F(\bm\Omega^F)^{-1}\\
-\bm\Omega^F\bm{\mathfrak L}^\prime(\bm\Omega^x)^{-1}&\mbf I_{rT}
\ea
\r)\nn\\
&=\l(\ba{cc}
(\bm\Omega^\xi)^{-1}&\mbf 0_{NT\times rT}\\
\mbf 0_{rT\times NT} &\l(\bm\Omega^F-\bm\Omega^F\bm{\mathfrak L}^\prime(\bm\Omega^x)^{-1}\bm{\mathfrak L}\bm\Omega^F\r)^{-1}
\ea
\r)
\l(\ba{cc}
\mbf I_{NT}&-\bm{\mathfrak L}\\
-\bm\Omega^F\bm{\mathfrak L}^\prime(\bm\Omega^x)^{-1}&\mbf I_{rT}
\ea
\r)\nn\\
&=\l(\ba{cc}
(\bm\Omega^\xi)^{-1}&-(\bm\Omega^\xi)^{-1}\bm{\mathfrak L}\\
-\l(\bm\Omega^F-\bm\Omega^F\bm{\mathfrak L}^\prime(\bm\Omega^x)^{-1}\bm{\mathfrak L}\bm\Omega^F\r)^{-1}\bm\Omega^F\bm{\mathfrak L}^\prime(\bm\Omega^x)^{-1}&\l(\bm\Omega^F-\bm\Omega^F\bm{\mathfrak L}^\prime(\bm\Omega^x)^{-1}\bm{\mathfrak L}\bm\Omega^F\r)^{-1}
\ea
\r);\nn
\end{align}
and the Woodbury identities:
\begin{align}
&\l(\bm\Omega^F-\bm\Omega^F\bm{\mathfrak L}^\prime (\bm\Omega^x)^{-1}\bm{\mathfrak L}\bm\Omega^F \r)^{-1}=
\l(\bm\Omega^F-\bm\Omega^F\bm{\mathfrak L}^\prime \l(\bm{\mathfrak L}\bm\Omega^F\bm{\mathfrak L}^\prime+\bm\Omega^\xi\r)^{-1}\bm{\mathfrak L}\bm\Omega^F \r)^{-1}
=(\bm\Omega^F)^{-1}+\bm{\mathfrak L}^\prime(\bm\Omega^\xi)^{-1}\bm{\mathfrak L}=\bm{\mathfrak P},\nn\\
&\bm\Omega^F\bm{\mathfrak L}^\prime (\bm\Omega^x)^{-1}
=\bm\Omega^F\bm{\mathfrak L}^\prime\l(\bm{\mathfrak L}\bm\Omega^F\bm{\mathfrak L}^\prime+\bm\Omega^\xi\r)^{-1}= 
\l((\bm\Omega^F)^{-1}+\bm{\mathfrak L}^\prime(\bm\Omega^\xi)^{-1}\bm{\mathfrak L}\r)^{-1}\bm{\mathfrak L}^\prime(\bm\Omega^\xi)^{-1}=\bm{\mathfrak P}^{-1}\bm{\mathfrak L}^\prime(\bm\Omega^\xi)^{-1}.\nn
\end{align}

\section{Identification of simulated factors and loadings}\label{app:LF}

Let $\bm L:=(\bm\ell_1\cdots \bm\ell_N)^\prime$ and define:
\begin{align}
\wt{\bm{ R}}:= \l( \bm L^\prime\wt{\mbf V}^{\chi}\wt{\mbf V}^{\chi\prime} \bm L\r)^{-1} \bm L^\prime \wt{\mbf V}^{\chi} \wt{\bm{\mathcal S}} (\wt{\mbf M}^\chi)^{1/2},\nn
\end{align}
First, notice that $\wt{\bm\Gamma}^\chi$ is a consistent estimator of ${\bm\Gamma}^\chi$ and so the eigenvectors and eigenvalues of $\wt{\bm\Gamma}^\chi$ are consistent estimators of the eigenvectors and eigenvalues of ${\bm\Gamma}^\chi$. Therefore, with probability tending to one as $N,T\to\infty$, $N^{-1}\wt{\mbf M}^\chi$ is finite and invertible and 
the columns of $\bm L$ and of $\wt{\mbf V}^{\chi}$ span the same space, thus
\begin{align}
&\bm L\l(\bm L^\prime \bm L\r)^{-1}\bm L^\prime\wt{\mbf V}^{\chi}=\wt{\mbf V}^{\chi},\label{eq:id1}\\
&\wt{\mbf V}^{\chi}\l(\wt{\mbf V}^{\chi\prime}\wt{\mbf V}^{\chi}\r)^{-1}\wt{\mbf V}^{\chi\prime}\bm L=\wt{\mbf V}^{\chi}\wt{\mbf V}^{\chi\prime} \bm L=\bm L,\label{eq:id2}\\
&\text{rk}\l(\bm L^\prime \wt{\mbf V}^{\chi}\r)=r\label{eq:id3}.
\end{align}
It follows that $\bm L^\prime \wt{\mbf V}^{\chi}$ is invertible, and $( \bm L^\prime\wt{\mbf V}^{\chi}\wt{\mbf V}^{\chi\prime} \bm L)^{-1}=(\bm L^\prime\bm L)^{-1}$ is well defined. 
Therefore, $\wt{\bm{ R}}=O_{\mathrm P}(1)$ and $(\wt{\bm{ R}})^{-1}=O_{\mathrm P}(1)$.

Finally, by \eqref{eq:id1}, \eqref{eq:id2}, and \eqref{eq:id3}
\begin{align}
&\bm L\wt{\bm{ R}} = \bm L \l( \bm L^\prime\wt{\mbf V}^{\chi}\wt{\mbf V}^{\chi\prime} \bm L\r)^{-1} \bm L^\prime \wt{\mbf V}^{\chi} \wt{\bm{\mathcal S}} (\wt{\mbf M}^\chi)^{1/2}= \bm L \l( \bm L^\prime \bm L\r)^{-1} \bm L^\prime \wt{\mbf V}^{\chi} \wt{\bm{\mathcal S}} (\wt{\mbf M}^\chi)^{1/2}= \wt{\mbf V}^{\chi} \wt{\bm{\mathcal S}} (\wt{\mbf M}^\chi)^{1/2}=\bm\Lambda,\nn \\
&\wt{\bm{ R}}^{-1}\bm f_t  = (\wt{\mbf M}^\chi)^{-1/2}\wt{\bm{\mathcal S}} \l(\bm L^\prime \wt{\mbf V}^{\chi}\r)^{-1}\bm L^\prime\wt{\mbf V}^{\chi}\wt{\mbf V}^{\chi\prime} \bm L \bm f_t=(\wt{\mbf M}^\chi)^{-1/2}\wt{\bm{\mathcal S}}\wt{\mbf V}^{\chi\prime} \bm L \bm f_t=(\wt{\mbf M}^\chi)^{-1/2}\wt{\bm{\mathcal S}}\wt{\mbf V}^{\chi\prime} \bm\chi_t=\mbf F_t.\nn
\end{align}

\addcontentsline{toc}{section}{References}
\singlespacing
{\small{
\setlength{\bibsep}{.2cm}
\bibliographystyle{chicago}
\bibliography{BL_biblio}
}}
\end{document}